%% file: paper.tex
\documentclass[a4paper,fleqn,usenatbib]{mnras}

\usepackage{graphicx}	
\usepackage{amsmath}	
\usepackage{amssymb}	
\usepackage{multicol}        
\usepackage{bm}		
\usepackage{pdflscape}	
\usepackage{natbib}
\usepackage{amsfonts}
\usepackage{placeins}
\usepackage{array,longtable}		
\usepackage{geometry}
\usepackage{multirow, booktabs} 				
\usepackage{caption}
\usepackage{subcaption}
\usepackage{acronym}
\bibpunct{(}{)}{,}{a}{}{;} 
\usepackage{url}
\usepackage{float}
\usepackage[export]{adjustbox}

\usepackage[utf8]{inputenc}

\newcommand{\kms}{\,km\,s$^{-1}$} 
\newcommand{\Hi}{\textsc{Hi}}

\usepackage[T1]{fontenc}
\usepackage{ae,aecompl}

\usepackage{newtxtext,newtxmath}

\acrodef{ghasp}[\textsc{ghasp}]{Gassendi H-Alpha survey of Spirals}
\acrodef{sdss}[\textsc{sdss}]{Sloan Digital Sky Survey}
\acrodef{ned}[\textsc{ned}]{NASA/IPAC Extragalactic Database}

\title[GHASP Mass Models]{GHASP: an H$\alpha$ kinematical survey of spiral
galaxies - XI. Distribution of luminous and dark matter in spiral and irregular nearby galaxies using WISE photometry.}

\author[Korsaga et al.]{M. Korsaga$^{1,2}$\thanks{marie.korsaga@lam.fr},
C. Carignan$^{1,3}$,
P. Amram$^{2}$,
B. Epinat$^{2}$,
T. H. Jarrett$^{1}$
\\
$^{1}$Department of Astronomy, University of Cape Town, Private Bag X3, Rondebosch 7701, South Africa\\
$^{2}$Aix Marseille Univ, CNRS, LAM, Laboratoire d’Astrophysique de Marseille, Marseille, France\\
$^{3}$Observatoire d'Astrophysique de l'Universit\'{e} de Ouagadougou, BP 7021, Ouagadougou 03, Burkina Faso}

\pubyear{2017}

\begin{document}
\label{firstpage}
\pagerange{\pageref{firstpage}--\pageref{lastpage}}
\maketitle


\begin{abstract}

We present the mass distribution of a sample of 121 nearby galaxies with high quality optical velocity fields and available infra-red {\it{WISE}} 3.4 $\mu$m data. Contrary to previous studies, this sample covers all morphological types and is not biased toward late-type galaxies. These galaxies are part of the Fabry-Perot kinematical {\it{GHASP}} survey of spirals and irregular nearby galaxies. Combining the kinematical data to the {\it{WISE}} surface brightness data probing the emission from the old stellar population, we derive mass models allowing us to compare the luminous to the dark matter halo mass distribution in the optical regions of those galaxies. 
Dark matter (DM) models are constructed using the isothermal core profile and the Navarro-Frenk-White cuspy profile.
We allow the M/L of the baryonic disc to vary or we keep it fixed, constrained by stellar evolutionary models (WISE W$_1$-W$_2$ color) and we carry out best fit (BFM) and pseudo-isothermal maximum disc (MDM) models. We found that the MDM provides M/L values four times higher than the BFM, suggesting that disc components, on average, tend to be maximal.
The main results are: (i) the rotation curves of most galaxies are better fitted with core rather than cuspy profiles; (ii) the relation between the parameters of the DM and of the luminous matter components mostly depends on morphological types. More precisely, the distribution of the DM inside galaxies depends on whether or not the galaxy has a bulge.

\color{black}
\end{abstract}




\newpage

\section{Introduction}

Current $\Lambda$CDM models teach us that baryonic matter evolved within large dark matter (DM) halo cocoons. It is known since the beginning of the '70s  
\citep{Freeman+1970}, and '80s 
\citep{Bosma+1981} that, without the presence of this dark halo, rotation curves of galaxies should be rapidly decreasing after a peak located around 2.2 disc scale lengths of the baryonic matter located in an exponential disc. Except if we consider alternative models like MOND, which postulates that gravity deviates from a pure Newtonian force at very low acceleration, DM halos are thus mandatory to explain the shape of the rotation curves of galaxies, which remain usually more or less flat in the outer regions out to the edge of the \Hi\, disc. 

Different predictions are classically used to describe the shape of the central density profiles. One family of models supposes that the central core density of the halo is almost constant within the first kpc or so. Those models show that the DM halo density distribution can be described by an isothermal \citep{Carignan+1985} or pseudo-isothermal sphere \citep{vanAlbada+1985}, this last one being used to avoid a central divergence of the density observed when a singular isothermal sphere profile is used. The other family of models, coming from $\Lambda$CDM simulations \citep{Navarro+1996}, supports that DM halos should display cuspy central density distributions. Unfortunately, the sensitivity of the rotation curves to the exact density profile of the halos is quite low and one must use the highest sensitivity and the highest resolution possible to arrive at useful comparisons \citep{Ouellette+1999}.

Understanding the formation and evolution of galaxies requires a fine knowledge of the nowadays luminous and dark matter distribution within local galaxies. One problem in the debate between the cuspy or core density profiles is that numerical simulations predict the shape of the halo density profile at the time of formation while galaxies are observed after many Gyr of evolution. Internal dynamics and interaction between the DM halo, the luminous disc and the environment may have modified the shape of the halo density profile. Despite this debate, studies using the two predictions produced acceptable  results about the distribution of the DM halo of galaxies \citep[e.g,][]{Ouellette+2001,Salucci+2001,Block+2002,Gentile+2004,Spano+2008}. 

\citet{Kormendy+2004} suggest that there are clear scaling laws between the DM halos parameters and the optical parameters (e.g. $M_B$) of the parent galaxies. For example, as suggested in previous studies \citep[e.g,][]{Carignan+1988}, halos in less luminous galaxies tend to have smaller core radius $r_0$ and higher central density $\rho_0$. 

To disentangle the central mass distribution of the DM halos, high spatial resolution rotation curves are requested in the inner disc. Such data are commonly provided by H$\alpha$ observations. This has been done by several authors \citep[e.g.][]{Ouellette+1999,Ouellette+2001,Spano+2008}. In the last few years, two different approaches have been used to try to discriminate between those two families of DM halos. 
The first one by \citet{Bottema+2015} was to limit the study to a small number of very well determined rotation curves, with very well known distances and spanning an as broad range of luminosities as possible ($\sim$ 2.5 orders of magnitude). 
For their observed high quality and extended rotation curves, four different schemes, each with two free parameters, were used. For the ``maximum disc fits'', the authors found that the rotation curves could be well matched, although this required a large scatter in the M/L values of the individual galaxies in their sample. 
Interestingly, they found that the rotation curves of low mass galaxies could not be reproduced when assuming the Navarro-Frenk-White cuspy profile, implying that one should prefer cores over cusps. They used a combination of the pseudo-isothermal sphere with a universal M/L ratio to successfully describe the observed rotation curves, leading to ``sub maximum disc'' mass contributions. Similar conclusions had been reached a year before by \citet{Toky+2014} using 3.6 $\mu$m mid-IR photometry instead of R band data. 

The second approach of \citet{Lelli+2016} was to use a large sample of 175 nearby galaxies (the SPARC sample) from previous \Hi\, and H$\alpha$ kinematical studies, combined with 3.6 $\mu$m (Spitzer) photometry. Using a unique M/L ratio of 0.5 for the whole sample, their main result is that the disc component is nearly maximal for high-mass, high-surface-brightness galaxies while it contributes less to the total mass in low-mass, low-surface-brightness galaxies. A mean value of M/L $\gtrsim$ 0.7 is statistically ruled out because it implies an over-maximum baryonic contribution in high surface brightness and high mass galaxies.This study gives a good idea of what to expect from a sample with well resolved rotation curves in the inner parts and with stellar mass distribution well traced using mid-infrared bands. Following the study of \citet{Lelli+2016}, \citet{Katz+2017} used almost the same SPARC sample to study the inner density profile dependence on the ratio of stellar to DM mass. They confronted the SPARC data to cosmological N-body simulations \citep{DiCintio+2014a,DiCintio+2014b} and found an agreement between DM halo profiles modified by baryonic processes and the $\Lambda$-Cold Dark Matter cosmology. They also showed that these halo profiles describe galaxy rotation curves better than halo models which do not account for baryonic physics.  

\citet{Bottema+2015} used a sample of 12 galaxies for which three out of four H$\alpha$ rotation curves have been computed from 2D H$\alpha$ velocity fields. 
 \citet{Kormendy+2004} used a sample of 55 galaxies taken from the literature, extending results already published in other conference reviews.
\citet{Lelli+2016} used a sample of 175 galaxies for which 15 out of 56  H$\alpha$ rotation curves have been computed from 2D H$\alpha$ velocity fields. The last relatively large sample (36 galaxies) using 2D kinematics was studied in \citet{Spano+2008}.  However, most of those studies \citep[e.g,][]{Kormendy+2004,Bottema+2015,Spano+2008} were focused on late rather than early morphological types galaxies. 
In the present work, we study mass models for a sample of 121 galaxies using high resolution H$\alpha$ rotation curves  computed from  2D H$\alpha$ velocity fields.  The sample covers morphological types ranging from Sa to Im types and the stellar component is modelled using mid-infrared photometry.   

This paper is organised as follows. 
In section \ref{sect:sample}, we present the sample and the data used to constrain mass models.
Section \ref{sect:photometry} describes the decomposition of the infrared luminosity profiles.
In section \ref{sect:model} the mass models are presented and the results of the fits are described in section \ref{sect:results}.
Section \ref{sect:discussion} presents the discussion of the DM halo parameters and section \ref{sect:conclusion} gives a summary and the general conclusions.
Appendix \ref{appendixA} lists the global properties and mass models parameters for 8 galaxies. The complete catalogue is available online.
In Appendix \ref{appendixB}, we present the photometry, the luminosity profiles and the mass models of two galaxies. The remaining sample is available online. In Appendix \ref{appendixC}, we compare the mass models derived from \Hi\, and optical rotation curves limited to the solid body rising parts for 2 out of 15 galaxies \citep[taken from][]{Toky+2014}; the remaining mass models are available online. 
Throughout this paper, we assume a Hubble constant H$_0 = 75$~km~s$^{-1}$~Mpc$^{-1}$.

\section{Data and Sample Selection}
\label{sect:sample}

We aim at studying a large galaxy sample, spanning a wide range in mass and morphological types, for which accurate constraints can be provided on mass models in the inner regions of the galaxies. We therefore need accurate optical rotation curves and photometry representative of the bulk of the stellar mass.

\subsection{Data}

Our parent sample is the {\it{GHASP}} (acronym for Gassendi HAlpha survey of SPirals) sample, which contains 203 spiral and irregular galaxies observed with the 1.93m telescope of the Observatoire de Haute Provence (France).
The H$\alpha$ line was scanned with a Fabry-Perot (FP) interferometer in photon-counting mode providing both high spectral and spatial resolution 3-D data cubes over a large field of view ($\sim 6\times 6$~arcmin$^2$). The typical spatial resolution is around 2 arcsec full width at half maximum and the spectral resolution is around 10000 with a spectral sampling corresponding to $\sim 15$~km~s$^{-1}$.
Velocity fields were derived from these data cubes before extracting $\sim 5$~km~s$^{-1}$ accurate rotation curves from the full 2D information for 170 galaxies.
Those data have been published in several papers \citep{Garrido+2002,Garrido+2003,Garrido+2004,Garrido+2005,Epinat+2008b,Epinat+2008a}.
In the present paper, we use the rotation curves published in \citet{Epinat+2008a}, who also provide a global view on the full GHASP sample. Those rotation curves were extracted using the method detailed in \citet{Epinat+2008b}.

We use photometry from the Wide-field Infrared Survey Explorer (WISE). WISE is a NASA-funded Medium-Class Explorer mission consisting in a 40 cm primary-mirror. This space infrared telescope mapped the entire sky in the following mid-infrared bands: W$_1$ (3.4 $\mu$m), W$_2$ (4.6 $\mu$m), W$_3$ (12 $\mu$m) and W$_4$ (22 $\mu$m). The science instrumentation includes a 1024 $ \times $ 1024 pixels Si:As and HgCdTe array and each band covers a field-of-view of 47 $\times$ 47 arcmin$^2$ with an angular resolution of 6 arcsec in the short bandpass and 12 arcsec in the longest one \citep{Jarrett+2012}. In the present study, we use the W$_1$ band luminosity profiles to describe the baryonic component in our mass models (cf. section \ref{subsect:stellarpop}).

\subsection{Sample Selection}

We use a sample of 121 spiral and irregular nearby galaxies for which accurate H$\alpha$ rotation curves, derived from 2D velocity fields, and WISE infrared surface photometry were available.

From the whole GHASP sample of 203 galaxies, 147 galaxies have both available H$\alpha$ rotation curves and WISE photometric data in the W$_1$-band (3.4 $\mu$m).
A quality flag was attributed to each galaxy to describe the quality of the rotation curve: 1 for very high, 2 for high and 3 for low quality. This flag is reported in column  (14) of Table \ref{tab:photometry}. In order to compute mass models with good constraints, we restricted our analysis to galaxies with flags 1 or 2, reducing the sample to 121 galaxies. 
In brief, we excluded 26 galaxies having ill-defined rotation curves due to various reasons: existence of a strong bar, strong signs of galaxy interaction, too high inclination (>75$^o$) or another peculiarity (e.g. strong lopsidedness) and/or very low SNR data.

\begin{figure*}
	\hspace*{-0.85cm} \includegraphics[width=5.87cm]{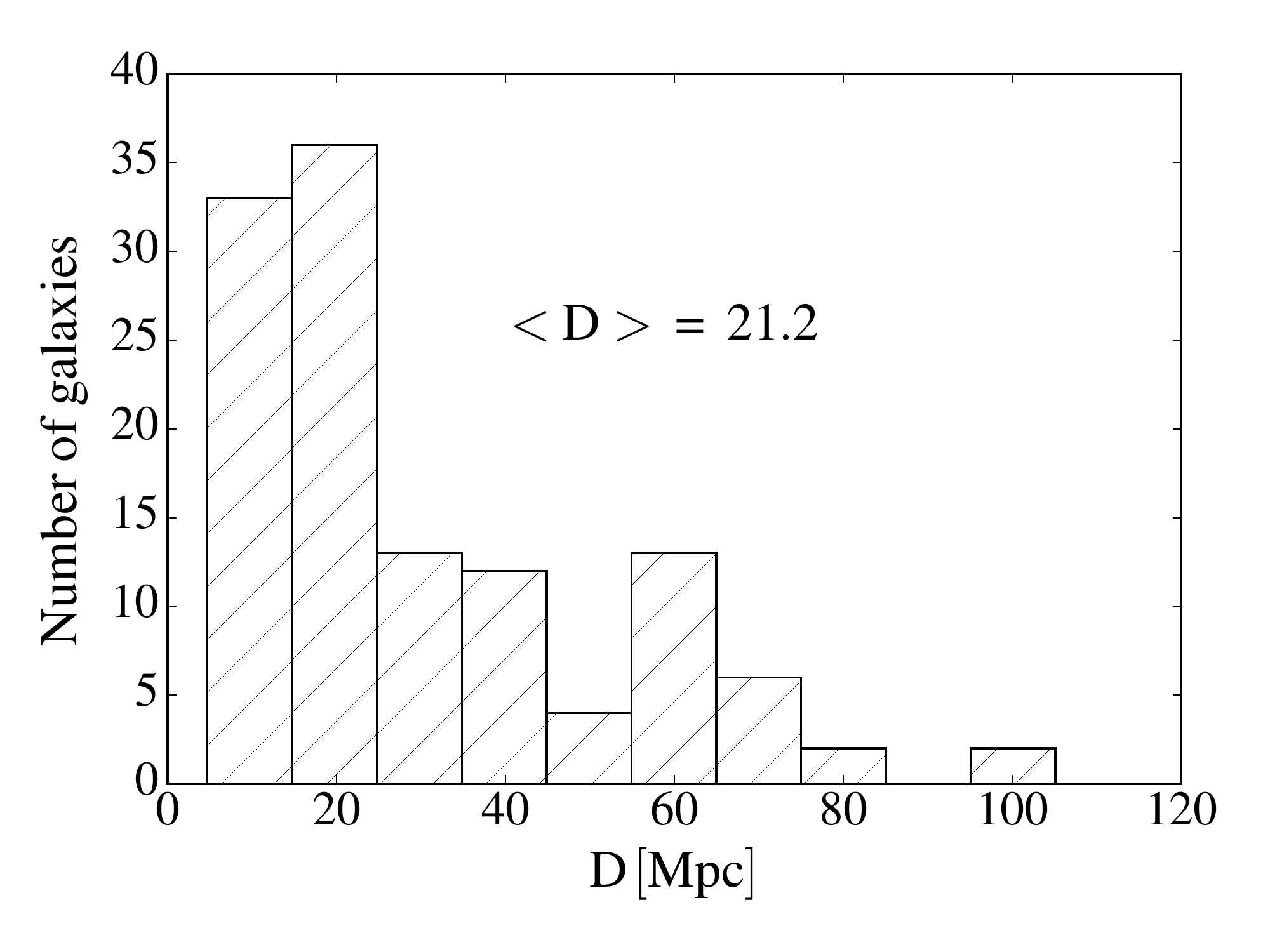}
	\hspace*{-0.35cm} \includegraphics[width=5.87cm]{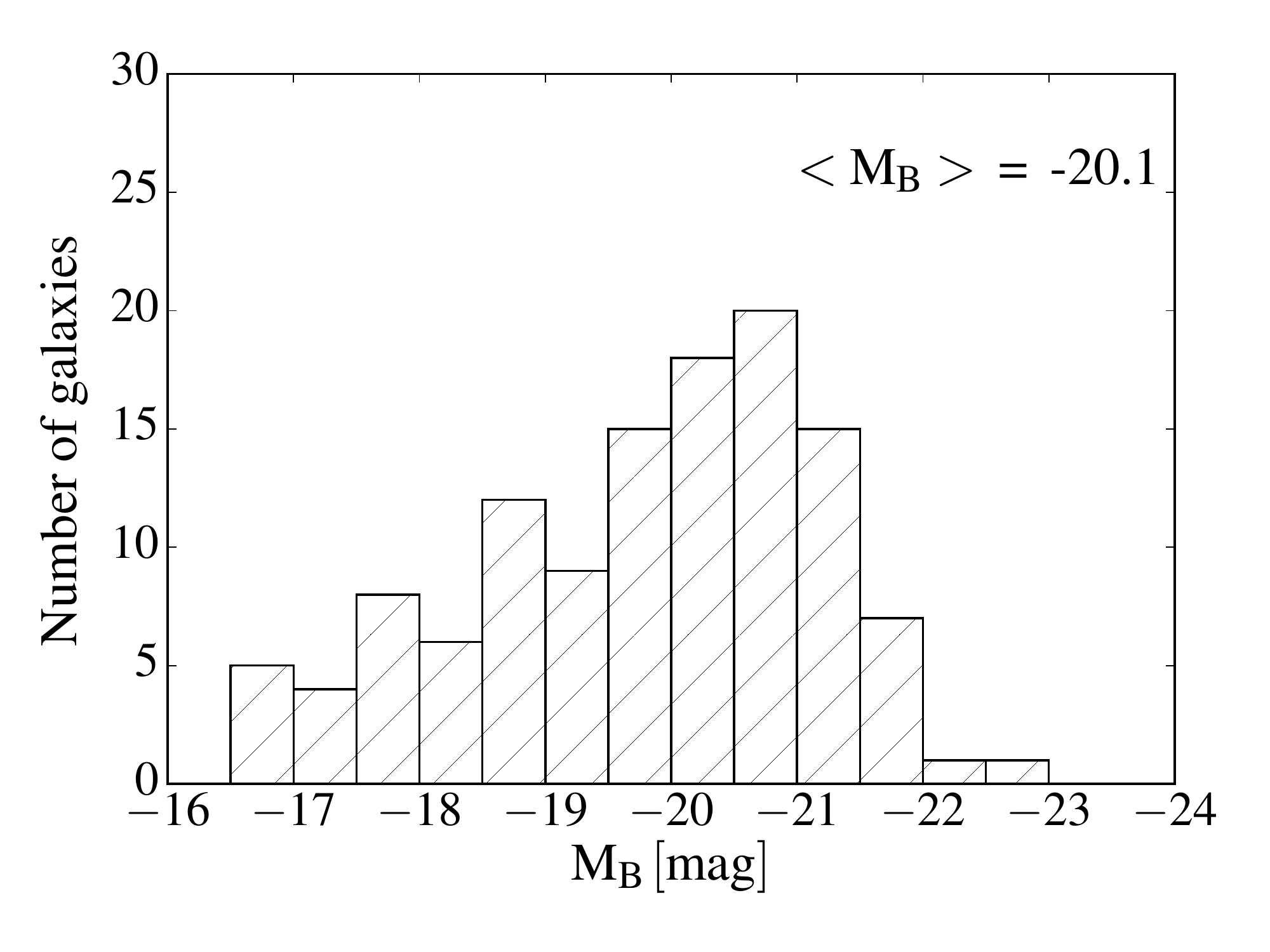}
	\hspace*{-0.35cm} \includegraphics[width=5.87cm]{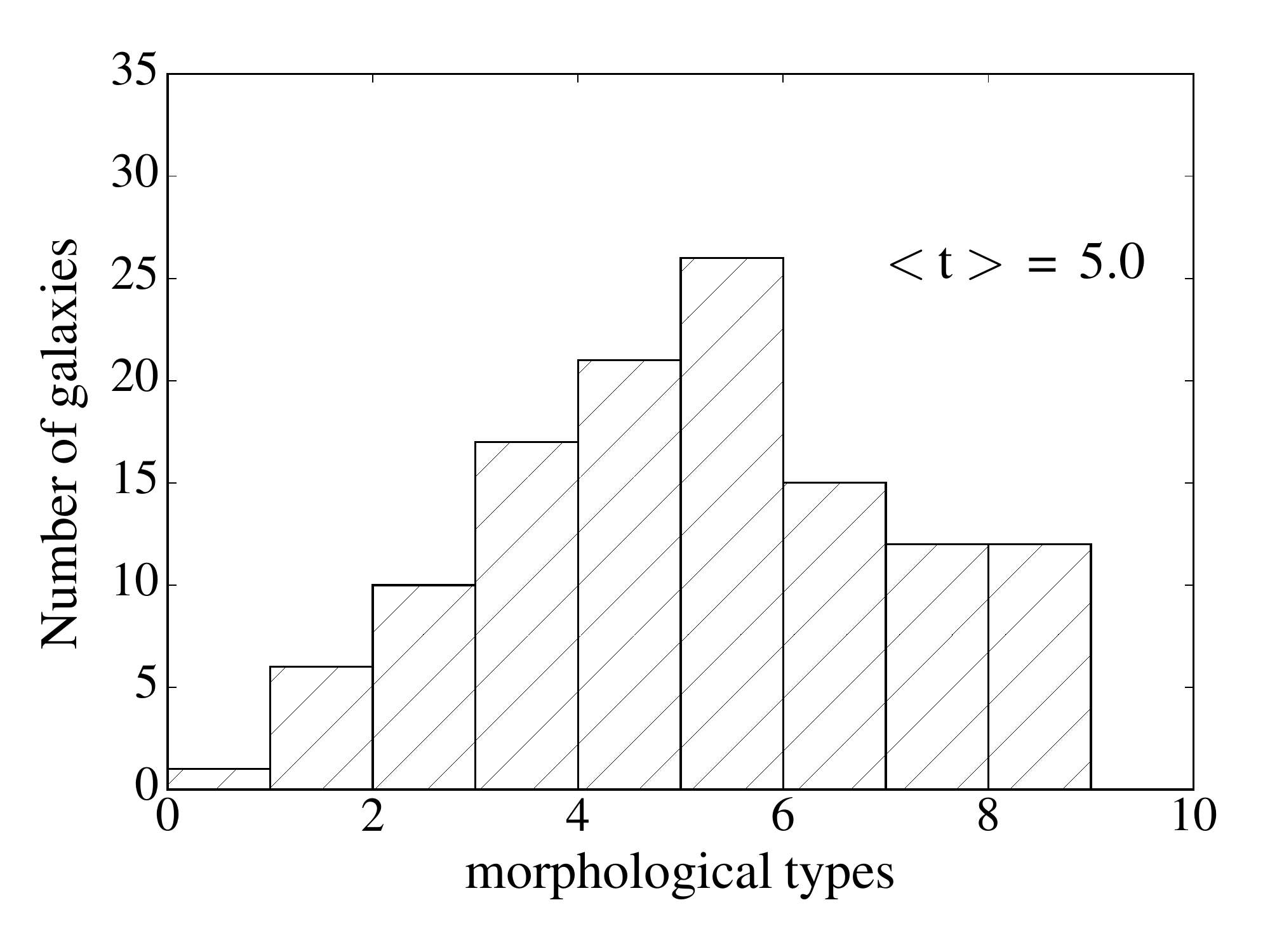}
	\hspace*{-0.85cm} \includegraphics[width=5.87cm]{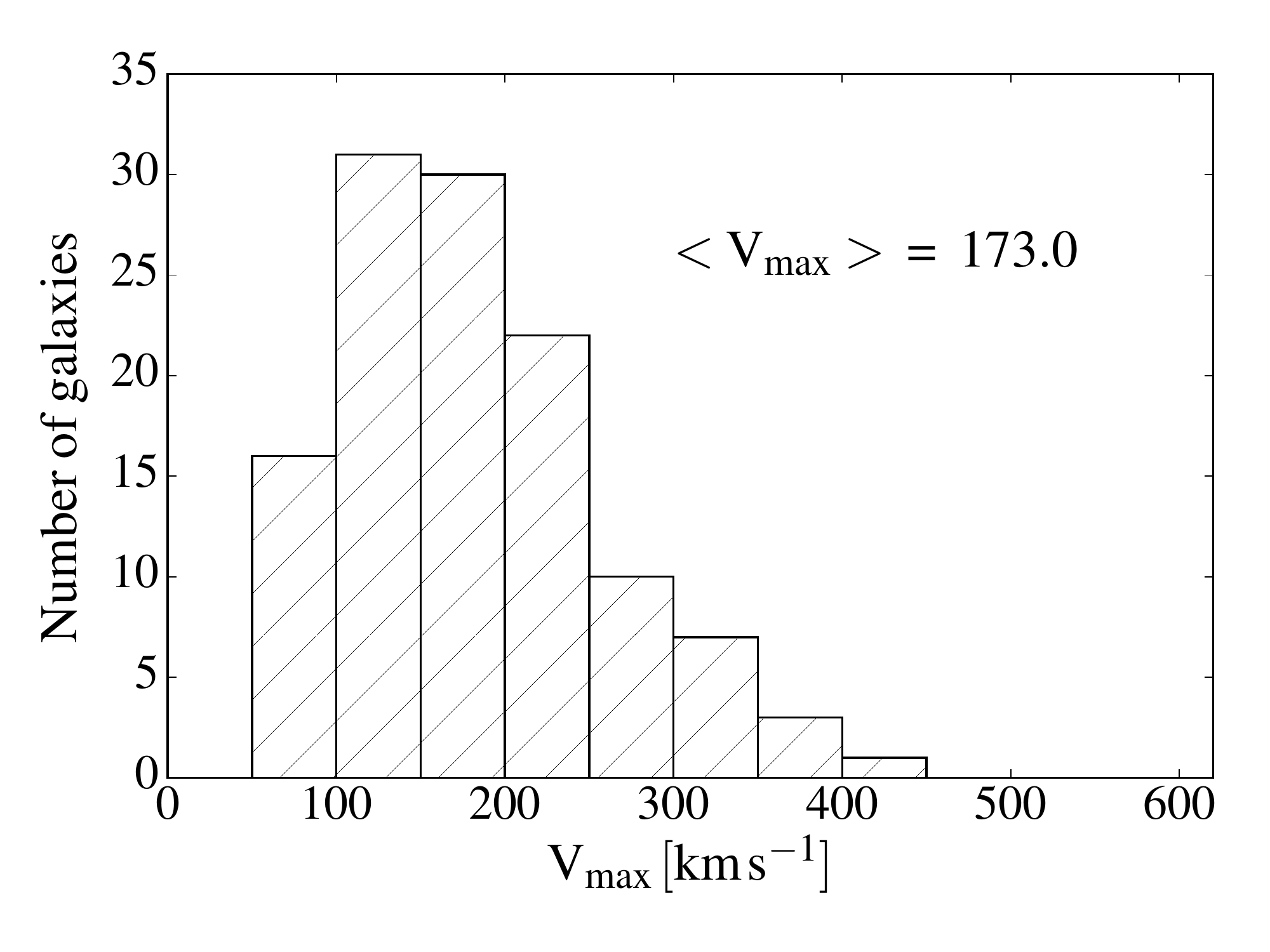}
	\hspace*{-0.35cm} \includegraphics[width=5.87cm]{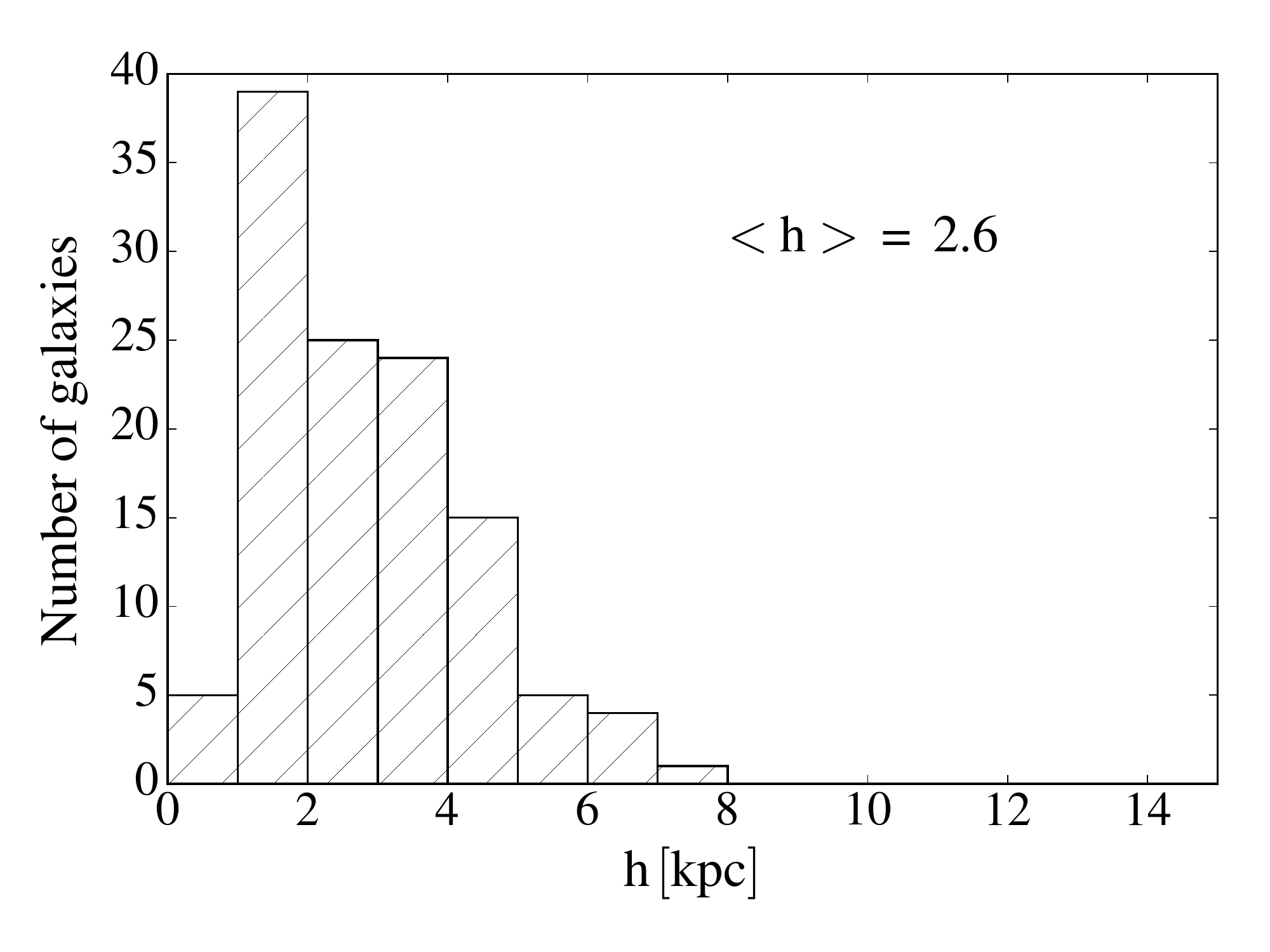}
	\hspace*{-0.35cm} \includegraphics[width=5.87cm]{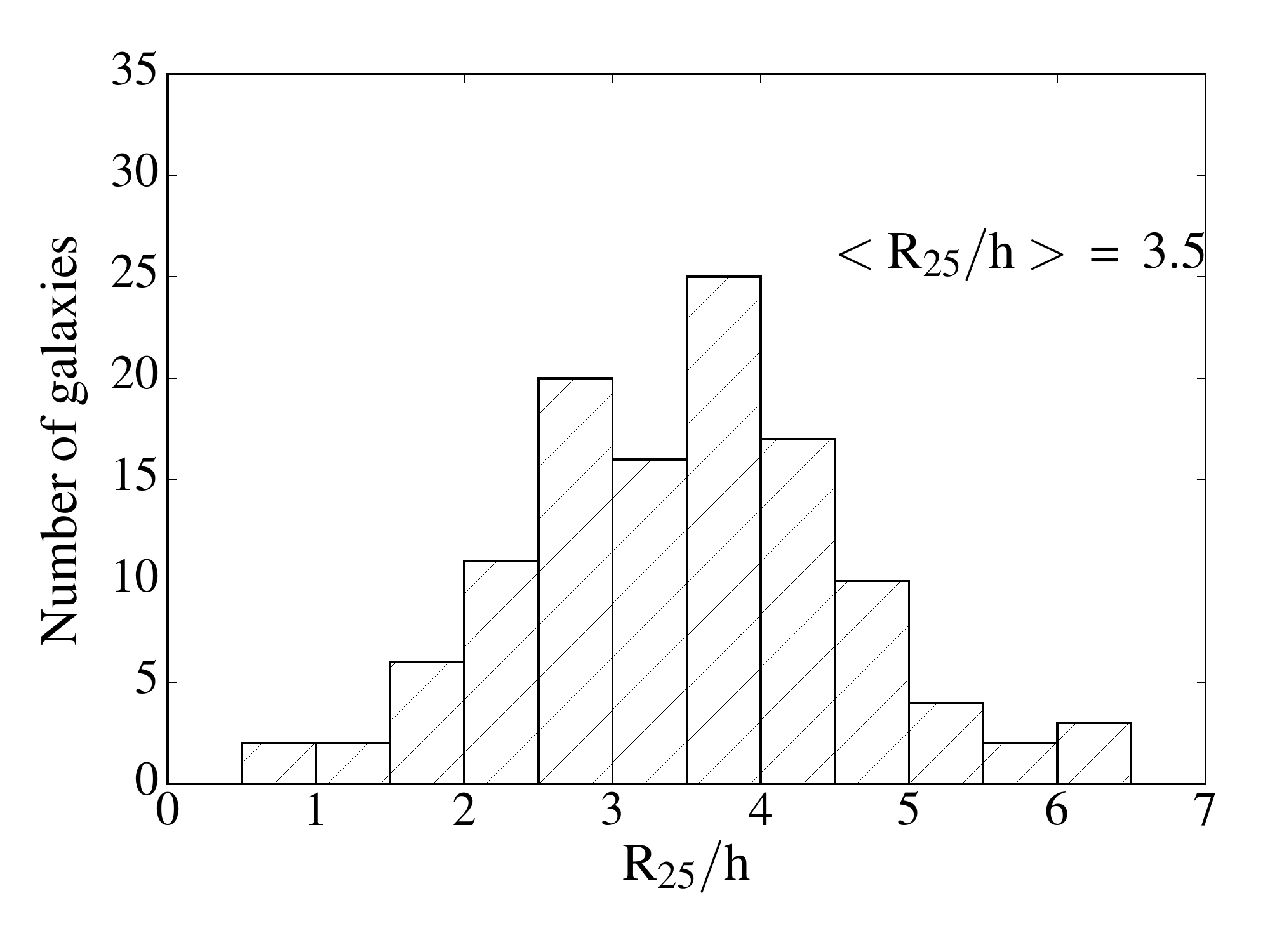}
\caption{Properties of our sample. First line; from left to right, we have respectively the distance of the galaxies, the absolute magnitude and the morphological type. Second line; from left to right, we have respectively the maximum rotational velocity, the disc scale length and the ratio of the isophotal radius R$_{25}$ to the disc scale length. The median of each parameter is shown in the middle of each plot.}
\label{fig:histogram}
\end{figure*}

The final sample of 121 galaxies we defined is not complete in volume but it samples the main galaxies parameters space.
Indeed, as illustrated by Fig. \ref{fig:histogram}, it samples {{\it (i)}} nearby galaxies (from 5 to 100 Mpc with a median distance of 21.2 Mpc); {{\it (ii)}} $M_B$ absolute blue magnitude ranging from -16 to -23 (with a median value of -20.1); {{\it (iii)}} morphological types from Sa to Irregular (the median value of t is 5.0, i.e. $Sc$ galaxies); {{\it (iv)}} maximum rotation velocity from 50 to 400 \kms\, (the median velocity is 173.0 \kms, a tail in the distribution is observed toward high mass galaxies), i.e. dynamical masses ranging from 0.5 to 2.5 10$^{11}\  M_{\odot}$ and finally {{\it (v)}} galaxy sizes, represented by their optical diameters R$_{25}$, which corresponds to the isophotal radius at the limiting surface brightness of 25 mag arcsec$^{-2}$, or by their disc scale lengths, ranging from 1 to 10 kpc (median value of 2.6 kpc). The disc scale lengths are discussed in section \ref{subsect:DBP}. The distances and velocities used in this work are taken from \citet{Epinat+2008b}. 

\section{The WISE surface brightness photometry}
\label{sect:photometry}

\subsection{Stellar population in the near/mid infrared}
\label{subsect:stellarpop}

According to observations and stellar population models, the light coming from discs is usually dominated by Population I stars (blue and young stars) and bulges by Population II stars (red and old stars) but the majority of the stellar mass is provided by red stars, both from the disc (old Pop I) and bulge, better seen in the near and mid-infrared bands. Those bands map the bulk of stellar masses and thus provide the most representative stellar mass distribution.
We therefore use mid-infrared bands to model the stellar populations.

Luminosity profiles were extracted from the WISE photometry \citep{Jarrett+2013}. Optimal profile Point Spread Function fitting has been used to measure the fluxes of stars and the calibration zero-point magnitude is derived from the measurements of calibration stars \citep{Jarrett+2013}. The adopted 1 $\sigma$ elliptical isophotal radius corresponding to the 1 $\sigma$ isophotal surface brightness is equal to 23.0  mag arcsec$^{-2}$ (Vega; and 25.7 mag arcsec$^{-2}$ in AB) for W$_1$ (as shown in Fig. \ref{massmodel}) and to 21.8 mag arcsec$^{-2}$ for W$_2$ \citep{Jarrett+2013}. Because W$_1$ is the shortest wavelength band of WISE and the most sensitive for almost all types of galaxies, yielding the largest isophotal radius, we use the 3.4 $\mu$m luminosity profile to describe the baryonic component in our mass models. The central surface brightness of the luminosity profiles are given in Table \ref{tab:photometry}, column (4).

\subsection{Photometric mass-to-light ratio}
\label{subsect:masstolight}

The WISE photometry can be used to infer the mass-to-light (M/L) ratio of galaxies in our sample.
Indeed, infrared bands are the best suited to constrain the values of the M/L ratio of the luminous matter. In addition, population synthesis models show that the relation between light and mass is more nearly constant in the near/mid-infrared than in the UV and in the optical \citep{deDenus-Baillargeon+2013}. Indeed, recent star formation mainly visible in the UV and in the optical produces large amount of light from little mass during short times, which impacts the M/L ratio of a galaxy. This is also observed in Tully-Fisher relations for which the scatter decreases from blue to near infrared bands \citep[e.g.,][]{Verheijen+2001}.

For each galaxy, the stellar M/L ratio can be calculated as a function of the mid-infrared color using the following relation from \citet{Cluver+2014}:
\begin{equation}
\log(M_{\rm stellar}/L_{W1}) = -2.54 (W_1-W_2) - 0.17
\end{equation}
where $L_{W1}$ is the inband luminosity in the W$_1$ band and where W$_1$ and W$_2$ correspond to magnitudes in the W$_1$ and W$_2$ bands respectively. W$_1$-W$_2$ corresponds to the rest-frame color of the source and ranges between -0.1 and 0.3 in our sample.
We will refer to this M/L ratio as the photometric M/L ratio and we will note it fixed M/L.

\subsection{Radial Profile Decompositions}
\label{Radial Profile Decompositions}

Since we aim at studying the inner distribution of dark matter halos, we need to constrain the contribution of stars to the dynamical potential in the inner parts of galaxies with care.

Among the 121 galaxies included in our final sample, the median morphological type is Sc-type galaxies, i.e. that 52 galaxies have a morphological type earlier than Sc, 29 are Sc's and 40 are of later types than Sc. Galaxies with a morphological type earlier than or equal to Sc potentially display a significant bulge in their radial profiles. Indeed 81 galaxies requested a decomposition into multiple components while the remaining 40 did not. Nevertheless we did not decide from the morphological type if the profile requested or not a decomposition but instead by visual inspection of each profile not to be subject to misclassification, which could happen for the most distant objects. 

To transform the radial light profiles into radial mass profiles we need to disentangle the different components because their distribution, hence their impact on circular velocities, is different.
Indeed, material in a plane (gas and stellar disc) globally follows circular orbits in the plane of the disc while material in spherical-like structures (stellar bulge) displays random motions.
The geometry of the mass distribution has an impact on the kinematics which has to be taken into account in mass models.
For instance, for a given mass, at a given radius, the rotational velocity expected for a flat component is higher than the one expected for a spherical component of the same mass. Even if the material (stars or gas) orbiting along bars, within spiral structures or in rings presents strong deviations from circular motions, that should be taken into account. To simplify the model and to minimise the number of free parameters, we only disentangle spherical components (bulges) from planar components (disc, bar, spiral arms, rings, etc.) that we call disc in the mass models.  

\begin{figure}
\begin{center}
	\includegraphics[width=8.5cm]{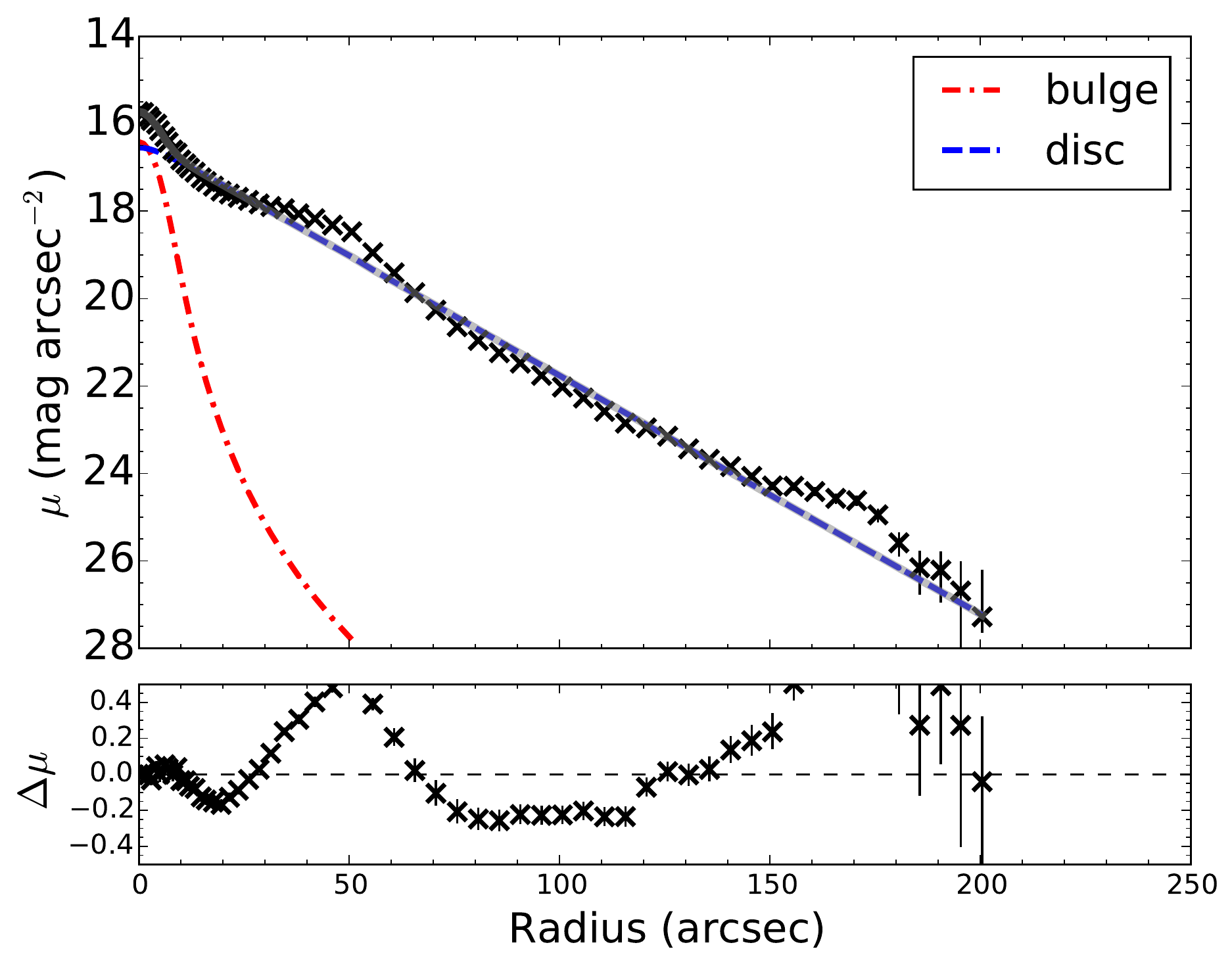}
	\includegraphics[width=8.5cm]{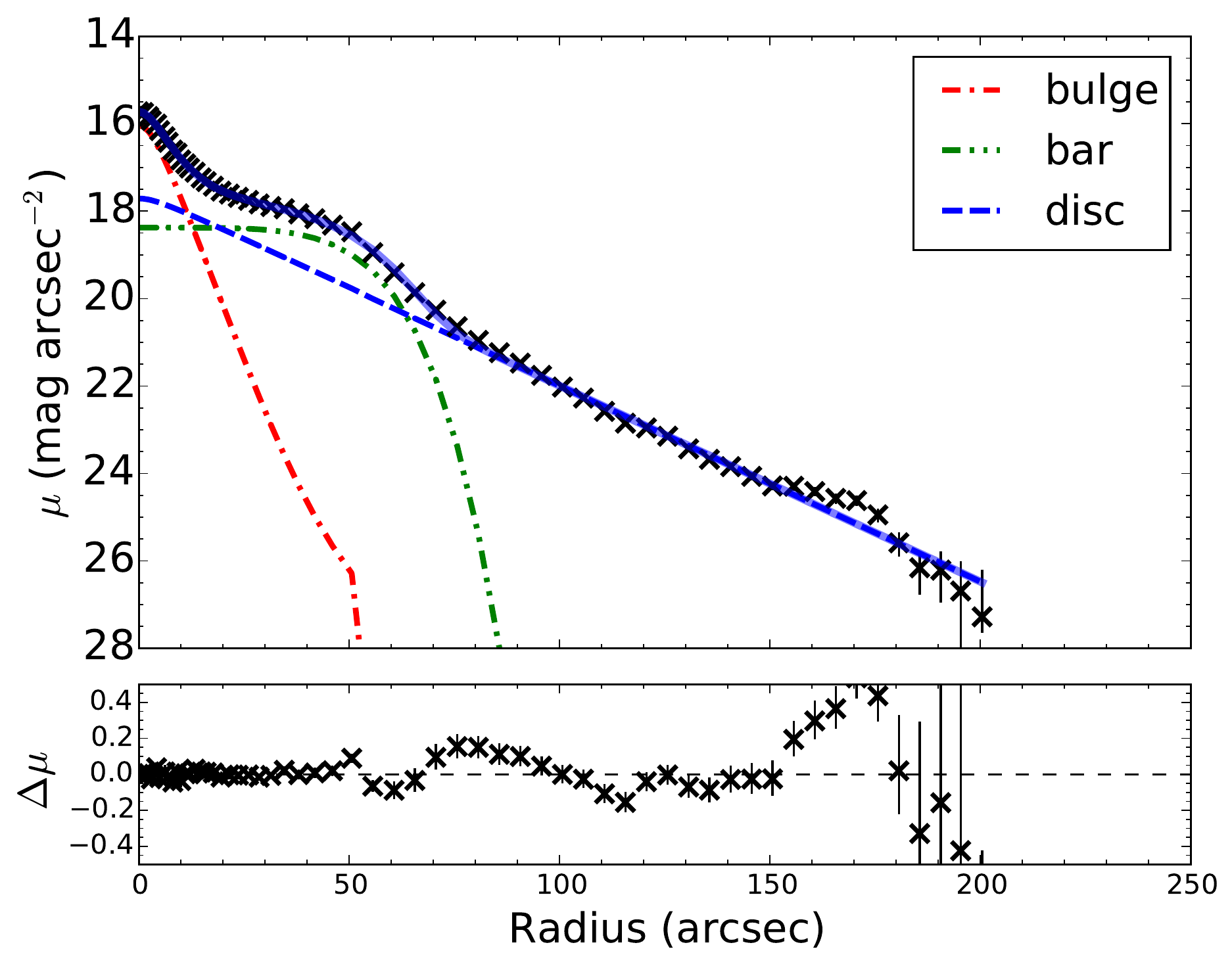}
\caption{Example of the structural decomposition of UGC 8900 in the W$_1$ band without including a bar component (top panel) and including a bar component (bottom panel). The observed surface brightness profile in plotted using black cross and the different components bulge, disc, bar (bottom panel) in red, blue and green dash lines respectively. The lower subpanel represents the fitting residuals, which are obviously reduced when a bar is included in the model.}
\label{fig:decomposition}
\end{center}
\end{figure}

In principle, the subtraction of the bulge component from the observed profile should provide a bulge-free profile. In practice, it only works if the galaxy does not contain any additional structure such as a bar and/or a ring.  If the profile contains a bar or a ring, this method fails and produces strong residuals in the very center. Thus, in order to minimise the residuals values in the central regions, we have to proceed to a full decomposition in modelling all the components together to end up with the different components used for the mass models. It is illustrated by Fig. \ref{fig:decomposition} for the galaxy UGC 8900 where strong residuals are visible when a bulge only is subtracted (top panel), while residuals within the uncertainties are obtained when the bar and the bulge have been modelled together. However, for galaxies with very little or no bulge (specifically late types or irregular galaxies), the luminosity profiles were not decomposed, the disc component is then directly the observed surface brightness profile. Furthermore, we used a parametric profile fitting procedure to separate the photometric components: bulges, discs, bars, arms and rings. 

The surface brightness profile were decomposed by using the method described in \citet{Barbosa+2015}.

The disc is modelled with an exponential profile defined by:
\begin{equation}
I_d( r ) = I_0 \exp{(-r/\rm h)}
\label{equation1}
\end{equation}
with $I_0$ the central intensity of the disc and h the disc scale length.
The other components like bulges, bars, rings, lenses are determined by using a S\'{e}rsic function given by :
\begin{equation}
I{_b}( r ) = I{_e} \exp{ \left( -b{_n} \left[(r/r_e)^{1/n} - 1\right]\right)}
\end{equation}
where $r_e$ is the effective radius, $I_e$ is the intensity at the effective radius, $n$ is the S\'{e}rsic shape parameter and $b_n$ a S\'{e}rsic index function due to the parametrization of the function at the effective radius. The total apparent magnitude is :
\begin{equation}
m_{\rm sersic} = -2.5 \log{\left(\frac{2\pi I_e\, r_e^2 \, e^{b_n} \, n \, \Gamma(2n)}{b_n^{2n}} \frac{b}{a}\right)}
\end{equation}
with $\Gamma(x)$ the complete gamma function of a variable $x$.

Examples of parameters of the bulge and disc components are shown in Table \ref{tab:photometry}.  This table represents only a fraction of our sample. The full table is provided online. The figures showing the disc and bulge components for the whole sample are also displayed online (see an example on the right top panel of Fig. \ref{massmodel}). 

\subsection{Disc and Bulge Parameters}
\label{subsect:DBP}

\begin{figure}
	\vspace*{-0.0cm}\includegraphics[width=7.0cm]{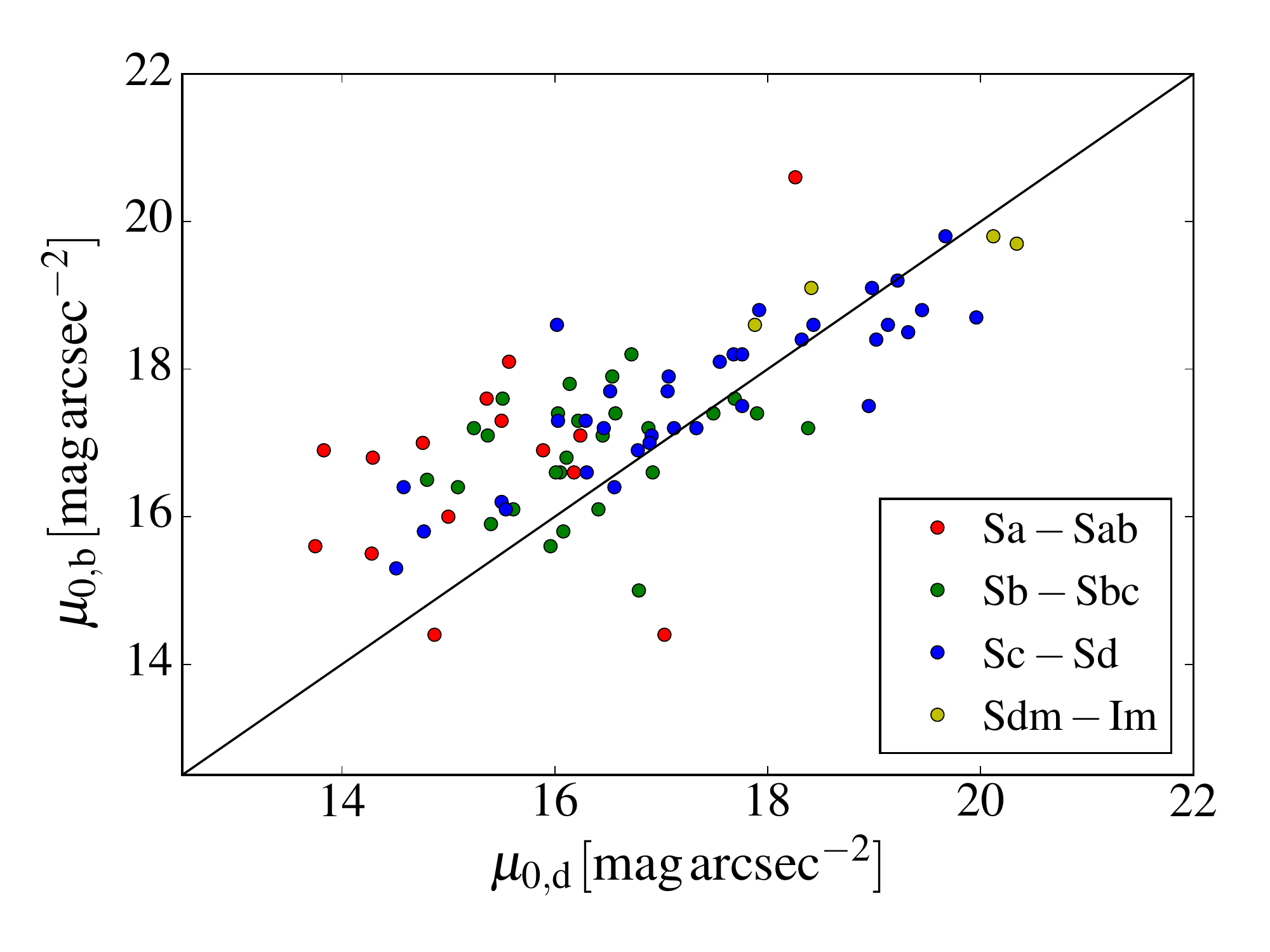}
	\vspace*{-0.0cm}\includegraphics[width=7.0cm]{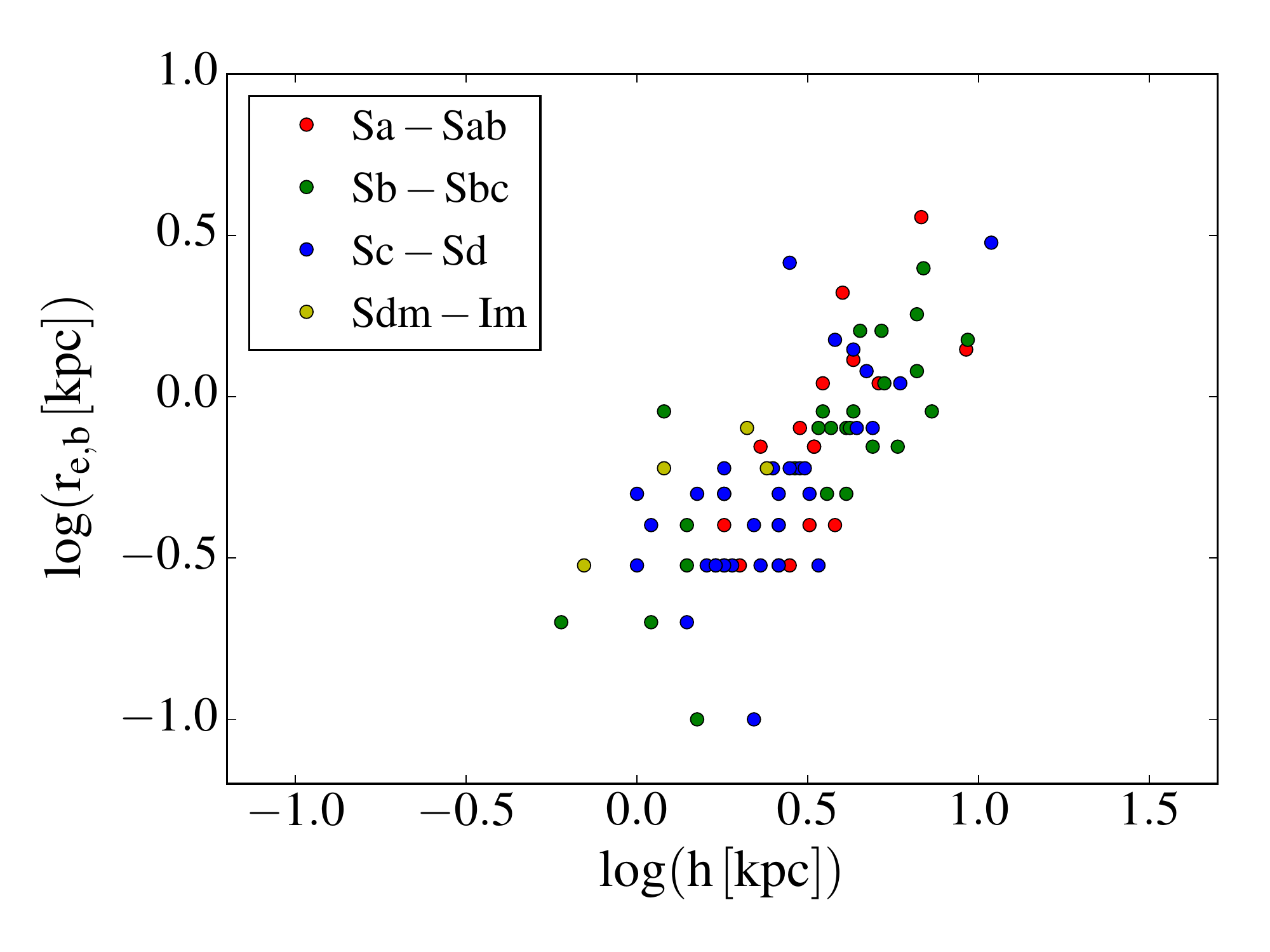}
	\vspace*{-0.0cm}\includegraphics[width=7.0cm]{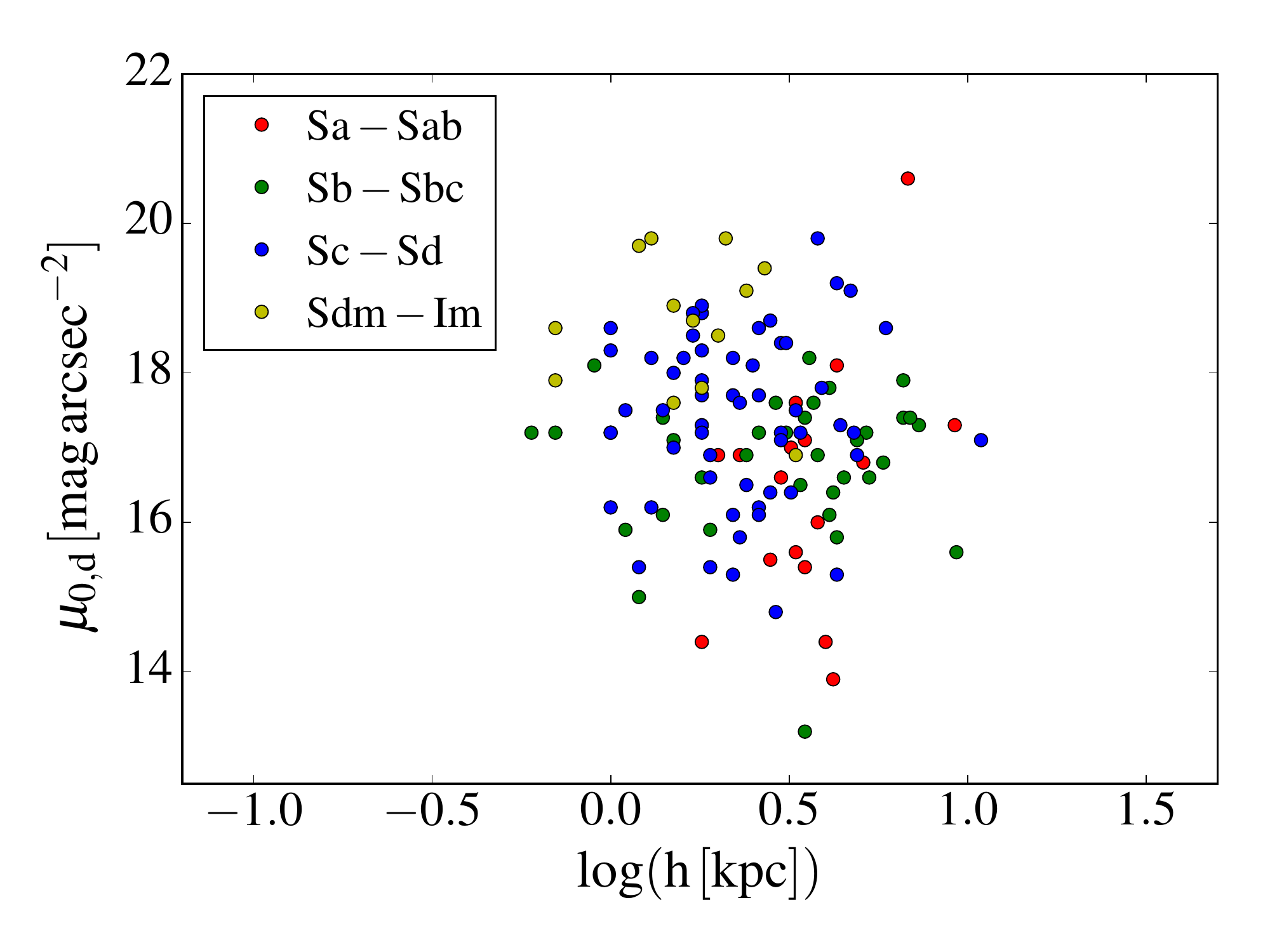}
	\vspace*{-0.3cm}\includegraphics[width=7.2cm]{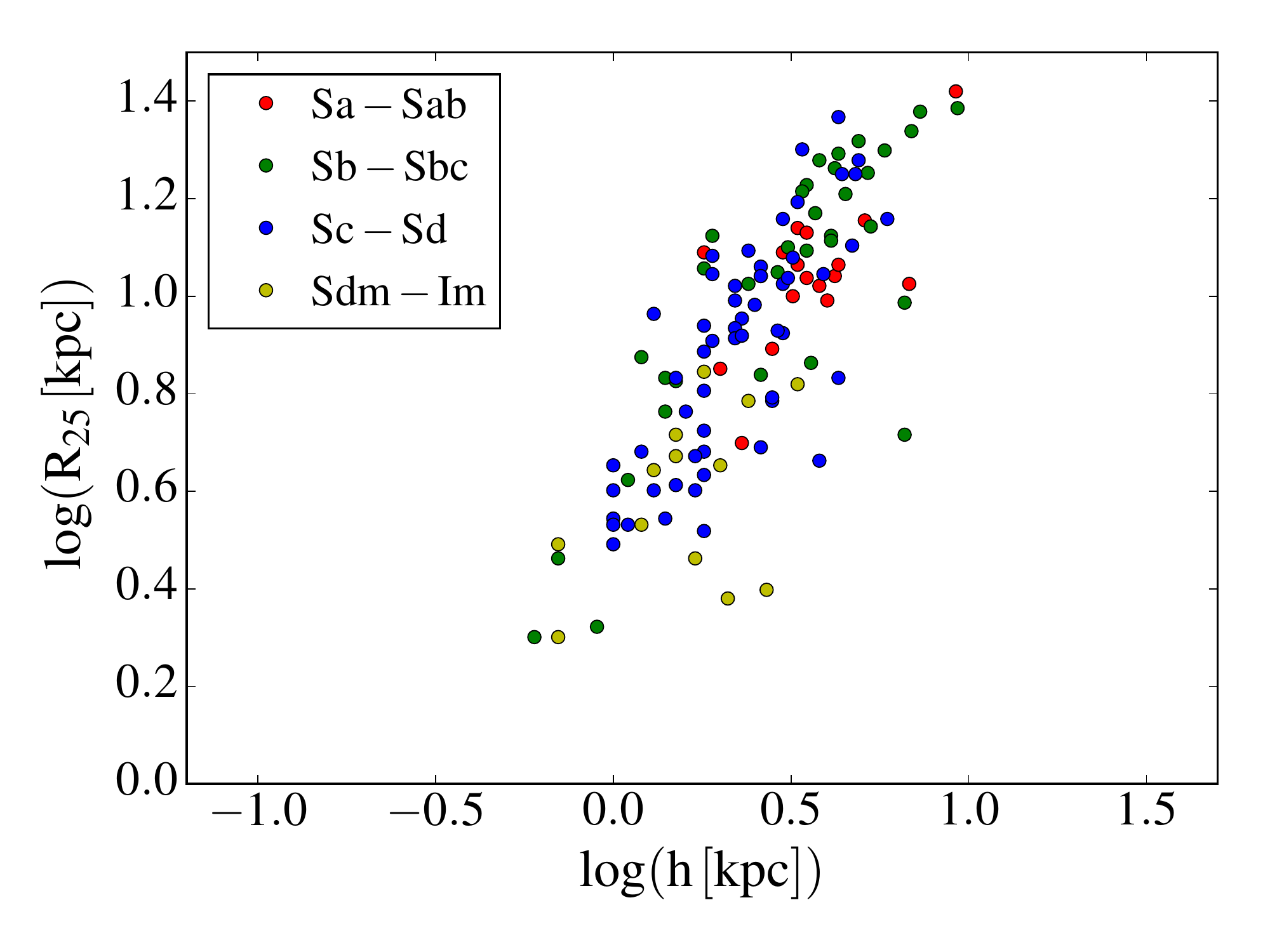}
\caption{Correlations between parameters derived from the W1 profiles. The top panel shows the bulge central surface brightness versus the disc central surface brightness. The two middle panels represent the bulge effective radius and the disc central surface brightness versus the disc scale length respectively. The bottom panel shows the isophotal radius R$_{25}$ versus the disc scale length.}
\label{fig:photometry}
\end{figure}

\begin{figure}
	\vspace*{-0.0cm}\includegraphics[width=7.0cm]{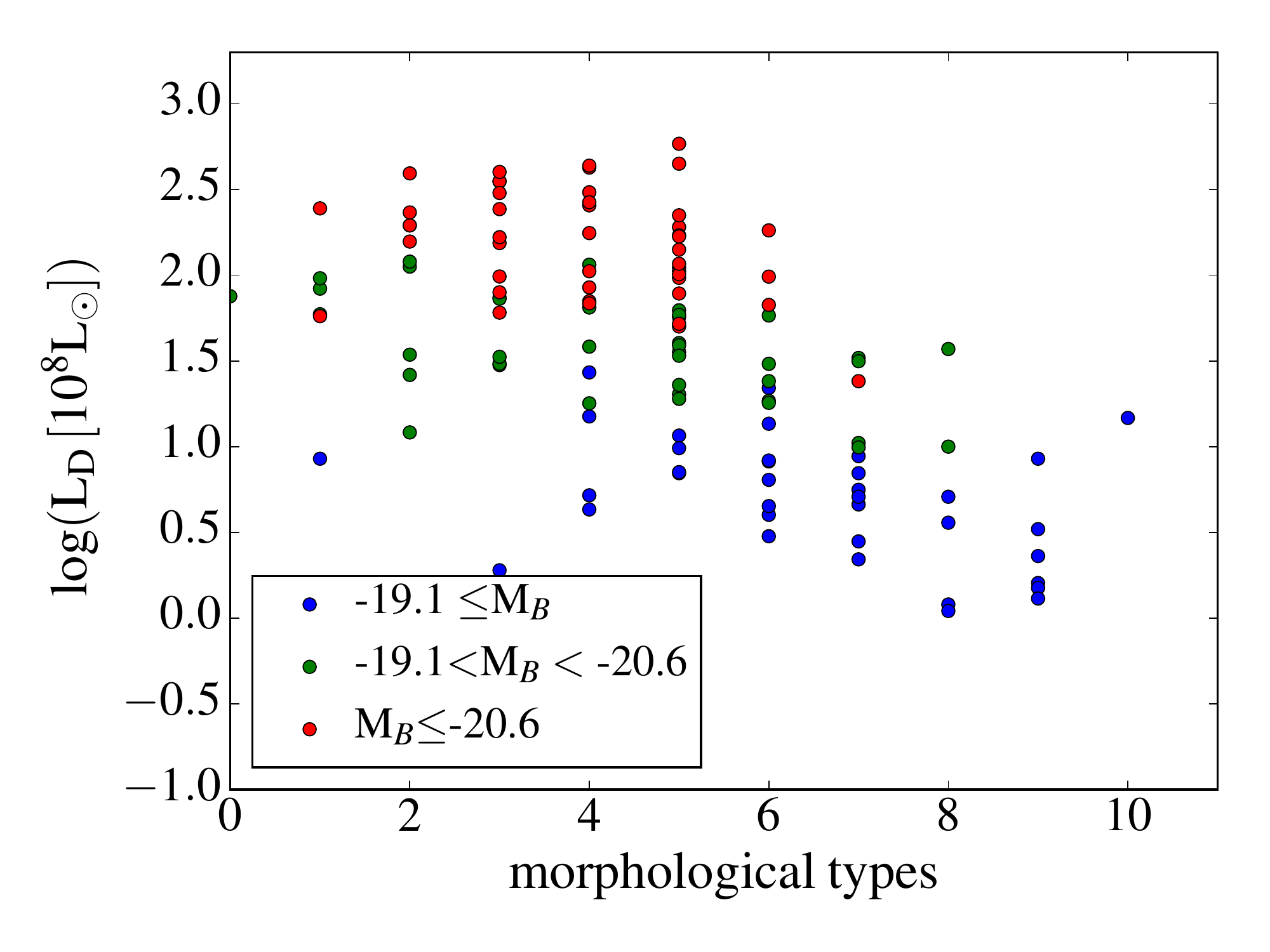}
	\vspace*{-0.3cm}\includegraphics[width=7.0cm]{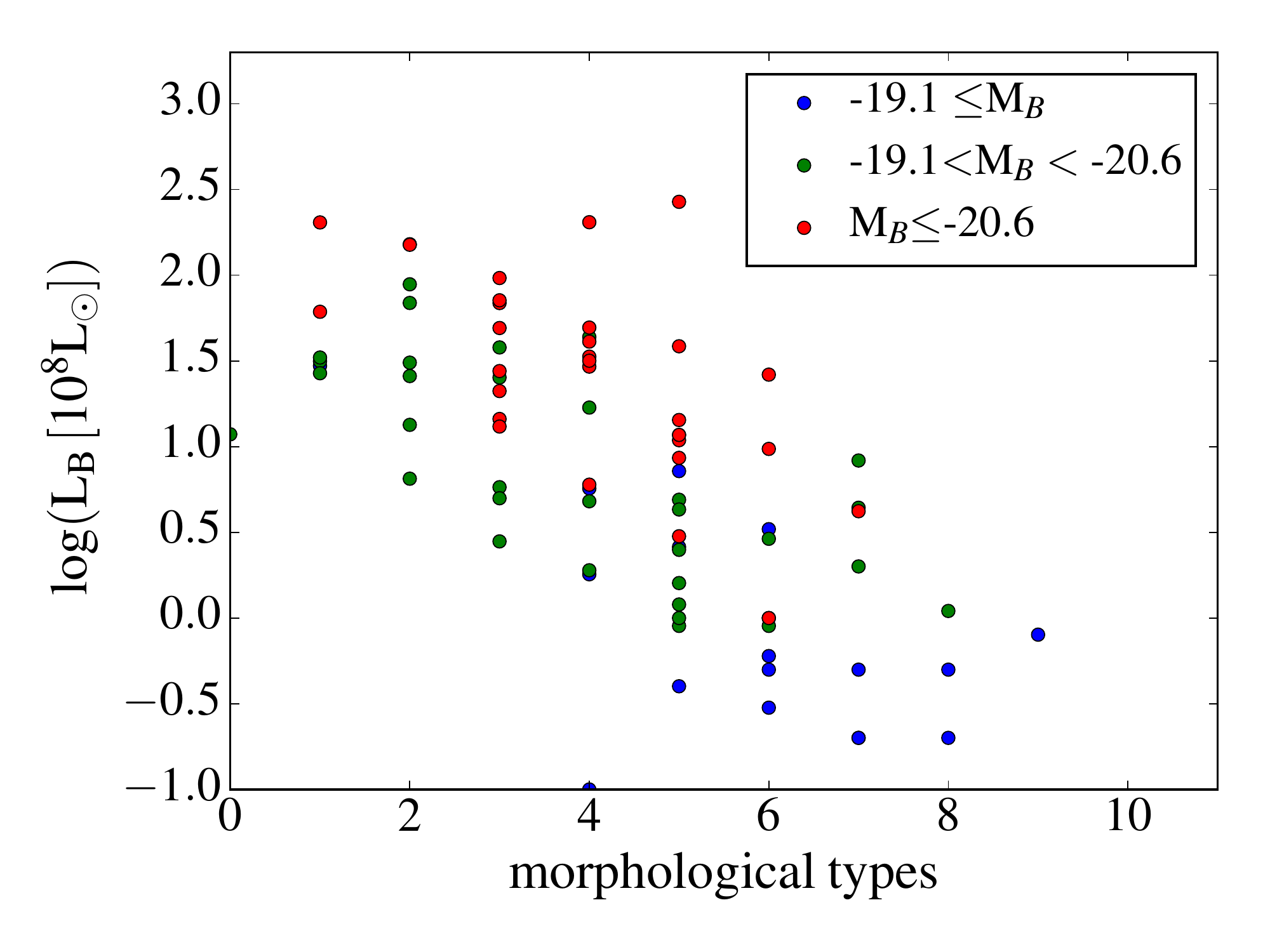}
	\vspace*{-0.3cm}\includegraphics[width=7.0cm]{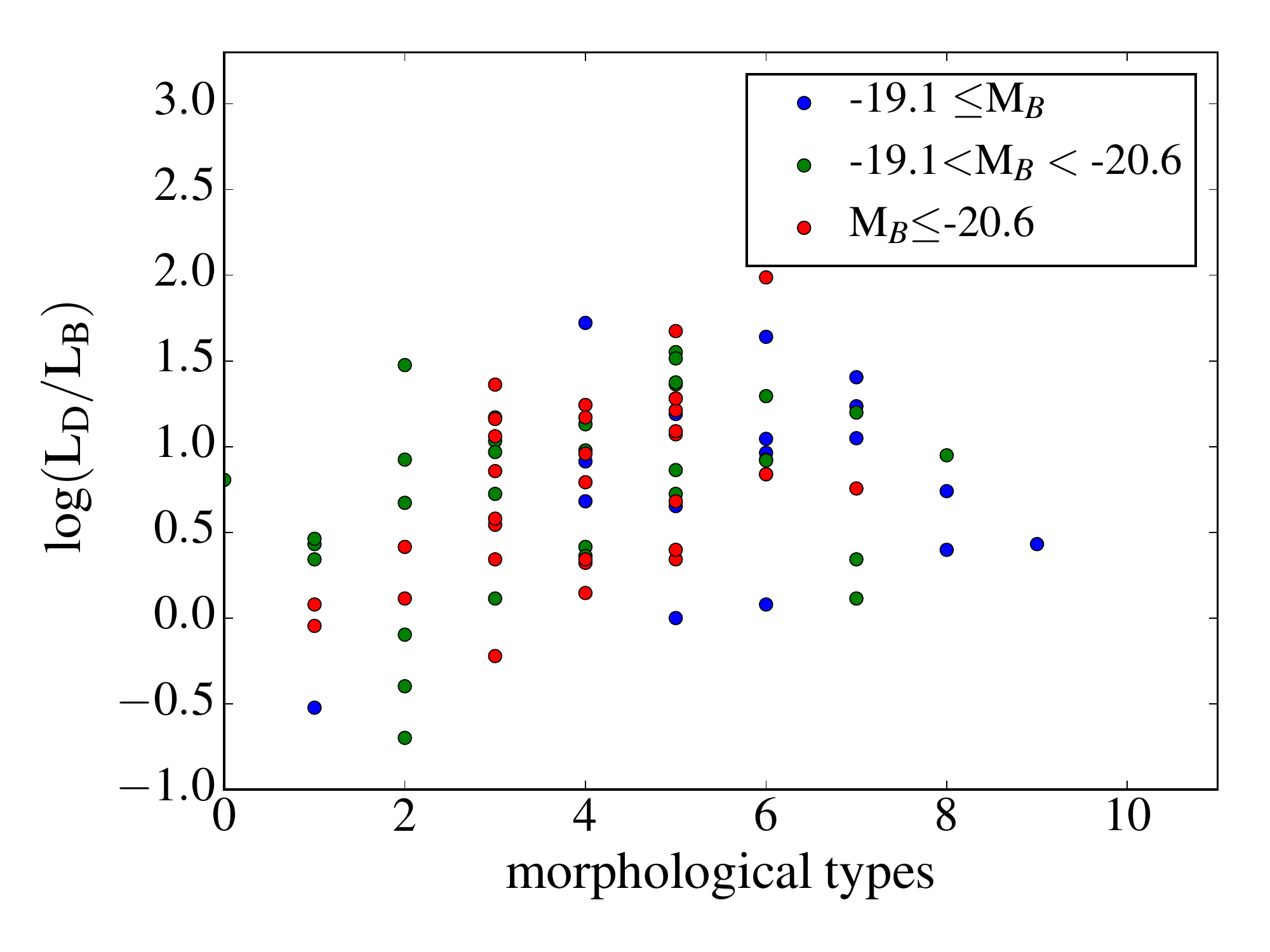}
	\vspace*{-0.3cm}\includegraphics[width=7.0cm]{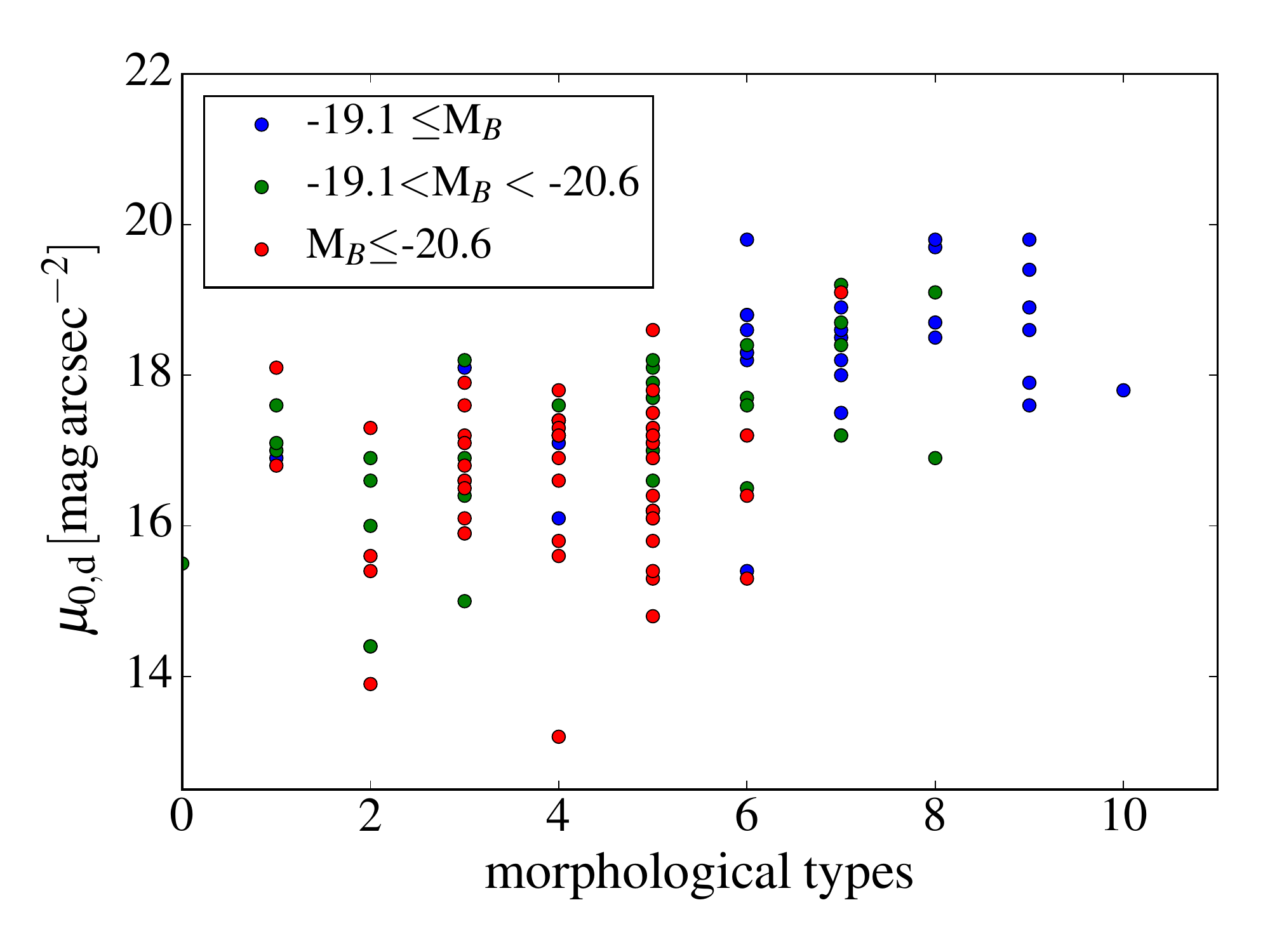}
\caption{From the top to the bottom panel:  disc luminosity, bulge luminosity, ratio of disc-to-bulge luminosity and the central disc surface brightness versus the morphological types.}
\label{fig:luminosity}
\end{figure}

\begin{figure*}
\includegraphics[width=10.0cm]{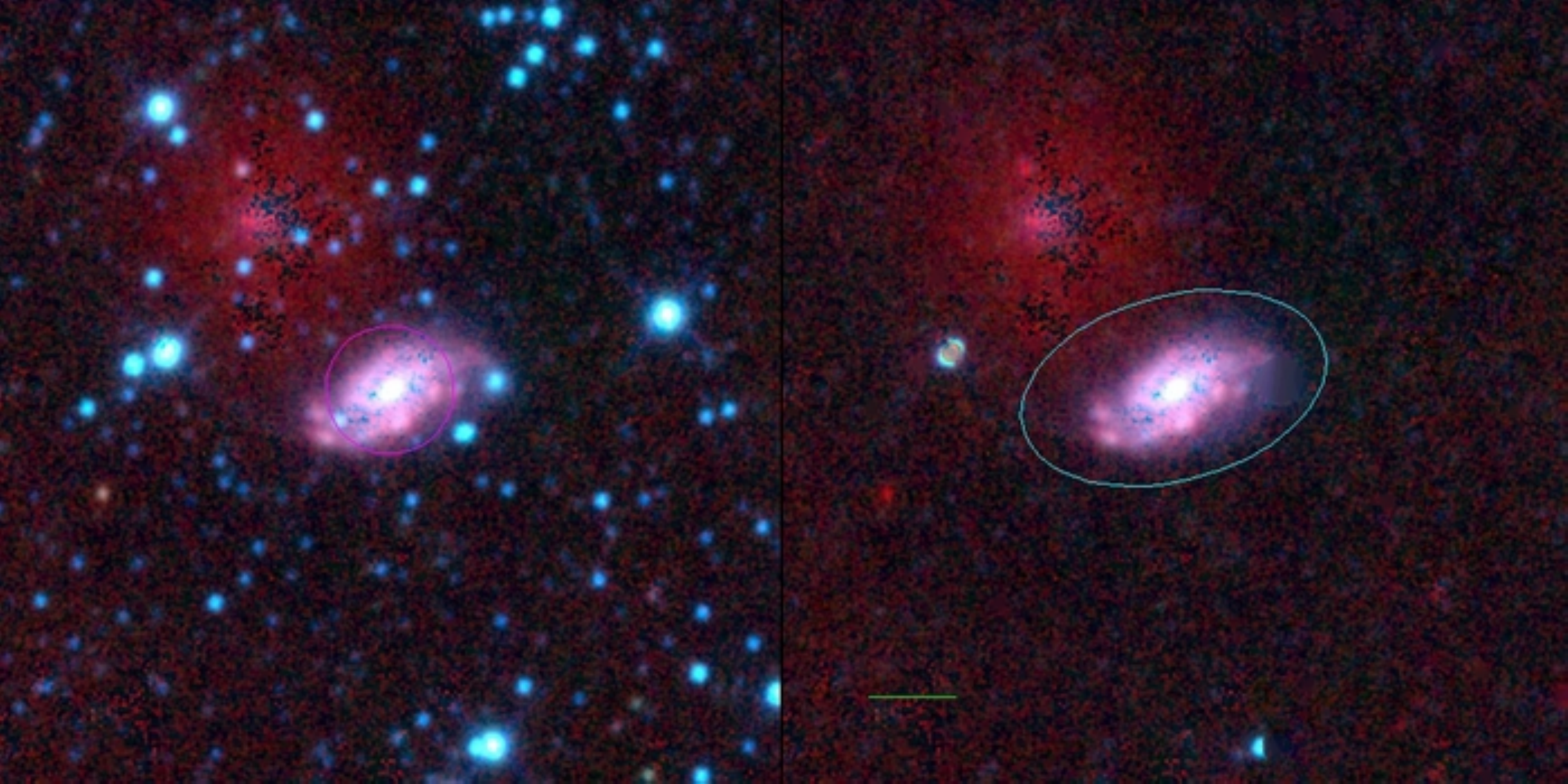}
\includegraphics[width=6.5cm]{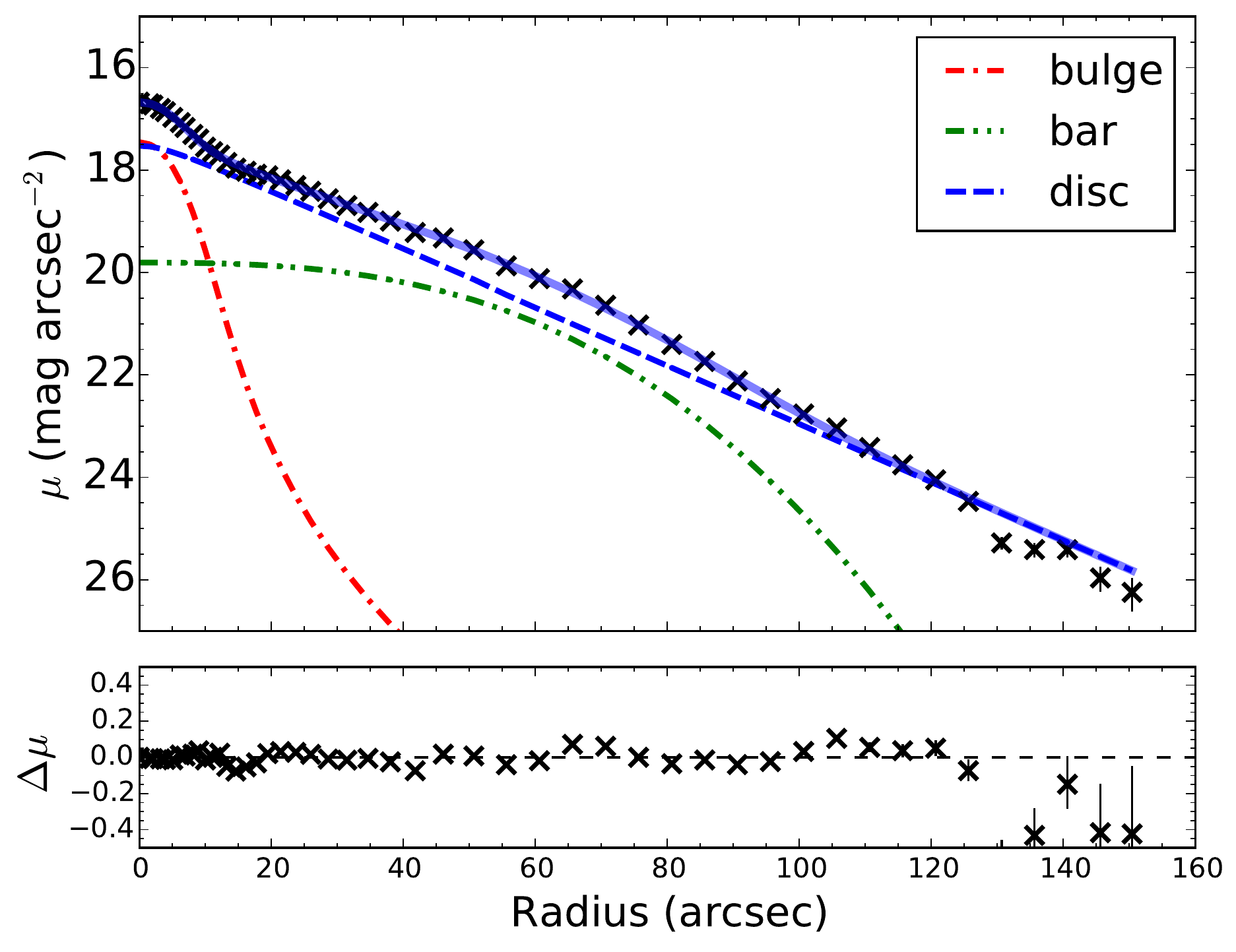}\\
\hspace*{-0.00cm} \includegraphics[width=0.35\textwidth]{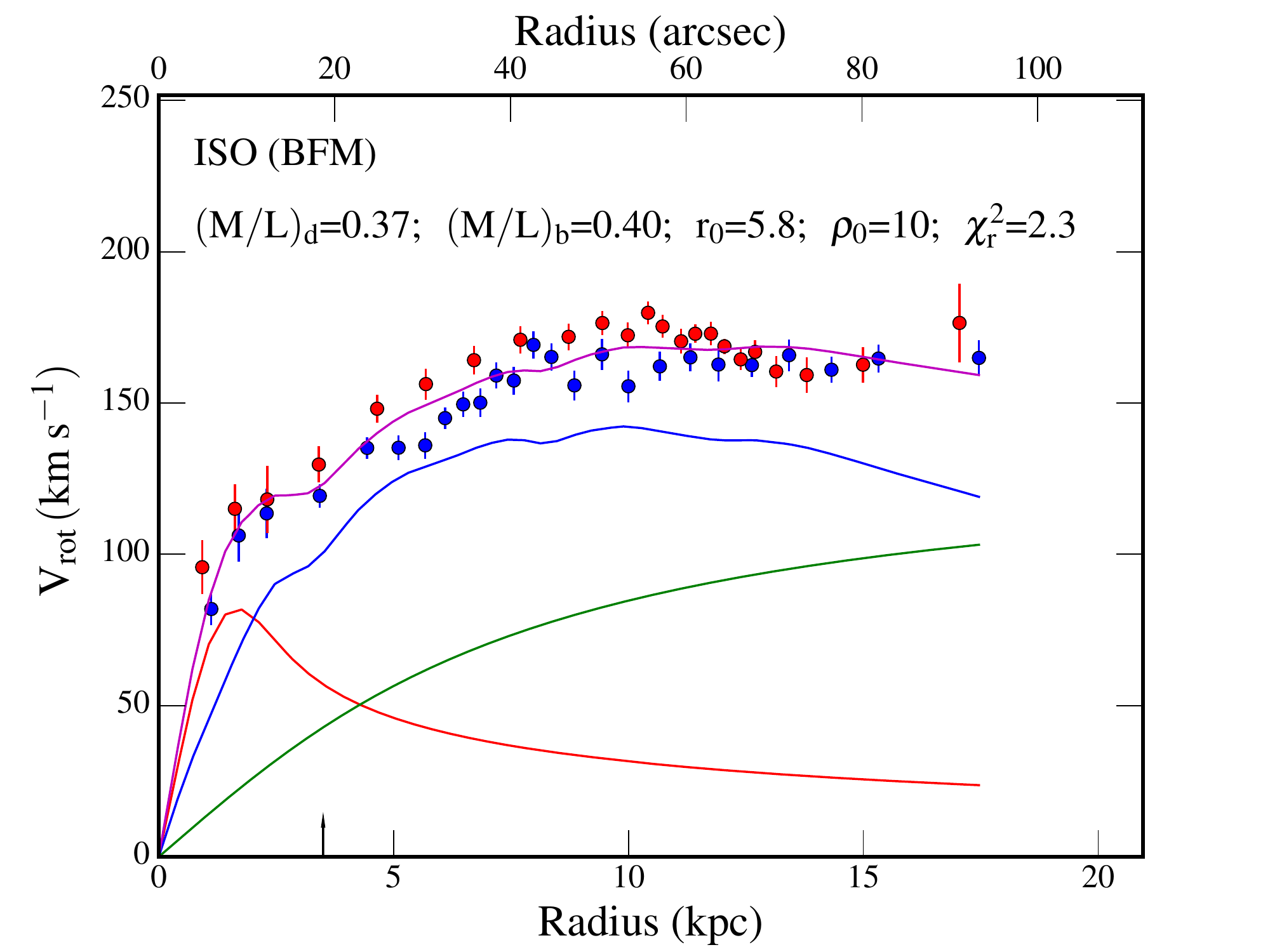}
\hspace*{-0.75cm} \includegraphics[width=0.35\textwidth]{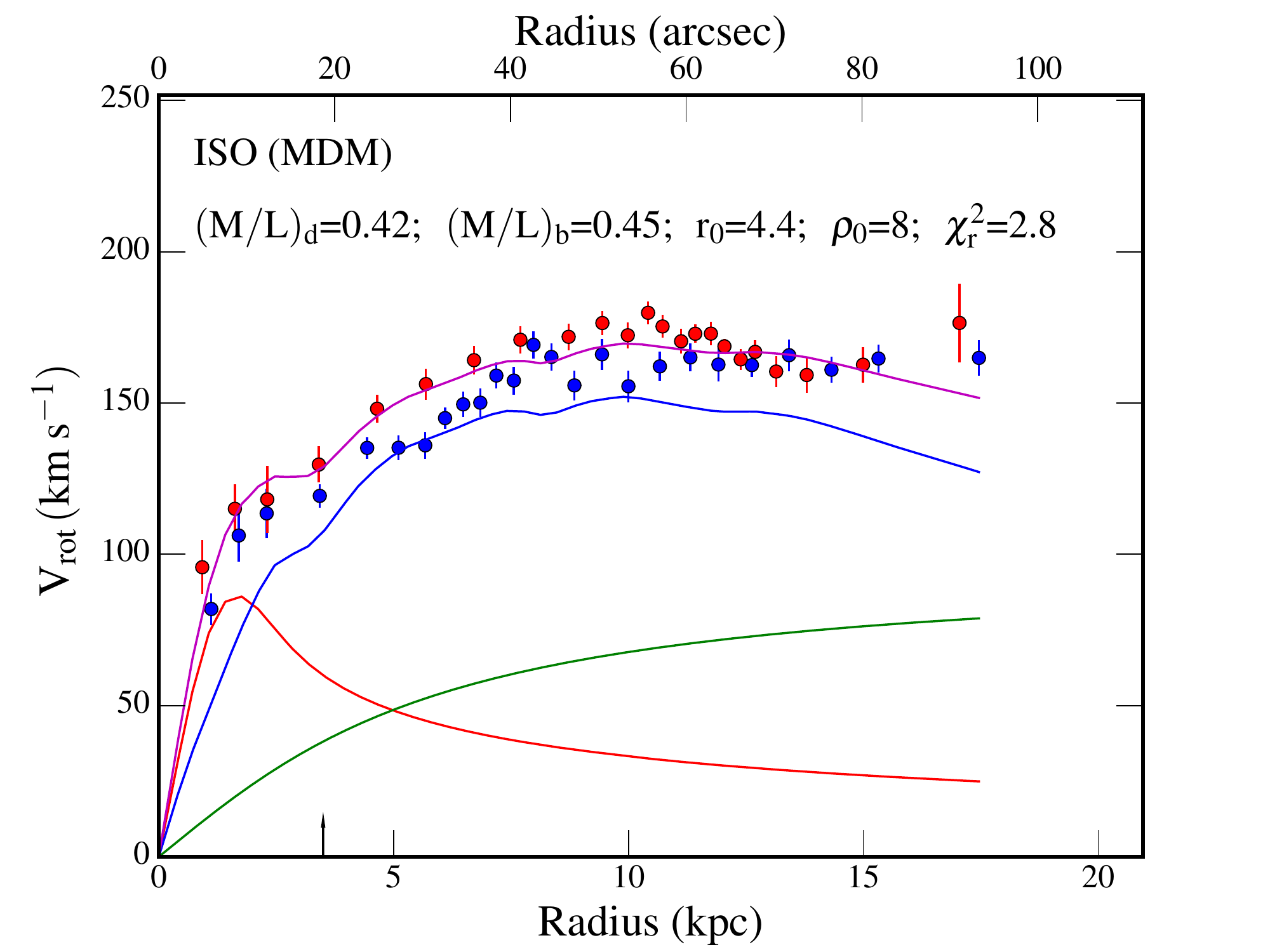}
\hspace*{-0.75cm} \includegraphics[width=0.35\textwidth]{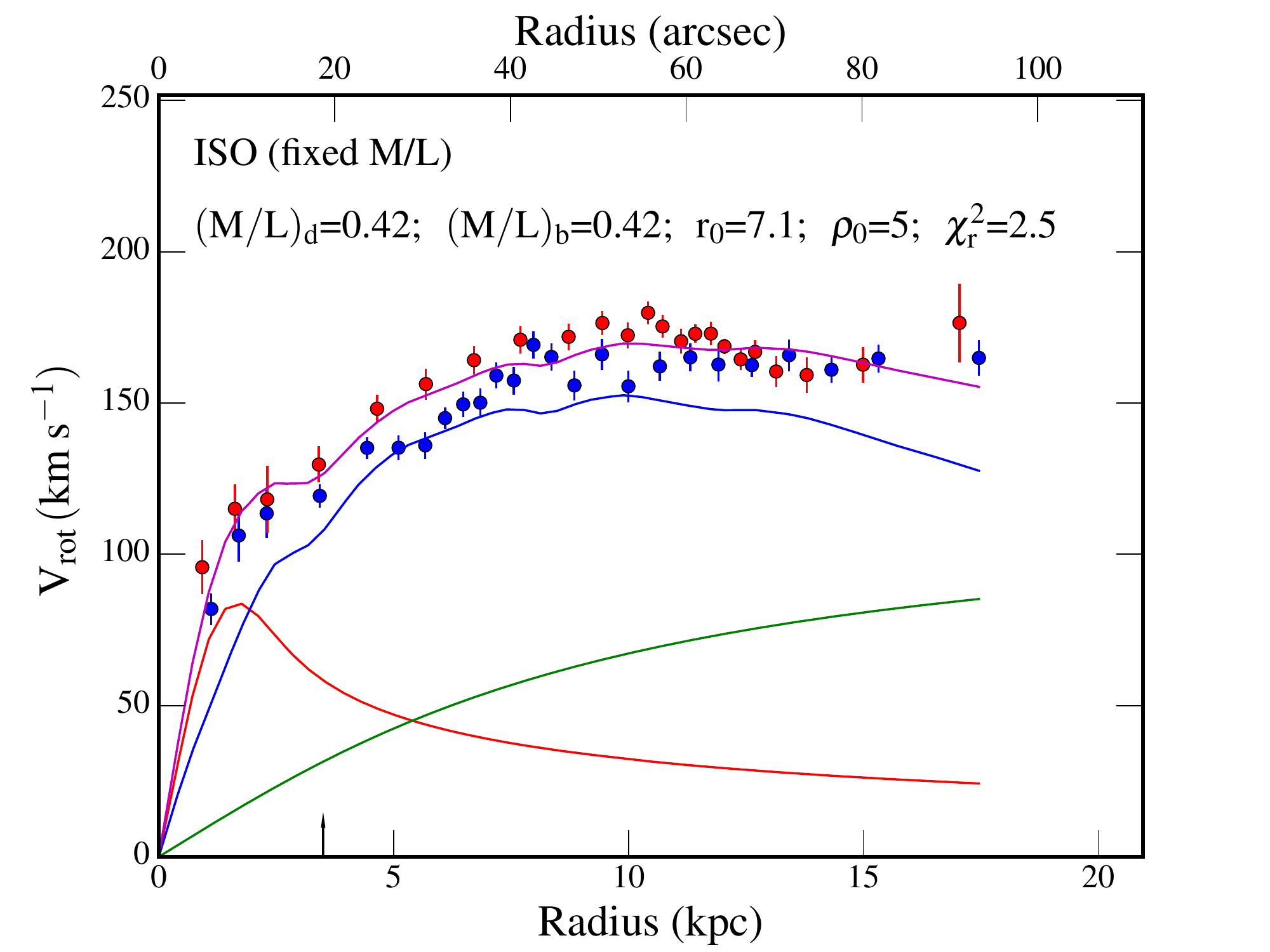}\\
\hspace*{-0.00cm} \includegraphics[width=0.35\textwidth]{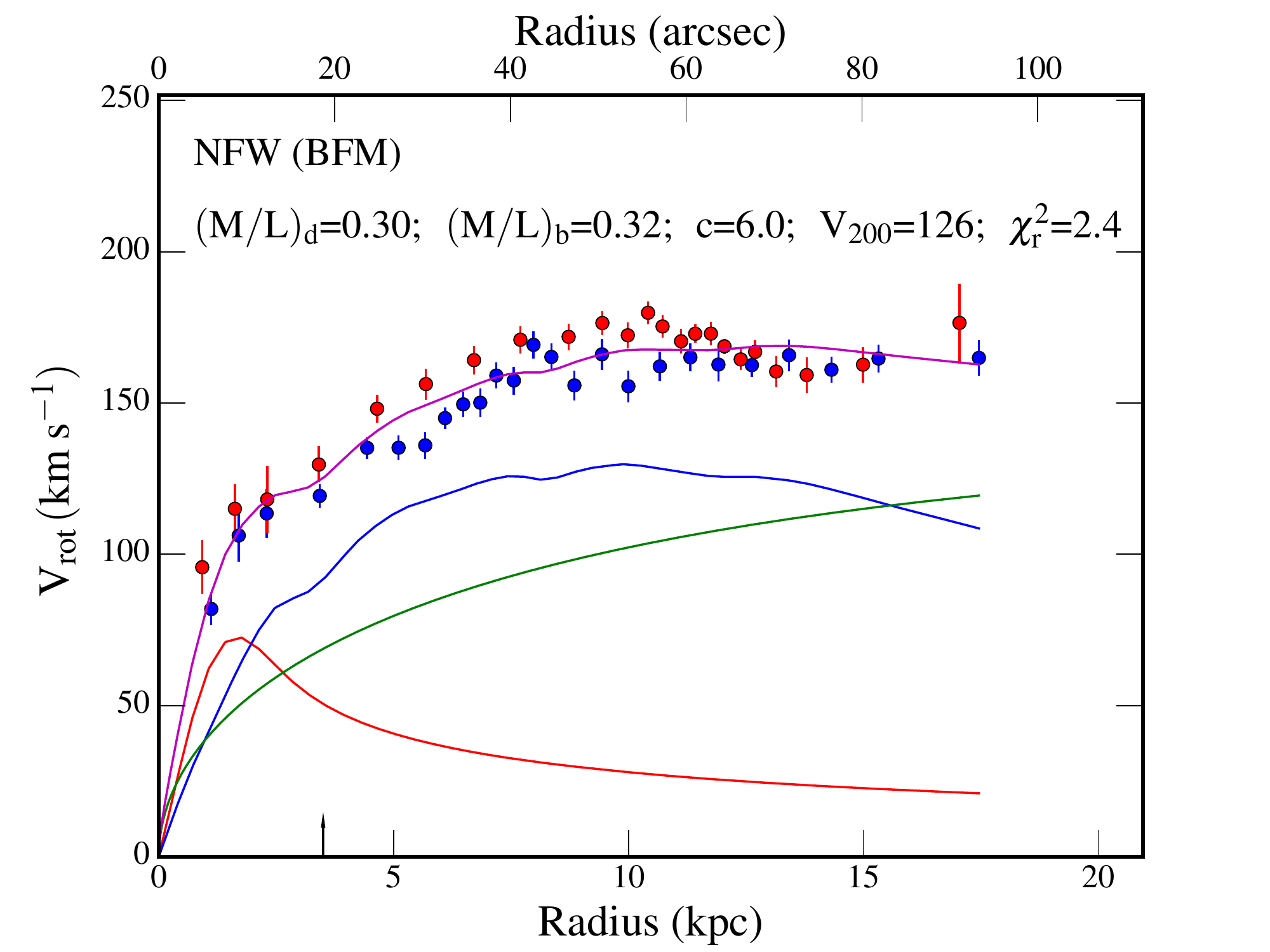}
\hspace*{-0.25cm} \vspace{-1.25cm} \includegraphics[width=0.31\textwidth]{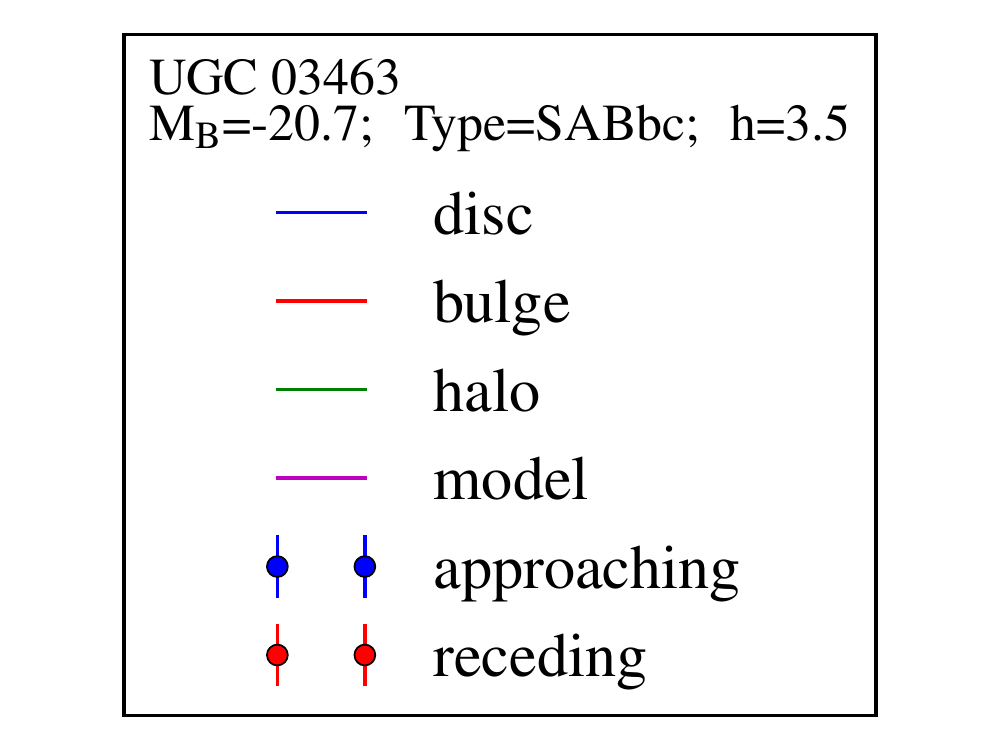} \vspace{1.25cm} \hspace*{-0.5cm}
\hspace*{-0.00cm} \includegraphics[width=0.35\textwidth]{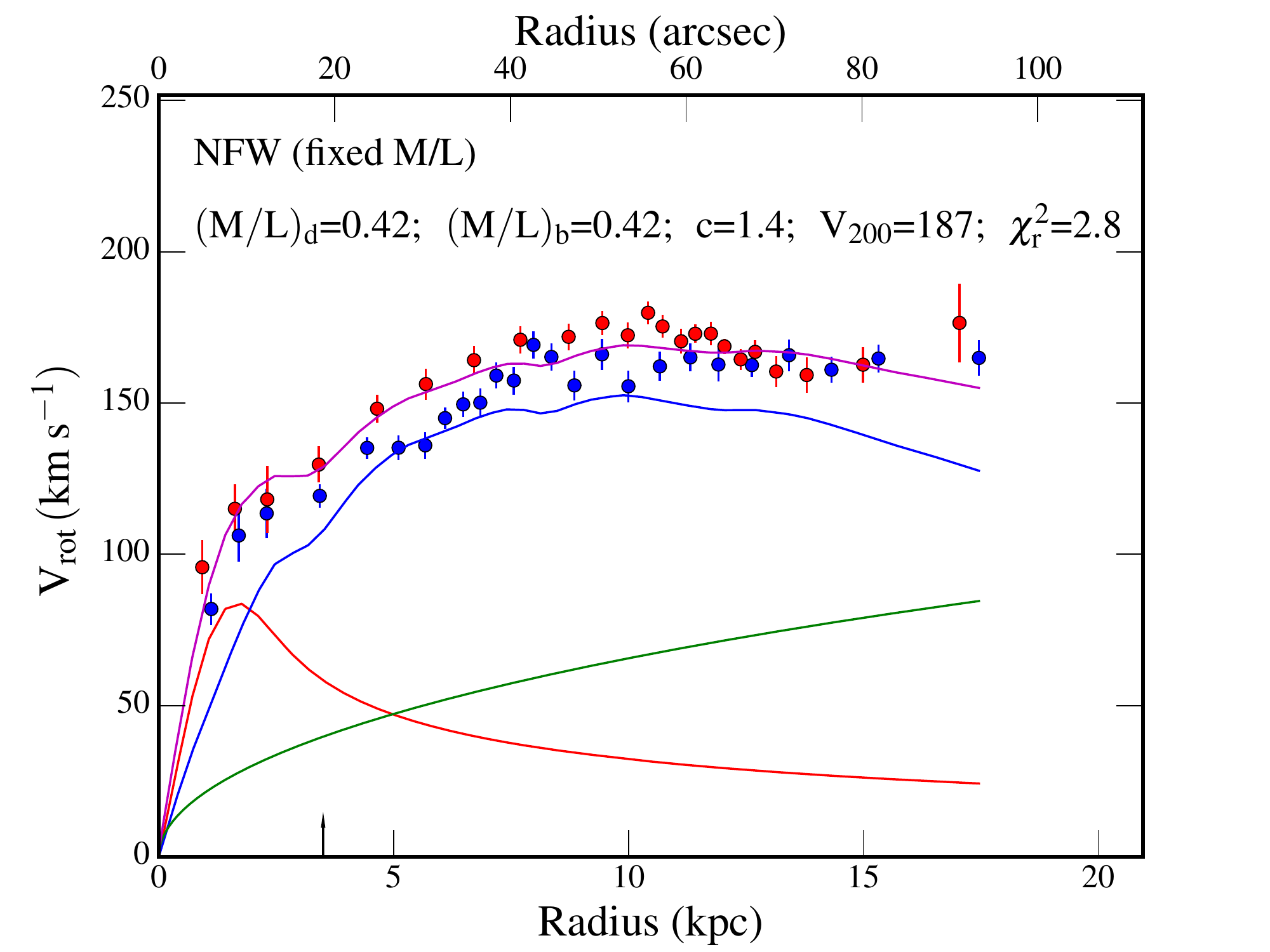}\\
\caption{Example of surface brightness profile decomposition and mass models for the galaxy UGC 3463. First line - (Left panel) WISE surface brightness image at 3.4 $\mu$m. Left side: image showing the field and the galaxy. Right side: image after the stars are removed; the circle represents the 1 $\sigma$ isophotal ellipse, used for integrated photometry and the green line shows 1 arcmin in length. (Right panel) Luminosity profile decomposition corresponding to the left image. Lines 2 and 3 - Mass models.  Second line: pseudo-isothermal sphere density profiles (ISO). Third line: Navarro, Frenk \& White density profiles (NFW). First column: Best Fit Model (BFM). Second column: Maximum disc Model (MDM) for line 2 (ISO model). Third column: Mass-to-Light ratio M/L fixed using WISE W$_1$-W$_2$ color.  The name of the galaxy, its B-band absolute magnitude, morphological type and disc scale length have been indicated in the insert located line 3-column 2. For each model, the fitted parameters and the reduced $\chi^2$ have been indicated in each sub-panel.}
\label{massmodel}
\end{figure*}

In Figs. \ref{fig:photometry} and \ref{fig:luminosity}, we illustrate several correlations between the photometric parameters displayed in Table \ref{tab:photometry}.

The top panel of Fig. \ref{fig:photometry} shows, as expected, that the brightest bulge central surface brightnesses are observed for the brightest disc central surface brightnesses, which are preferentially early type spirals while fainter later type galaxies display the weakest central surface brightnesses.  The same trend is seen in the first middle panel where the strongest bulges (largest effective radius $r_{e,b}$) tend to display larger discs (larger disc scale lengths h). The $y=x$ line of this plot emphasises that discs are on average brighter than bulges. The second middle panel of Fig. \ref{fig:photometry} shows that there is a correlation between the disc central surface brightnesses and the disc scale lengths only for the earliest type (Sa-Sab), for which large discs correspond to high central surface brightness, or for the latest type (Sdm-Im), for which small discs exhibit low central surface brightness, while for the intermediate types (Sb-Sd) there is no straightforward correlation, the central surface brightness does not depend on the disc scale length. The bottom panel of Fig. \ref{fig:photometry}  and the last panel of Fig. \ref{fig:histogram}  show a clear average relation between the isophotal radius R$_{25}$ and the disc scale length h. The median value of R$_{25}$/h is 3.5.

The two top panels of Fig. \ref{fig:luminosity} show that discs have slightly increasing luminosities from Sa to Sc before decreasing toward later types while bulges see their luminosity decreasing from early to late types, which is not surprising since this is one of the important criteria being used for morphological classification.
 A wide luminosity dispersion is noted within each morphological type meaning that we sample galaxies with different luminosities, sizes and thus masses.  In order to compare and to quantify the light coming from the disc and the bulge, the third panel shows the ratio of luminosity between the two components.  Most of the galaxies have their disc brighter than their bulge ($\log\ L_D/L_B>0$) with only 7 galaxies being bulge-dominated.  The dispersion in those three panels also points out that morphological classification should be taken with care when comparing only the disc to the bulge components. Later morphological types might have stronger bulges than earlier types. For instance, a $\log\ L_D/L_B \sim$ 0.5 can be observed from Sa (t=1) to Sd (t=8) galaxies; some t=6-7 galaxies might have a bulge almost as luminous as the disc. This justifies the fact that disc/bulge decomposition has been done independently of the galaxy morphological type. The bottom panel shows that, while the disc central surface brightness stay more or less constant for the early-types, it decreases with  from types Sc to Sm.

\section{Mass Models}
\label{sect:model}

Rotation curves of spiral and irregular galaxies need a DM halo component to be fitted  \citep[e.g,][]{Carignan+1985,vanAlbada+1985} and this is even mandatory to model the mass distribution in the outskirts of the \Hi\, disc. In the present study, we do not consider the \Hi\, gas component (densities and velocities), which are not available for a large part of the galaxies in our sample.  This will be done in a following paper. This nevertheless does not prevent us from studying the inner mass distribution in spiral and irregular galaxies. Indeed, even if rotation curves within the optical size of galaxies are largely dominated by the stellar components \citep[e.g,][]{Kalnajs1983, Kent1986}, it has already been shown in previous works  \citep[e.g,][]{Carignan+1985,Ouellette+2001,Spano+2008} that a dark halo is nevertheless often also mandatory to fit high resolution rotation curves limited to the stellar disc, meaning that not all discs are maximal. 

From the theoretical side, a dark halo component can be necessary to prevent discs having bar instabilities, as can be seen in analytic calculations and in numerical simulations of stellar discs with  initially exponential mass distribution \citep[e.g,][]{Hohl+1971,Ostriker+1973}. Amongst other considerations, it has also been shown that the presence of a bar might decrease the halo central density  distribution \citep{Weinberg+2002}.  In summary, observational and theoretical reasons justify that we can limit this study to optical rotation curves. Nevertheless, when rotation curves are not defined to sufficiently large radii, rotation curves could sometimes be modelled using disc-only or disc+bulge-only systems, without the need for DM \citep{Kalnajs+1983}. In such cases, those galaxies will be taken out of the final sample that will be used to study the relations between the dark halos' parameters and optical parameters.

We use two different models of dark matter distributions: the pseudo-isothermal sphere (ISO) distribution (section \ref{subsect:isothermal}), which is observationally motivated and that predicts a cored density profile, and the Navarro-Frenk-White (NFW) one (section \ref{subsect:NFW}), which is theoretically motivated (cosmological numerical simulations) and that features a cuspy halo density profile. These two distributions assume that the DM distribution is spherically symmetric.

In these models, we include two components for the luminous matter (disc and bulge, if present) and one for the DM halo. The final rotation curve is the quadratic sum of the individual contributions of these components:
\begin{equation}
V_{\rm cir}(r)= \sqrt{V_{\rm disc}^2 + V_{\rm bulge}^2  + V_{\rm halo}^2} \\
\label{eq4}
\end{equation}
where $V_{\rm cir}$ is the circular velocity, $V_{\rm disc}$ and $V_{\rm bulge}$  are the velocity of the disc and  bulge components respectively and $V_{\rm halo}$ is the DM halo velocity.

We will end the section by presenting the different strategies used to fit the different profiles (section \ref{subsect:strategies}).

\subsection{Pseudo-Isothermal density profile (ISO)}
\label{subsect:isothermal}

This profile features a flat core in the center \citep{Begeman+1987,Kravtsov+1998,Blok+2001}. The density distribution $\rho_{\rm iso}(r)$ of this DM halo is given by: 
\begin{equation}
\rho_{\rm iso}(r) = \frac{\rho_0}{\left[1+\left(\frac{r}{r_0}\right)^2\right]}
\label{eq5} 
\end{equation}
where $\rho_0$ is the central density of the DM, $r_0$ is the scaling radius. The corresponding velocity contribution $V_{\rm iso}(r)$ is given by: \\
\begin{equation}
V_{\rm iso}(r)^2= 4\pi G\rho_0 r_0^2 \left[1-\frac{r_0}{r} \arctan\left(\frac{r}{r_0}\right)\right]
\label{eq6}
\end{equation}
which is an increasing function of $r$, asymptotically reaching $V_{\rm max}=V(r=\infty)=\sqrt{4\pi G\rho_0 r_0^2}$.\\ 

This model has two parameters to describe its rotation curve, the central density $\rho_0$ and the scale radius $r_0$.

\subsection{$\Lambda$CDM density profile (NFW)}
\label{subsect:NFW}

The formation of galaxies is a non-linear process and therefore a direct computation through N-body simulations is required. Those simulations predict cuspy halo profiles, peaked in the center
\citep{Navarro+1996,Cole+1996,Moore+1998}.
The density distribution of the NFW halo is given by:\\
\begin{equation}
\rho_{\rm NFW}(r) = \frac{\rho_i}{\left(\frac{r}{r_s}\right)\left(1+\frac{r}{r_s}\right)^2}
\label{eq8}
\end{equation}
where $\rho_i$ is the density of the universe at the time of collapse and $r_s$ is a scale radius.
The resulting circular velocity is: \\
\begin{equation}
V_{\rm NFW}^2(r)= V_{200}^2\left[\frac{\ln(1+cx)-cx/(1+cx)}{x[\ln(1+c)-c/(1+c)]}\right]
\label{eq9}
\end{equation}
where $V_{200}$ is the velocity at the virial radius $R_{200}$, $c=R_{200}/r_s$ is the concentration parameter of the halo and $x$ is defined as $r/r_s$. The relation between $V_{200}$ (km~s$^{-1}$) and $R_{200}$ (kpc) is given by: 
\begin{equation}
V_{200}=\frac{R_{200}\times \text{H}_0}{100}
\end{equation}
where H$_0$ is the Hubble constant in km~s$^{-1}$~Mpc$^{-1}$.

In order to describe the velocity contribution associated to the NFW profile, we use the two following parameters: the concentration $c$ and the velocity $V_{200}$ at the virial radius.

\subsection{Fitting strategies}
\label{subsect:strategies}

It is usual to model the rotation curves of spiral and irregular galaxies with multi-components mass models consisting of discs (stellar and gaseous), bulge (if present) and dark halo components. The M/L ratio of the stellar components (disc and bulge) can be estimated from the stellar light distributions (see section \ref{subsect:masstolight}). The faint radial color gradients and low thickness variations observed in galaxies allow to use a radially constant M/L ratio \citep{Jong+1995} in the red, near-infrared and mid-infrared bands. At the opposite, color index is an imperfect tool to assign masses in bluer bands, i.e. to young stellar populations \citep{deDenus-Baillargeon+2013}. In this section, we describe the three different strategies used to deal with M/L: Best Fit Model (BFM), Maximum disc Model (MDM) and M/L fixed by colors.

The MDM consists in maximising the disc component, and therefore to use the highest M/L ratio and the shallower dark halo component, allowing a reasonable fit of the rotation curve. It has been very successful in reproducing the inner features due to spiral arms or bars observed along the rotation curves of galaxies  \citep[e.g,][]{Buchhorn1992,  Amram+1996}. For optical rotation curves, the ``maximum disc'' solution is favoured in several previous studies \citep[e.g,][]{deDenus-Baillargeon+2013, Spano+2008}. The Tully-Fisher relation obviously indicates that the stellar luminosity alone seems to determine the rotation velocity of spiral galaxies, favouring the maximum disc solution. Nevertheless, compared to the Tully-Fisher empirical evidences, models based on dark halo contraction due to adiabatic infall of baryons lead to the need of a major dark halo contribution in the central regions \citep{Courteau+1999,Dutton+2005}.
In addition ``maximum discs'' do not provide good solutions when NFW-like profiles are used.
In summary we do not know yet if discs are maximum or not. This is why we still have to consider alternatives like BFM or M/L ratio fixed using additional constrains like colors.

\subsubsection{Best Fit Model (BFM)}

The most open way to fit the different components of a mass model is to let all the parameters free. This gives the same weight to all the parameters.  The M/L ratio(s) is(are) the free parameter(s) of the stellar component(s). For galaxies without a bulge, only the M/L ratio of the disc is considered and it is supposed to be independent of the galacto-centric distance, assumption that cannot be made in the optical \citep{deDenus-Baillargeon+2013}. For galaxies with a bulge, we have to consider the M/L ratio of the bulge as a second baryonic free parameter.  We nevertheless set the constraint that the bulge M/L ratio has to be equal to or larger than that of the disc since bulges have older stellar populations and their stars are on average fainter. In addition to the baryonic free parameter(s), the various models (ISO and NFW) have additional free parameters.
The best fit model (BFM) technique consists of selecting the set of free parameters that minimises the $\chi^2$.  

Its main advantage is that it does not need any prior but its main inconvenient is the degeneracy of the solutions (see e.g. Fig. 4 of \citealp{Carignan+1990} or Fig. 13 of \citealp{Jobin+1990}, where different combinations of the free parameters can give equally acceptable solutions). It is indeed well known that the disc-halo decomposition is affected by the degeneracy of the possible solutions and for some galaxies, possible mass model solutions could range from "the minimum disc" to the maximum disc solutions \citep{vanAlbada+1985}.  The "minimum disc" solution means that in fact "no stellar disc" component is necessary. It could be used e.g. for modelling low surface brightness galaxies where the stellar component is negligible with respect to the cold gas disc and the dark halo components \citep{Carignan+1988}. This is not the case for most of the galaxies presented in the present sample.

We use the BFM technique for the three models. This leads to three/four free parameters for both the ISO ($r_0$, $\rho_0$ and M/L$_{\rm disc}$, M/L$_{\rm bulge}$ for bulgeless/bulge galaxies) and NFW ($c$, $V_{200}$ and M/L$_{\rm disc}$, M/L$_{\rm bulge}$ for bulgeless/bulge galaxies) models. To avoid some non physical values, we fixed minimal values on the parameters; the M/L ratio of the disc and bulge are limited at 0.1.

\subsubsection{Maximum disc Models (MDM)}
\label{MDM}

The definition of a MDM is somewhat confusing. We expect that a MDM provides a maximum rotation velocity for the disc that is comparable to the maximum rotation velocity of the observed rotation curve. In that case, an operational definition for the maximum disc hypothesis could be, for instance, that the stellar disc provides 85\% $\pm$ 10\% of the total rotational support of the galaxy at a radius equal to 2.2 disc scale lengths \citep{Sackett+1997}.  However, this definition cannot be used for fitting rotation curves.  Indeed, the disc rotation curve, computed using a model from the stellar light distribution, must match the observed rotation curve and not overestimate it.  But in practice the constraint comes from the rising part of both rotation curves, where they must match. If the rising shape of the disc rotation curve is steeper than the rising shape of the observed rotation curve, the maximum rotation velocity of the stellar disc could be much smaller than the maximum velocity of the rotation curve (see for instance UGC 2800, 10075, 11466, 11861), despite the fact that the disc rotation curve is literally scaled to its highest amplitude.

For the MDM, the disc M/L ratio is obtained using the following procedure: we first run the BFM to estimate the M/L ratio corresponding to the minimal $\chi^2_{\rm BFM}$. Then, we impose the disc M/L ratio to be higher than that obtained with the BFM but letting the $\chi^2$ to increase up to $1.3\times \chi^2_{\rm BFM}$. We chose this limit from the intrinsic dispersion of the data and from the degenerescence of the fits. For bulge  galaxies, the final model is the one having the highest disc M/L ratio and the bulge M/L ratio which provides the lowest $\chi^2$ while remaining equal or larger than that of the disc. 

The MDM technique was used for the ISO model only, with three or four free parameters ($r_0$, $\rho_0$ and M/L$_{\rm disc}$, M/L$_{\rm bulge}$) depending if galaxies are bulgeless or have a bulge.
We also ran but do not present the MDM in the case of the NFW profile because both the baryonic and the cuspy dark halo components tend to be equally maximised in the rising part of the rotation curve, furthermore the BFM is almost identical to the MDM in the case of the NFW dark halo.

\subsubsection{M/L ratio fixed by the colors}

The third method we used in this paper consists in constraining the stellar M/L ratio using the photometry, as described in section \ref{subsect:masstolight}. Like the MDM, the M/L ratio estimated from spectrophotometric evolution models give an upper limit to the baryonic mass 
\citep[e.g,][]{Carignan+1985,Bell+2001,Kassin+2006a,Kassin+2006b}.
Because the spectrophotometric models does not allow to disentangle a different M/L ratio for the disc and for the bulge, we use the same M/L ratio for the disc and for the bulge. We refer to this method as the fixed M/L ratio method.

This method was used for the three models. It leads to two free parameters for both ISO ($r_0$ and $\rho_0$) and NFW ($c$ and $V_{200}$) models.

\section{Results of the mass models}

\label{sect:results}

In this section, we present the mass models using the two DM profiles, ISO and NFW.
We have applied the various decompositions and mass models described in section 3 and 4 on our sample of 121 galaxies. As described in section \ref{sect:sample}, our sample of nearby galaxies covers a broad range of morphological types (from Sa to Im) and luminosities (masses). 
An example illustrating the mass models  is shown in Fig. \ref{massmodel} for the galaxy UGC 3463. For this case, all models give equivalently reasonable good fits. The profile decompositions and the different mass models for each galaxy are given in Appendix \ref{appendixB}. The whole catalogue is available online. 

When looking at the results and comparing with similar studies in the literature, one has to be careful and compare between themselves similar models. For example, ISO models can vary a lot from one author to the other. In the earlier models \citep[e.g,][]{Carignan+1985}, a nonsingular isothermal sphere model was used, charaterized by the core radius $r_c$, while later \citep[e.g,][]{vanAlbada+1985} a pseudo-isothermal sphere model having a scaling radius $r_0$ was used, where e.g. $r_c \simeq 3 r_0 / \sqrt{2}$.  \citet{Spano+2008}, who also modeled GHASP FP data, used a Hubble-modified density profile to model the DM halo, which means we cannot directly compare our results with theirs. Indeed, the shape of the Hubble-modified and of the ISO profiles do not have the same behaviour before and beyond their scaling radius \citep{Kormendy+2004, Binney+2008}. 

In what follows, we will compare our results to the ISO results of \citet{Kormendy+2004} and of \citet{Toky+2014}, who used exactly the same formalism. In addition, as we will see, we will only be able to compare bulgeless galaxies since the sample of \citet{Kormendy+2004} is limited to late-type galaxies from Sc to Irr, while our sample covers the whole range from Sa to Irr.


 
\subsection{Pseudo-isothermal models (ISO)}

\begin{figure*}
            \includegraphics[width=5.8cm]{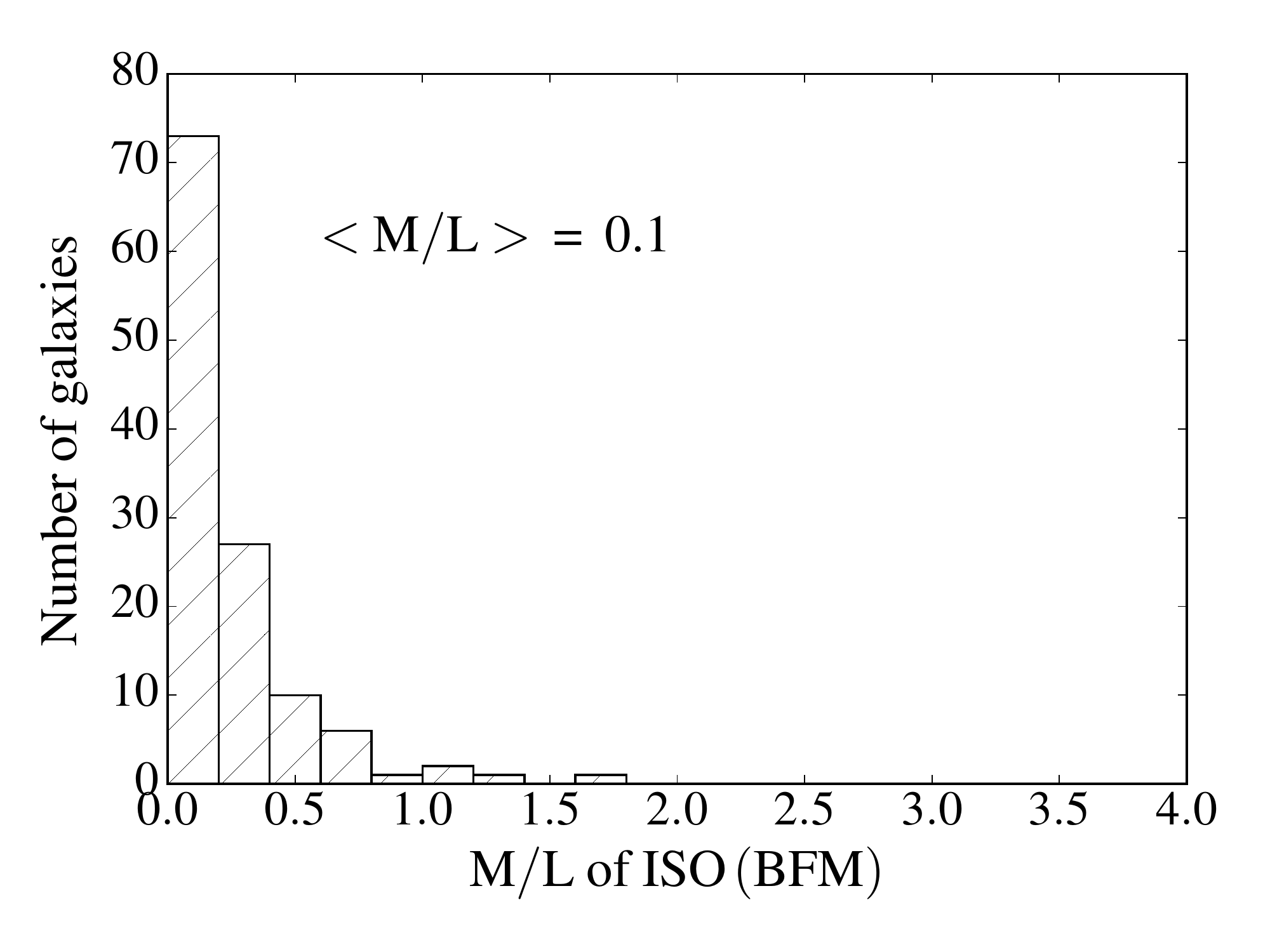}
            \includegraphics[width=5.8cm]{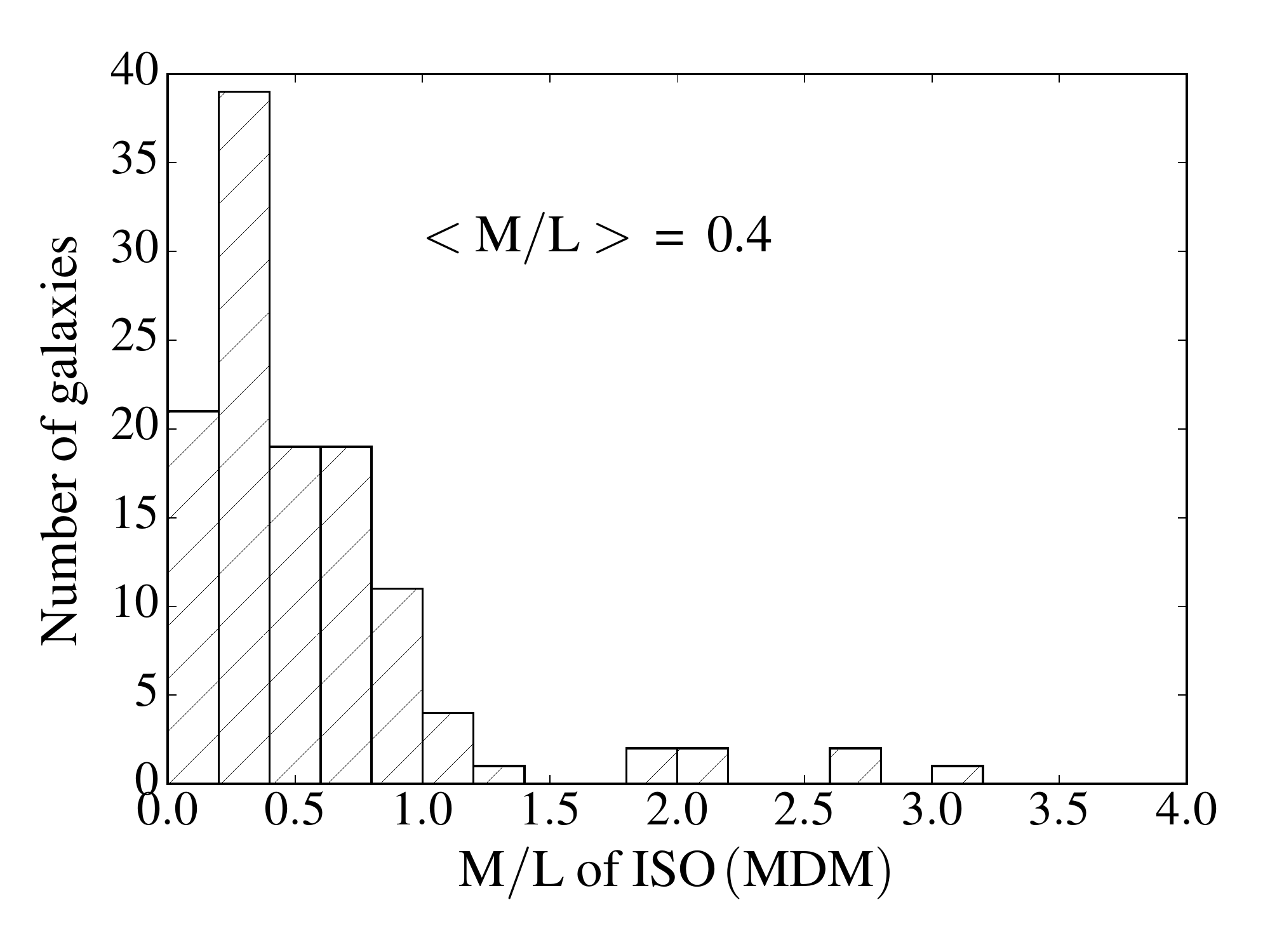}
            \includegraphics[width=5.8cm]{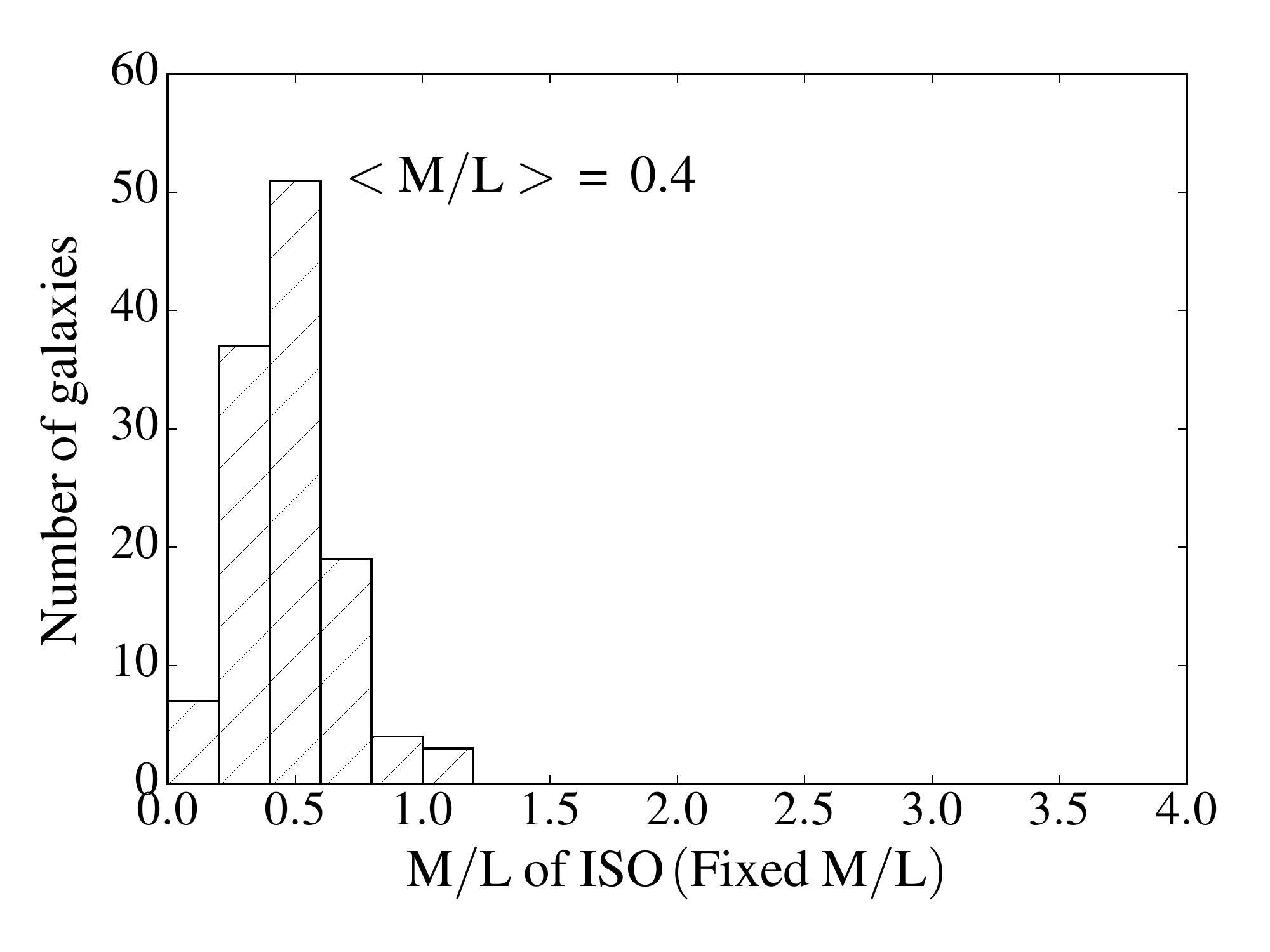}
\caption{Mass-to-light ratio distribution from the pseudo isothermal sphere model (ISO). From left to right: for the best fit model (BFM), for the maximum disc model (MDM) and for the M/L fixed using the WISE color (W$_1$-W$_2$).  The median <M/L> values are indicated in the plots for each case.}
\label{fig:M/L}
\end{figure*}

For the luminous disc, \citet{Lelli+2016} found that when modeling galaxies, the values of M/L in the 3.6 $\mu$m band are in the range of 0.2 to 0.7 M$_{\odot}$/L$_{\odot}$. As seen in Fig. \ref{fig:M/L}, we find a median value of M/L = 0.1 when using the BFM (left); 0.4 for the MDM (center) and also 0.4 for the fixed M/L (right).
 The values of the latter two models are in the range of Lelli's results. Recently, \citep{Kettlety+2018} found that the M/L can be fixed at a value of 0.65 when using the W1 band of WISE,  this is close to the average value we found for the fixed M/L and the MDM. We note that the average M/L ratio for the BFM is four times smaller than for the MDM, meaning that the BFM does not provide  maximum disc solutions.  Despite the fact that their M/L distribution are quite different, it is noticeable that the MDM gives the same average M/L value than the models for which the M/L ratios are fixed by the WISE colors. This would suggests that discs tend, on average, to be maximal, as defined in Sec. \ref{MDM}.

\begin{figure}
\begin{center}
	\includegraphics[width=8.5cm]{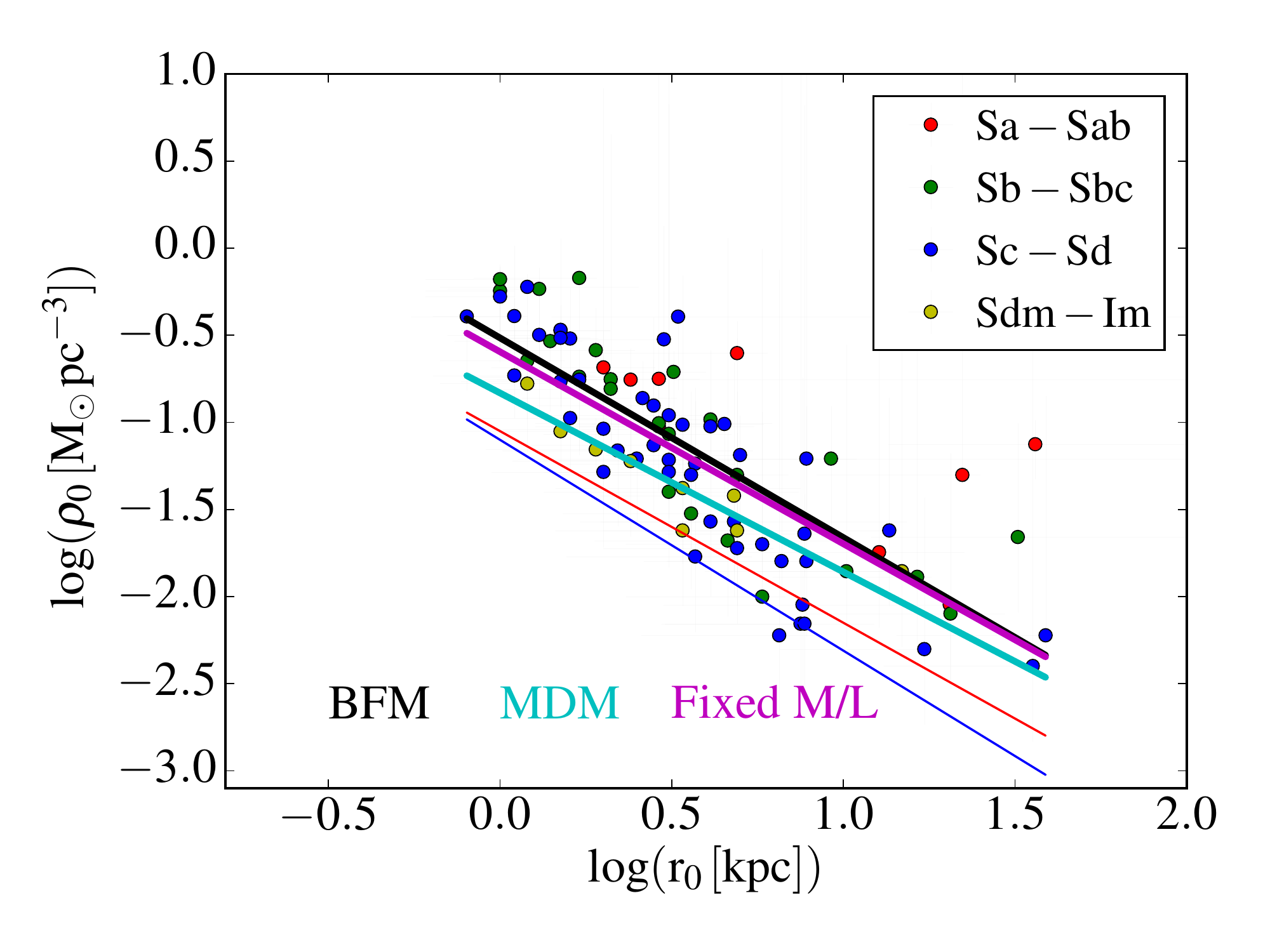}
	\includegraphics[width=8.5cm]{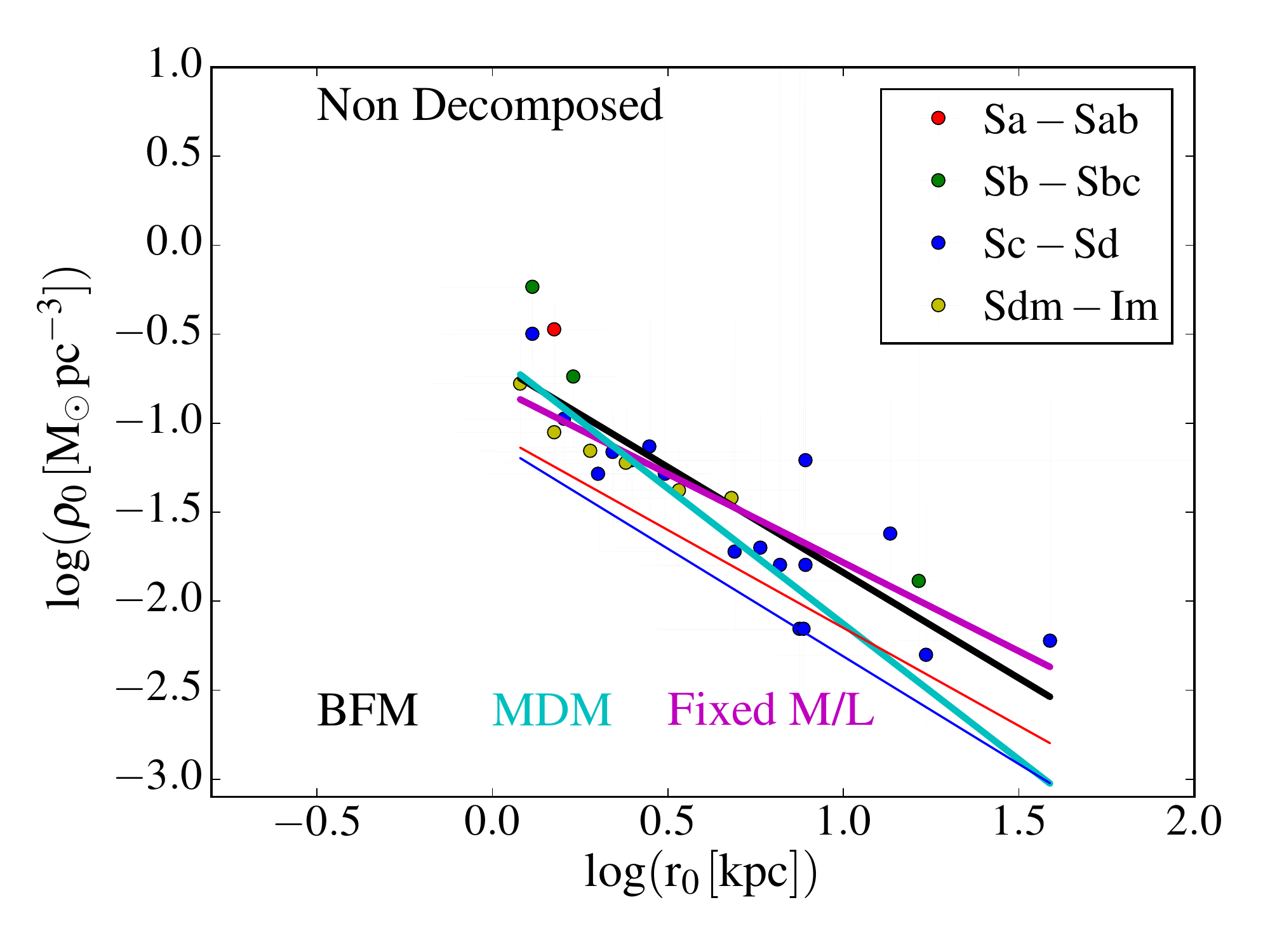}
	\includegraphics[width=8.5cm]{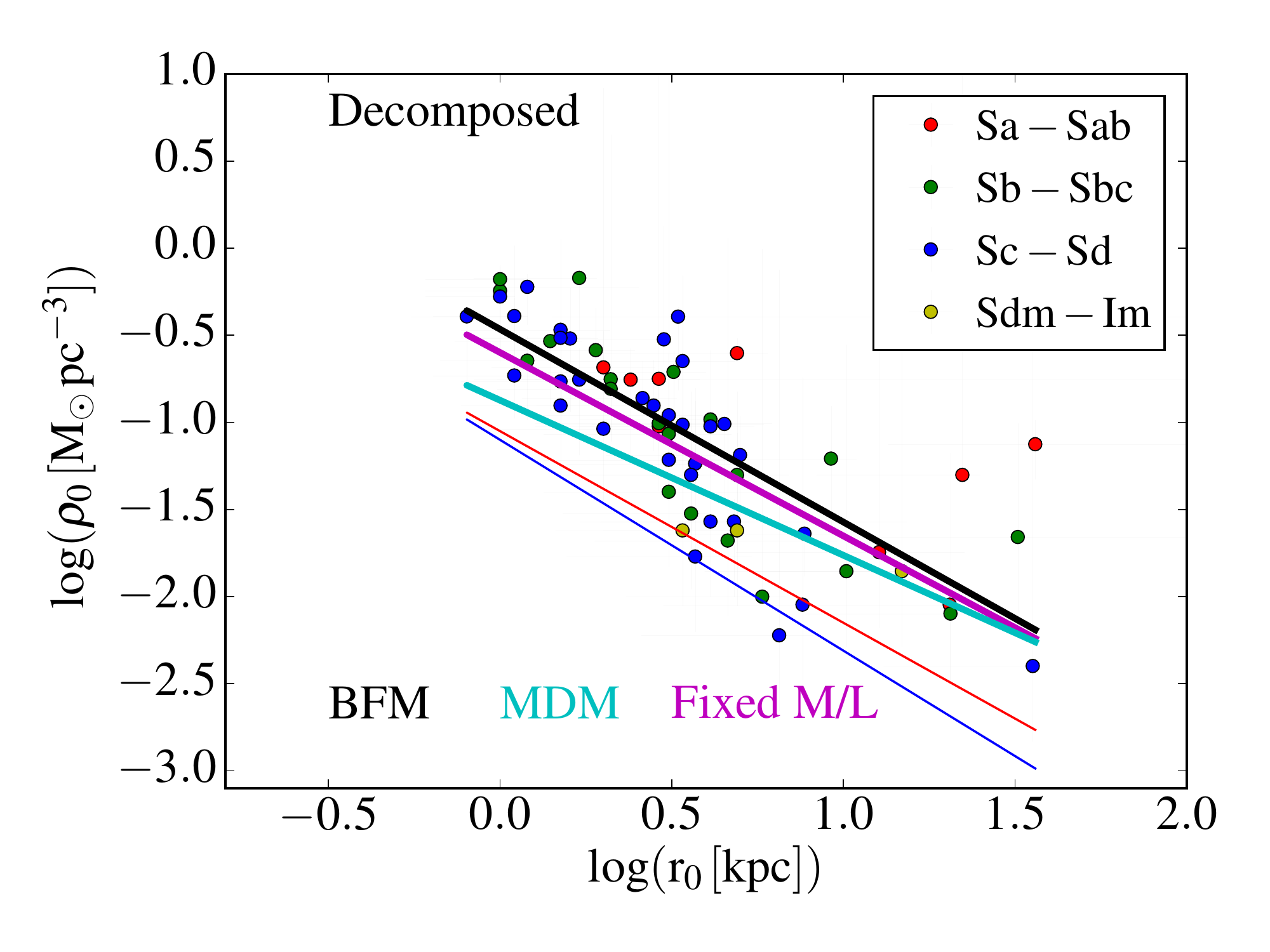}
\caption{Central halo density versus halo core radius for the ISO best fit models (BFM), maximum disc models (MDM) and fixed M/L models. It is given for the whole sample at the top, for the galaxies with no luminosity profile decomposition (later types) in the middle and for the galaxies for which we did a profile decomposition (earlier types) at the bottom. The thin dark blue and red lines represent respectively the fits found by \citet{Kormendy+2004} and \citet{Toky+2014}}
\label{fig:rho0r0}
\end{center}
\end{figure}

For the ISO models, we observe in Fig. \ref{fig:rho0r0} a clear anti-correlation between the central halo density and the halo core radius for the three models (BFM, MDM, Fixed M/L). This means that higher central density dark halos have smaller core radius. We show the results for the full sample (top panel) and for the sub-sample of galaxies which had no profile decomposition (middle panel) and those for which we did a profile decomposition (bottom panel). The points are for the BFM but we overlay the results for the 3 models. The different linear regressions found for the different models and for the full sample are:
\begin{equation}
\begin{array}{l}     
\rm \log\ \rho_0 = (-1.14 \pm 0.09)\,   \log\ r_0  - (0.51 \pm 0.06)\qquad[BFM]  \\
\rm \log\ \rho_0 = (-1.03 \pm 0.12)\,  \log\ r_0  - (0.83 \pm 0.08)\qquad[MDM]   \\
\rm \log\ \rho_0 = (-1.10 \pm 0.08)\,  \log\ r_0  - (0.59 \pm 0.08)\qquad[fixed\ M/L] 
\end{array} 
\end{equation}
The results for the sub-sample with no decomposition (middle: later types) and the sub-sample with decompositions (bottom: earlier types) give very similar results, but with different zero points. If we compare with the litterature for authors using the same ISO formalism, we find:
\begin{equation}
\begin{array}{l}       
\rm \log\ \rho_0 = -1.10\,   \log\ r_0  -1.05\qquad[BFM: R\&C2014]    \\
\rm \log\ \rho_0 = -1.21\,   \log\ r_0  -1.10\qquad[MDM: K\&F2004]    \\
\end{array} 
\end{equation}
\noindent \citet{Kormendy+2004} used MDM models while \citet{Toky+2014} used BFM models. Overall, our relations are closer for our sub-sample with no decomposition (later types) than for the sub-sample with decomposition.
The fact that the relations found are similar to those of the literature, with a similar slope but slightly different zero points,  could come from using different datasets. In the case of  \citet{Kormendy+2004}, their sample covers only late-type galaxies (Sc to Irr) while we cover all morphological types and they are using mainly B-band photometry. For  \citet{Toky+2014}, their results are based on  extended \Hi\, data while we mainly model the inner parts with our optical FP data but using also mid-infrared photometry.  
All those different approaches make direct comparisons very difficult.

\begin{figure}
	\begin{center}
            \includegraphics[width=8.5cm]{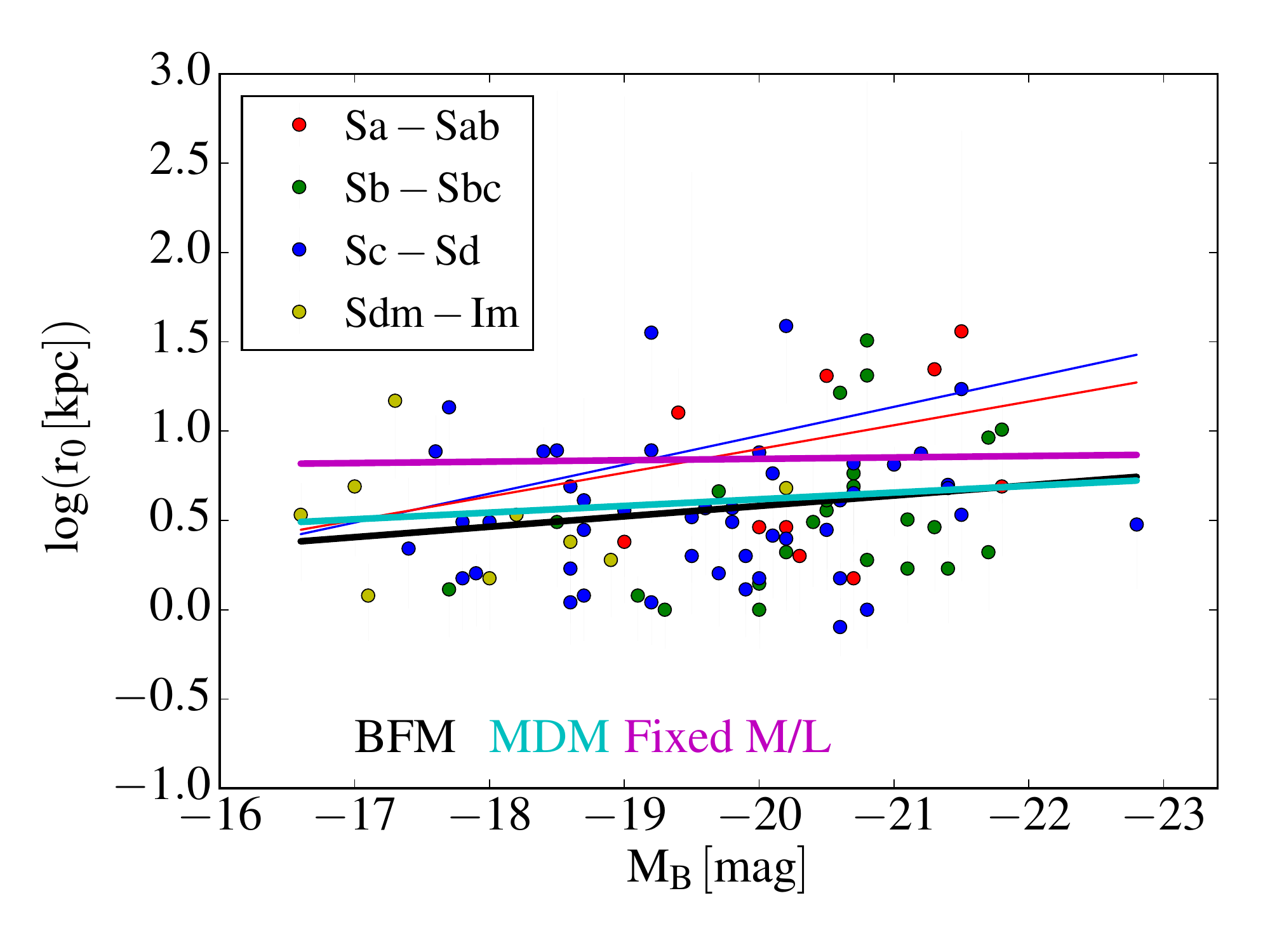}
             \includegraphics[width=8.5cm]{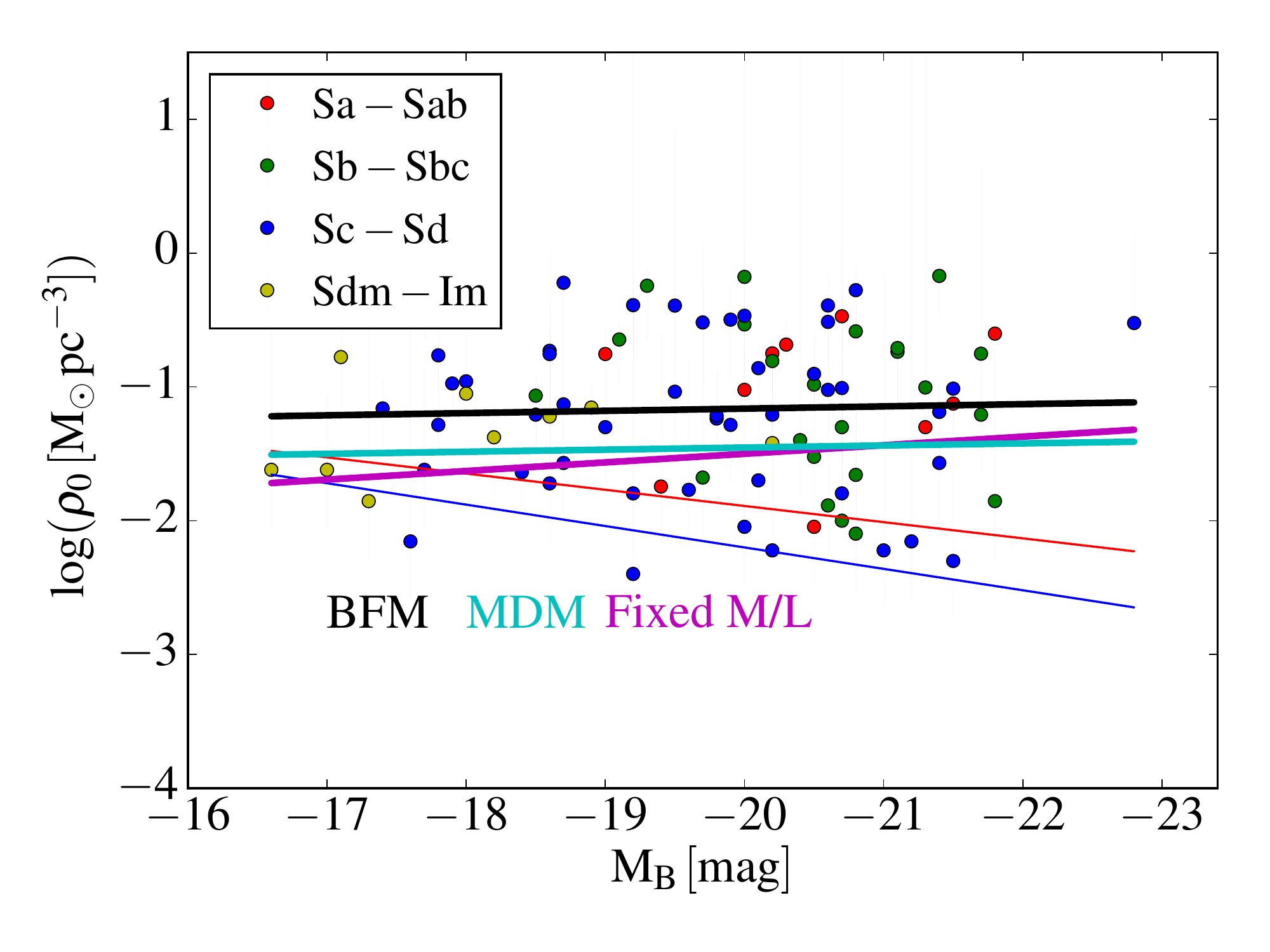}
\caption{Halo scaling radius (top panel) and central halo density (bottom panel) versus the absolute magnitude for the whole sample from the pseudo-isothermal (ISO) models. The black, cyan and magenta lines are the best fit (BFM), the maximum disc (MDM) and the fixed M/L  models respectively. The thin blue  and red lines represent respectively the fit found by \citet{Kormendy+2004} and \citet{Toky+2014}.}
\label{fig:all}
\end{center}
\end{figure}

Fig. \ref{fig:all}  shows the halo parameters (halo core radius and central density) as a function of the B-band luminosity M$_B$. To avoid crowding or multiple figures, we do not plot symbols representing individual galaxies for the MDM and the fixed M/L but only the fit to the data points of these models: a cyan line for the MDM and a magenta line for the fixed M/L model. Running BFM without fixing any lower limit for the core radius ($r_0$) and higher limit for the core density ($\rho_0$) show that
10 galaxies have a core radius $r_0 < 1$ kpc and 7 galaxies have a core density $\rho_0 > 0.75 $ M$_{\odot} \rm kpc^{-3}$. In addition, core densities $\rho_0$ > 0.75 M$_{\odot} \rm kpc^{-3}$ are only obtained where core radius $r_0$ are lower than 1.5 kpc. The combination of very small $r_0$ and high $\rho_0$ leads to DM halo with constant rotation velocities at all radius (except in the very center of the galaxies).  They add a kind of offset to the disc component(s).  In all cases, the MDM shows that a DM halo is not necessary to fit the rotation curves. Therefore, we found 25 galaxies for which we do not need halos parameters (shown with an asterisk in Appendix \ref{tab:iso}). These galaxies are obviously best fitted by a maximum disc model until almost the end of the optical radius (no DM halo) and those are removed in the fits between the core radius and the central density, and the core radius or the central density as a function of $\rm M_B$. On the other hand, only dwarf galaxies having a dynamical mass less than  $2\times 10^{9}\rm M_{\odot}$ and a rotation velocity smaller than  40 km/s might eventually have core radius lower than 1 kpc \citep[e.g, Holmberg I and II, DDO 53,][]{OH+2011} and our sample does not contain such low mass galaxies.  Furthermore we do not allow BFM to have core radius lower than 0.5 kpc and a core density larger than  0.75 M$_{\odot} \rm kpc^{-3}$, we do not take into account galaxies which reach the lower value 0.5 kpc or the upper value 0.75 M$_{\odot} \rm kpc^{-3}$ to fit the halos parameters.  Typically no clear correlation is found between the core radius or the central density and \rm $\rm M_B$ but the dispersion is rather high (0.50 dex and 0.75 dex respectively) around the average of $\sim 3$ kpc and $\sim 0.1$ M$_{\odot}$.pc$^{-3}$.

We do not confirm previous studies suggesting that less luminous dwarf galaxies tend to have smaller halo core radius and larger central density \citep[e.g,][]{Carignan+1988}. However, we observe a weak correlation between the core radius and the luminosity, less luminous galaxies tend to have smaller core radius. At the opposite, a weak correlation between the central density and the luminosity is observed: the less luminous galaxies tend to have shallower central density than brighter galaxies. Similarly, we note again that the mean central halo density is lower for the fixed M/L than for the BFM and the MDM models.

\begin{figure}
	\begin{center}           
             \includegraphics[width=8.5cm]{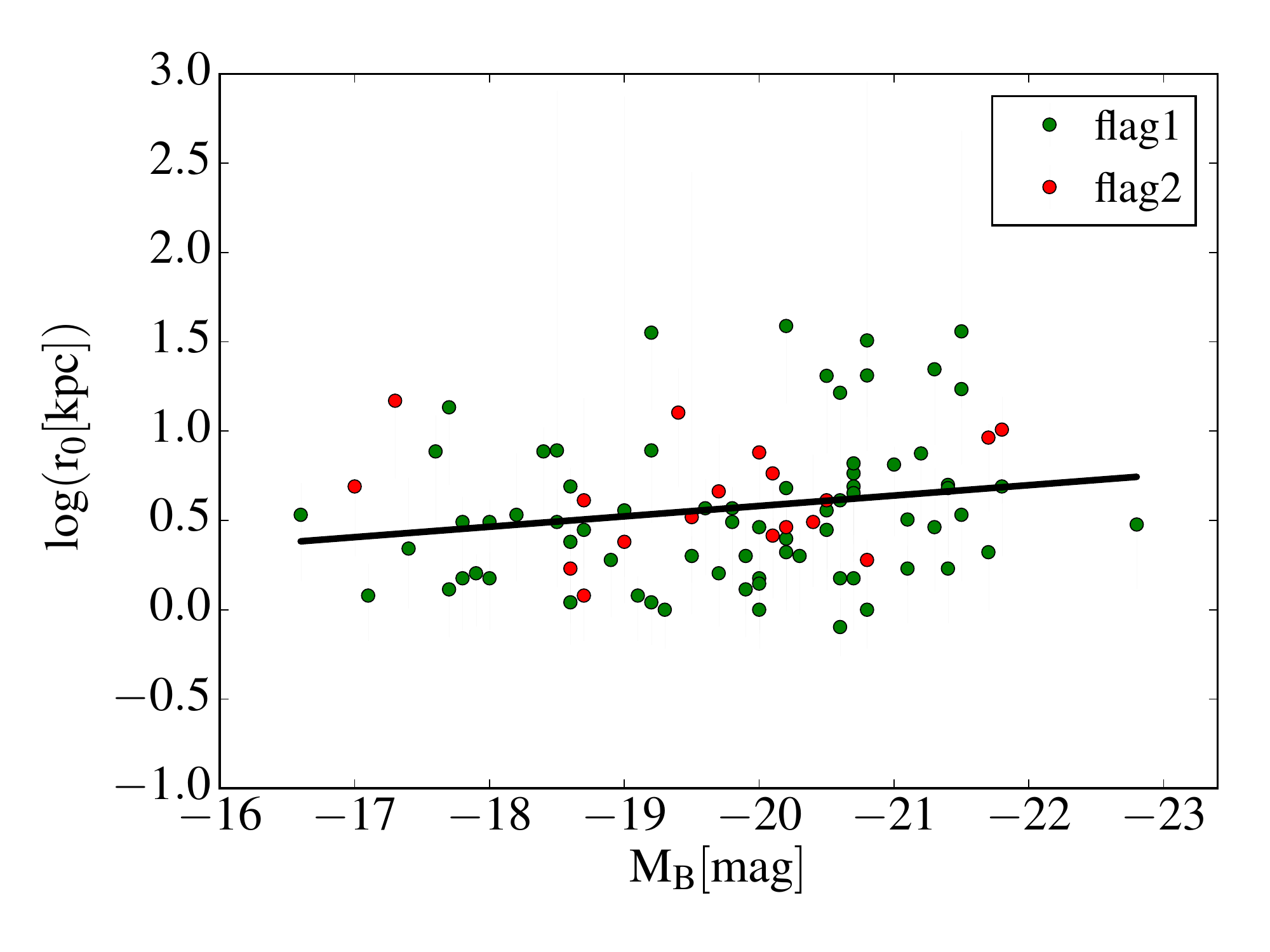} 	 
            \includegraphics[width=8.5cm]{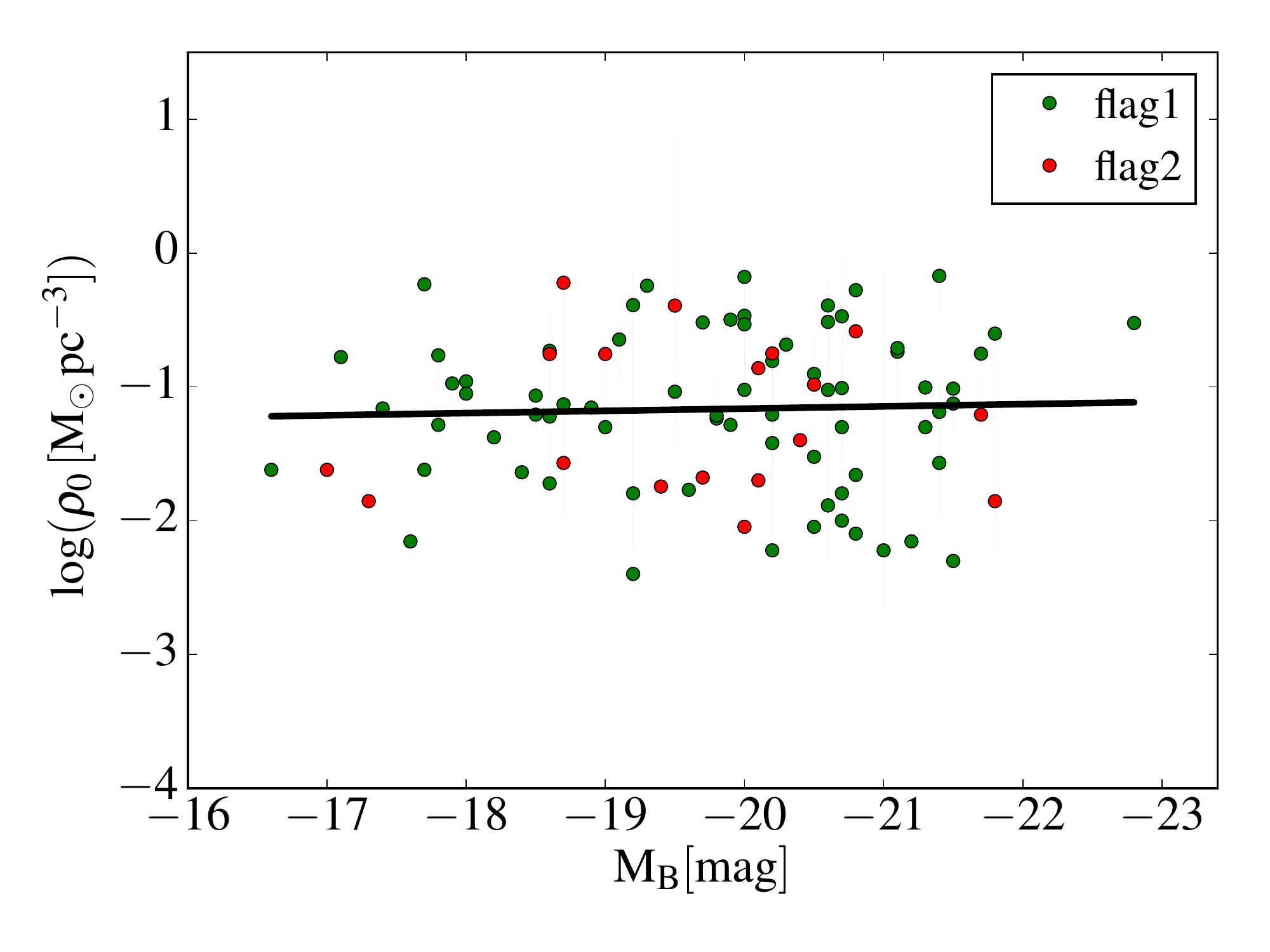}
\caption{Halo scaling radius (top panel) and central halo density (bottom panel) versus the absolute magnitude for all the sample from the ISO (BFM) models. The green symbols represent the higher quality rotation curves (flag 1) while the red circles show the lower quality rotation curves (flag 2). The black line linearly fits the dots.}
\label{fig:qualityrc}
\end{center}
\end{figure}

In order to understand the impact of the quality of the rotation curves on the lack of correlation, we have separated the higher quality rotation curves (flag 1) from the lower quality ones (flag 2) in Fig. \ref{fig:qualityrc}. The weak trends we found are not a result of the quality of the data. This will be discussed further in section \ref{sect:discussion}.

\subsection{NFW dark matter halos}

Cosmological numerical simulations \citep[e.g.,][]{Navarro+1996A} show a strong correlation between the velocity at the virial radius (i.e $V_{200}$) and its concentration in the sense that low mass halos are more concentrated. \citet{Blok+2008} emphasised that large values of the concentration parameter $c$ indicate a larger collapse factor while $c$ = 1 indicates no collapse. Due to the fact that concentration $c$ lower than unity makes no sense in the CDM context \citep{Blok+2008}, we do not allow $c$ \ < 1. This naturally lead to the fact that 4 and 31 galaxies have a concentration equal to the lower limit $c$ = 1, respectively for the BFM and fixed M/L model. Examples of parameters obtained for the NFW fits are listed in Table \ref{tab:nfw} and the full table can be found online. 

\begin{figure}
\includegraphics[width=9cm]{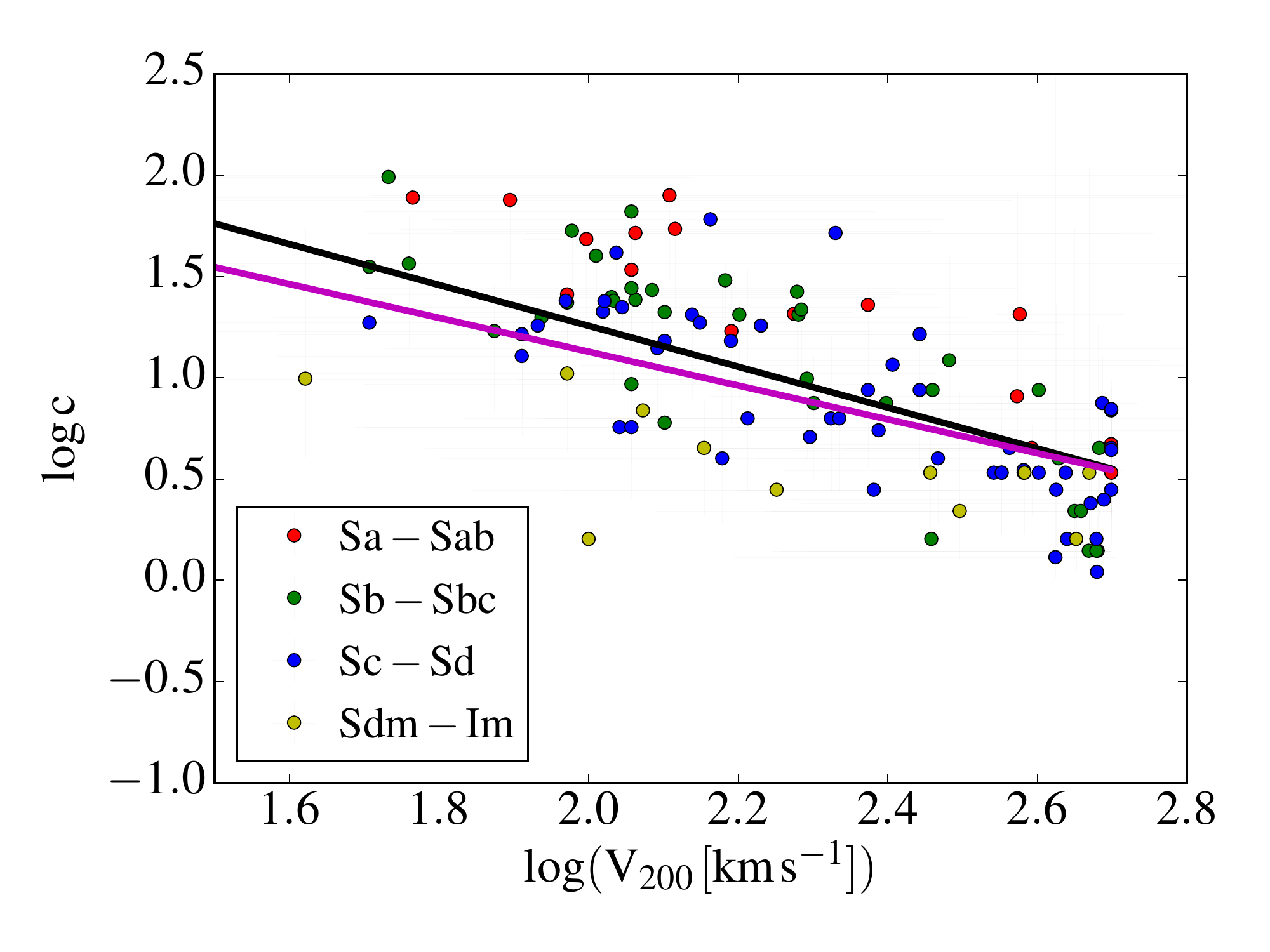}
\caption{Distribution of $\rm \log\ c $ as a function of $\rm \log\ V_{200}$ for the NFW best fit models (BFM) points. The black and magenta lines represent respectively the best fit model (BFM) and the fixed M/L models.}
\label{fig:nfw}
\end{figure}
In Fig. \ref{fig:nfw}, we plot the distribution of $\rm \log\ c$ as a function of $\rm \log\ V_{200}$. For the BFM, the correlation between c and V$_{200}$ is given by:
\begin{equation}       
\log\ c = (-1.01 \pm 0.09) \ \log\ V_{200}  + (3.27 \pm 0.22)
\end{equation}
We find an average value of concentration c = 8.71 $\pm$ 1.71, which iagrees within the error with the values usually found in the literature \citep[e.g. c $\simeq$ 10,][]{Bullock+2001} and with the value of $c$ = 6.9 derived by \citet{Noordermeer+2006} for a sample of early-type galaxies. 

When the M/L ratio is calculated using WISE colors, the correlation found between c and V$_{200}$ is:
\begin{equation}       
\log\ c = (-0.83 \pm 0.09) \ \log\ V_{200}  + (2.79 \pm 0.20)
\end{equation}
we find an average value c = 6.46 $\pm$ 1.59, closer to the results of \citet{Noordermeer+2006}. 

The mean concentration derived from the BFM is slightly higher than the one found using fixed M/L ratio, meaning that baryons may be responsible for the lower concentration inferred in the later case.
More interestingly, the slope of the linear regression linking $\log\ c$ to $\log\ V_{200}$ is also shallower in the later case, indicating that the concentration parameter is less dependent on the velocity at the virial radius when the M/L ratio is set by the baryonic content of the galaxy. However, this is a very small effect. More importantly, Fig. \ref{fig:nfw} shows that late-type galaxies tend to be located below the best fit line and the early type galaxies above, pointing out that early-type galaxies tend to display a more concentrated halo than later-type ones in the NFW paradigm. The same trend is also observed when M/L is fixed.

\begin{figure}
            \includegraphics[width=8.5cm]{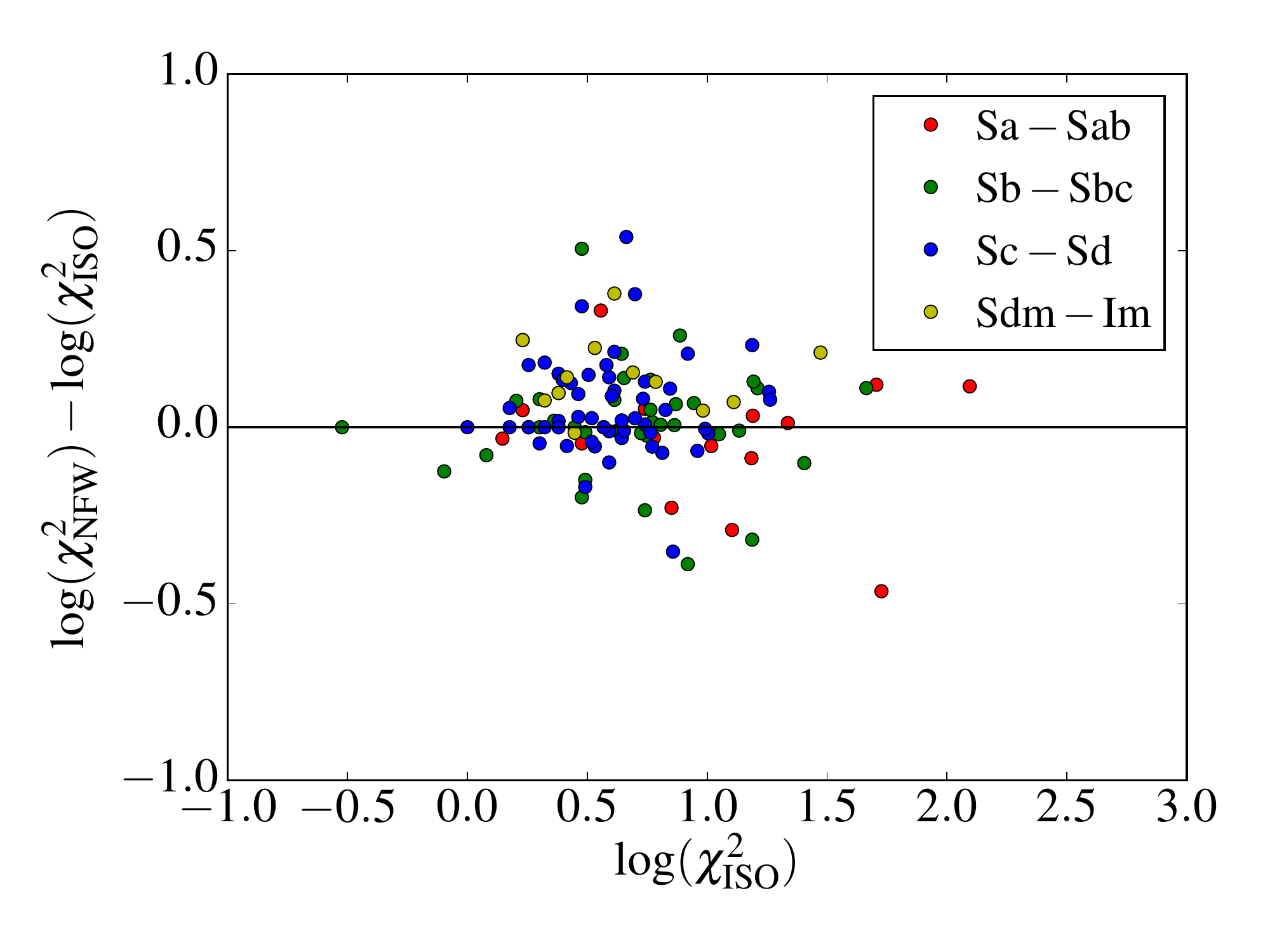}
            \includegraphics[width=8.5cm]{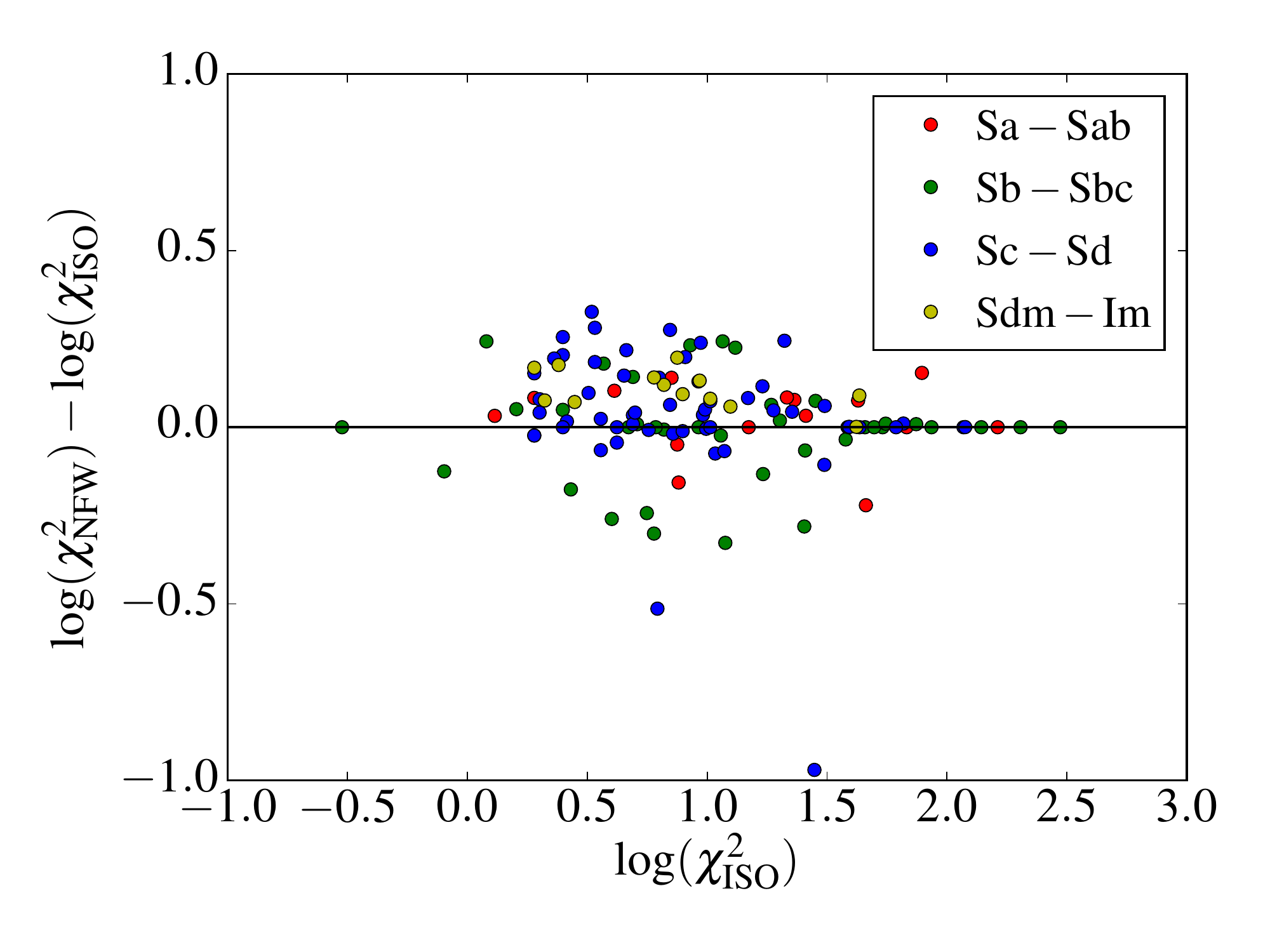}
\caption{The top panel presents the comparison between the reduced $\chi^2$ of NFW and ISO for the best fit model (BFM). The bottom panel shows the comparison between the reduced $\chi^2$ of NFW with fixed M/L and ISO with fixed M/L.}
\label{nfwiso}
\end{figure}

\subsection{ISO vs NFW profiles}

Fig. \ref{nfwiso} shows that, on average, smaller  $\chi^2$ values are found for the ISO fits than for NFW models. It can been seen in Tables \ref{tab:iso} to \ref{tab:nfw} listing all the parameters that, for a majority of galaxies, the observed rotation curves are better fitted by an ISO than by a NFW halo profile. Confirming previous works, our sample is thus better represented by a central cored than a cuspy DM density profile.

\section{Discussion}
\label{sect:discussion}

The presence of a bulge or of a bar may have an influence on the derived halo contribution. 

The previous studies on the mass distribution from high-resolution rotation curves, described earlier in this paper, have focused on late-type galaxies where the absence of a bulge makes these objects easier to analyse. We have here a sample that allows us to tackle the problem of earlier types bulge galaxies. 

The presence of a bar is also a major difficulty as it induces non-circular motions due to the streaming of the gas along the bar.   If a bar is perpendicular to the major axis of a galaxy, the measured rotational velocities  are overestimated and underestimated if the bar is parallel to the major axis \citep{Dicaire+2008,Randriamampandry+2015}, which may impact the mass models.

\subsection{Bulge influence}
\label{sect:irregular}

\begin{figure}
\begin{center}
	\includegraphics[width=8.5cm]{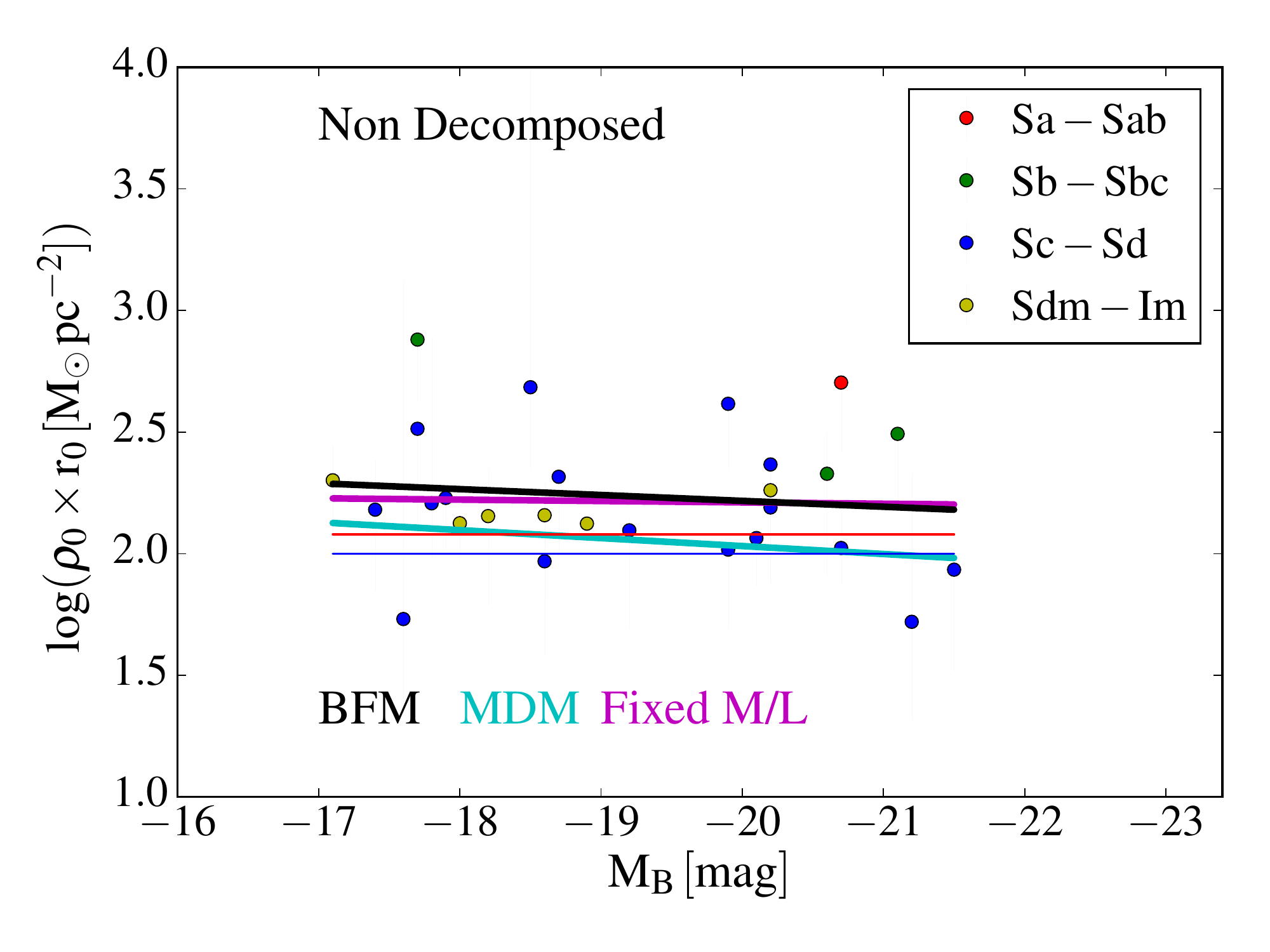}
	\includegraphics[width=8.5cm]{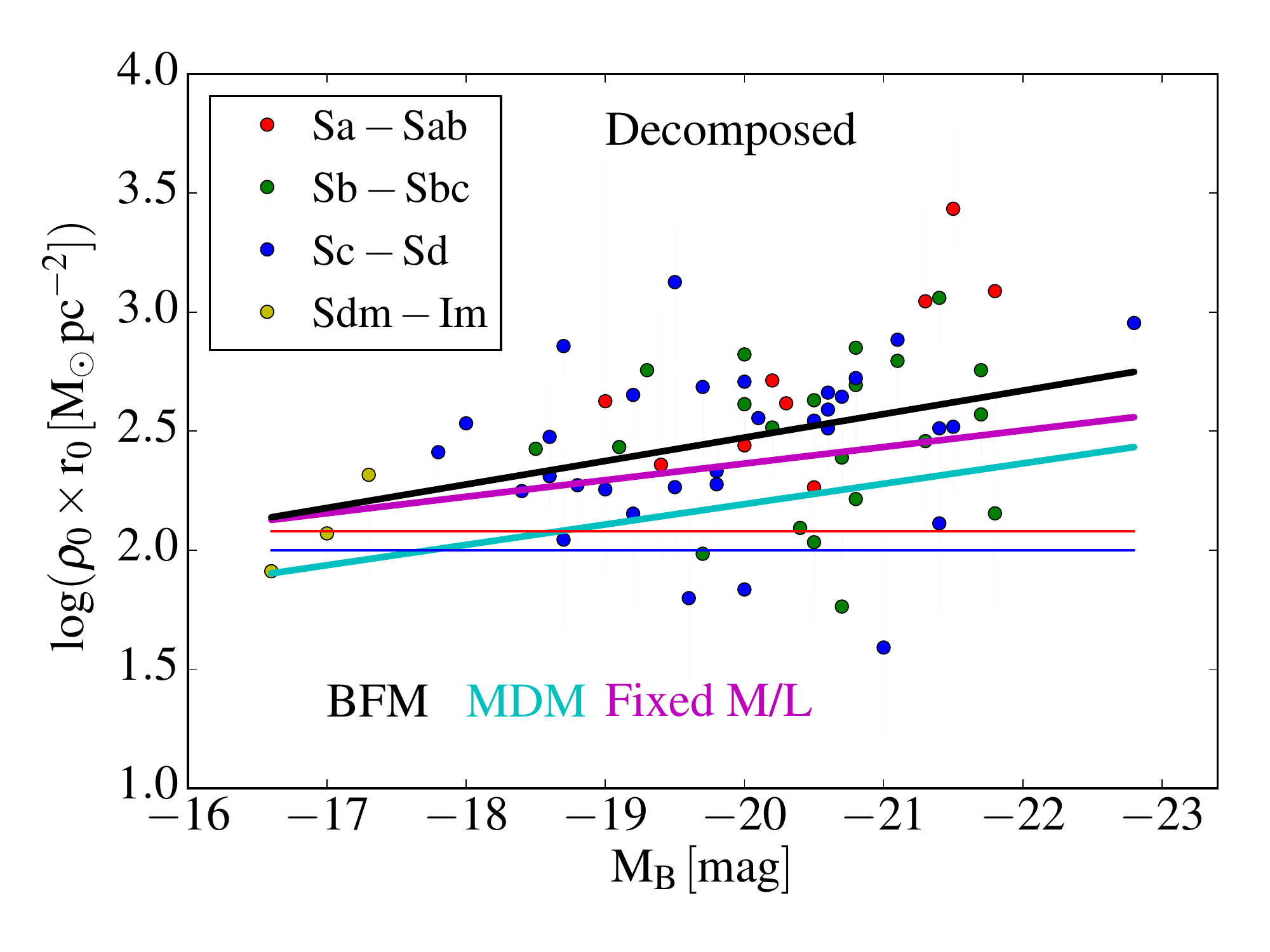}
	\includegraphics[width=8.5cm]{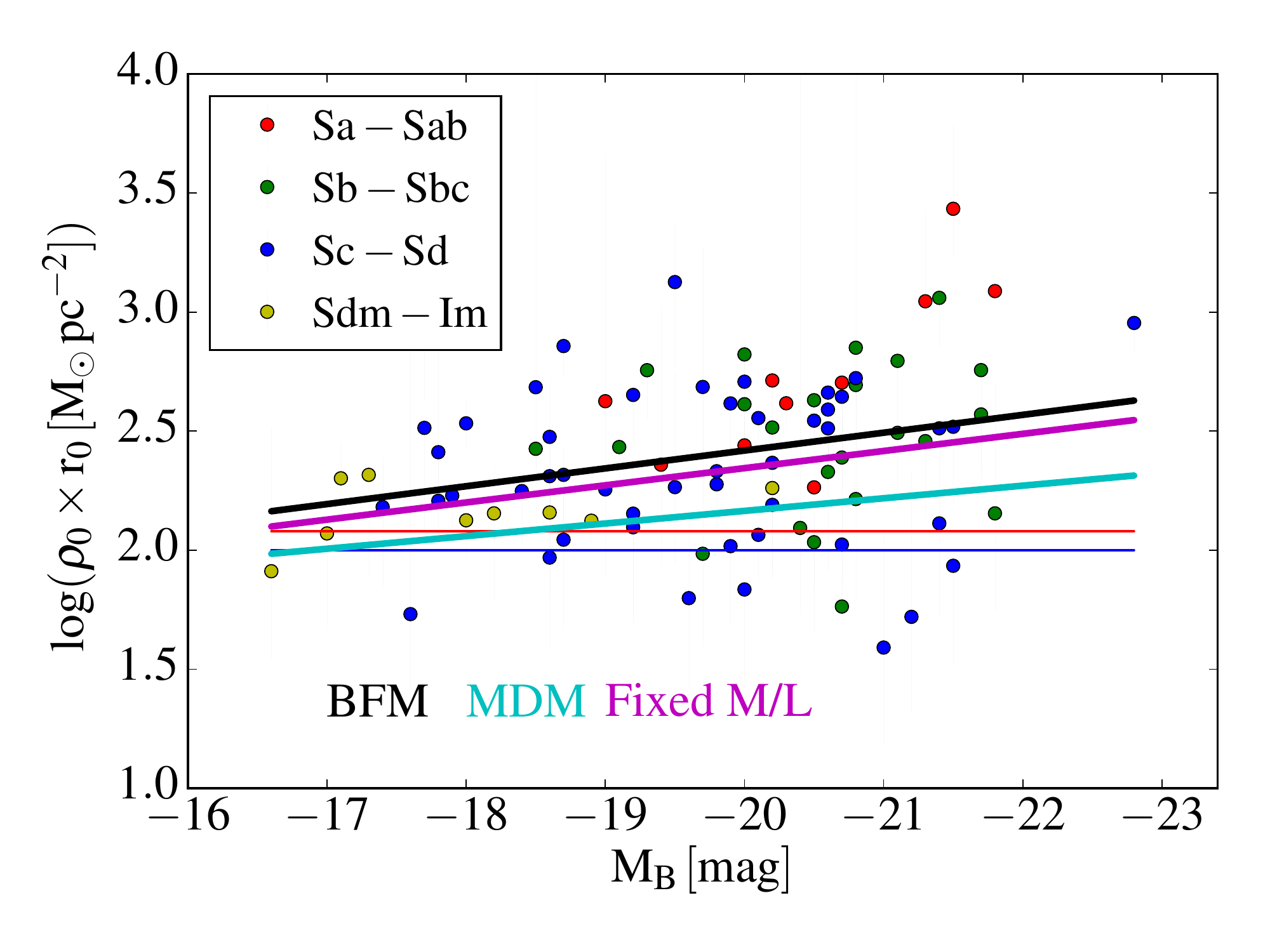}
\caption{$\rho_0 \times r_0$ as a function of $M_B$. The black, cyan and magenta lines represent the fit using the BFM, the MDM and the fixed M/L respectively. The thin blue and red lines represent respectively the fits found by \citet{Kormendy+2004} and \citet{Toky+2014} respectively. Top panel for galaxies with no luminosity profile decomposition (later types), middle for the galaxies for which we did a profile decomposition (earlier types) and the bottom panel for the whole sample.}
\label{product}
\end{center}
\end{figure}

Probably the most interesting scaling laws proposed by the previous studies \citep[first by][]{Kormendy+2004} is the fact that the product $\rho_0 \times r_0$ appears to be independent of luminosity. To understand the bulge influence on this result, we split our sample in two parts based on the luminosity profile decomposition presented in section \ref{Radial Profile Decompositions}: a first sample of 40 is composed of galaxies with very little or no bulge such that no decomposition was performed and the second set of 81 galaxies is composed of bulge galaxies for which we did a proper decomposition. The first group is essentially composed of the late type and irregular galaxies, but not only (see Fig. \ref{fig:luminosity}, third panel).
Using this classification scheme, we reproduce $\rho_0 \times r_0$ as a function of luminosity in Fig. \ref{product}. The galaxies with no decomposition (top) agree well with what was found before. However, if we take the whole sample including the early-type galaxies (bottom), this relation does not hold anymore and even less when using the galaxies for which we did a decomposition (middle).

In order to investigate this further, we used a slightly different approach to separate late and early types than the one used above, based on decomposing or not the luminosity profiles. We produced two sub-samples: (i) a {\it bulgeless} sample defined as the galaxies with $\rm L_{bulge}/L_{total} < 0.02$ and (ii) a {\it bulge} sample with $\rm L_{bulge}/L_{total} > 0.07$. So, the first sub-sample is mainly composed of late-type spirals while the second of earlier types galaxies. The advantage of this approach is that it better quantifies the bulge relative importance and enables to have two extreme classes with a comparable sample sizes, the bulgeless and bulge sub-samples having respectively 34 and 54 galaxies. 

\begin{figure*}
\begin{center}
	\includegraphics[width=8.5cm]{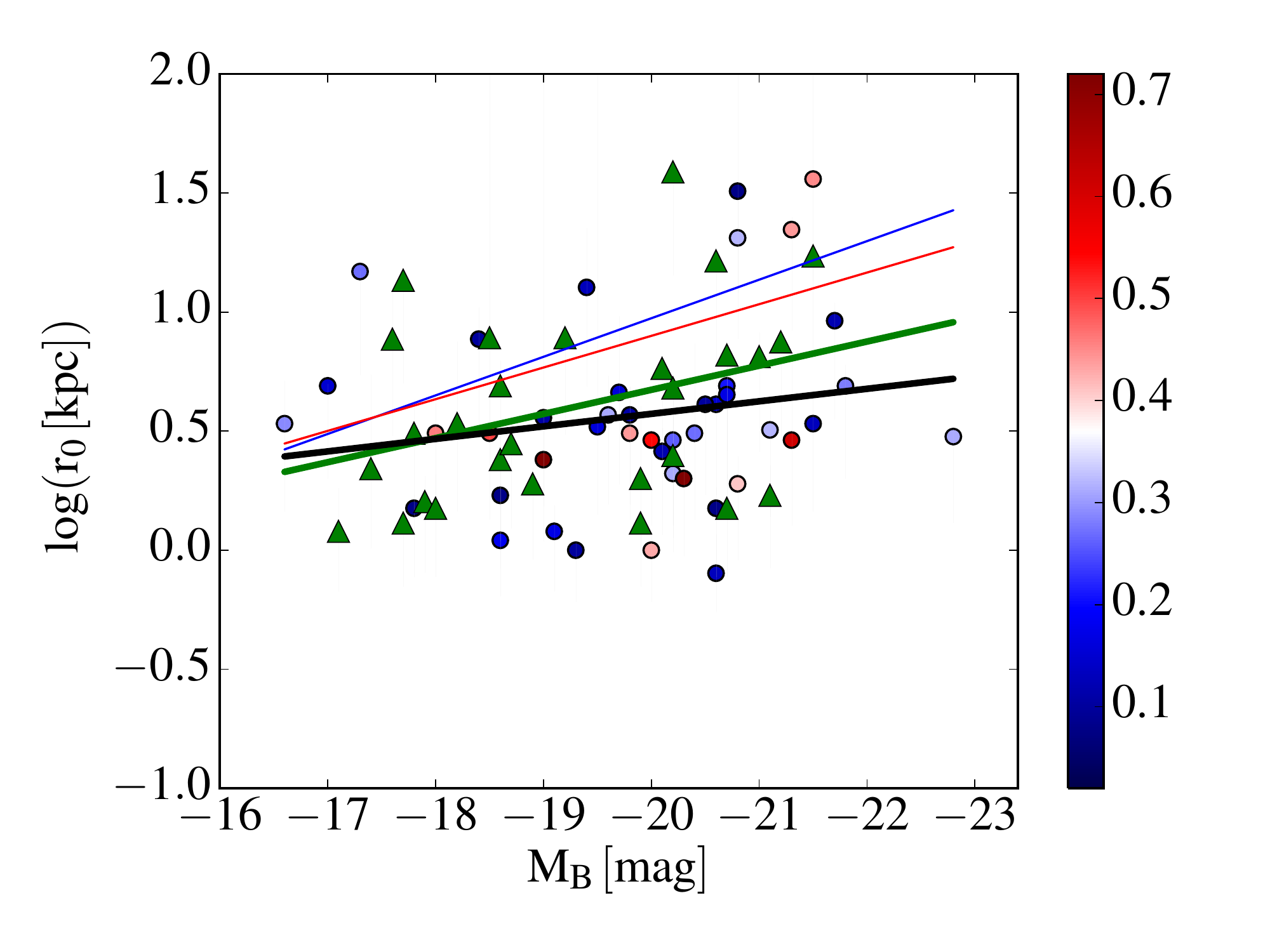}
	\includegraphics[width=8.5cm]{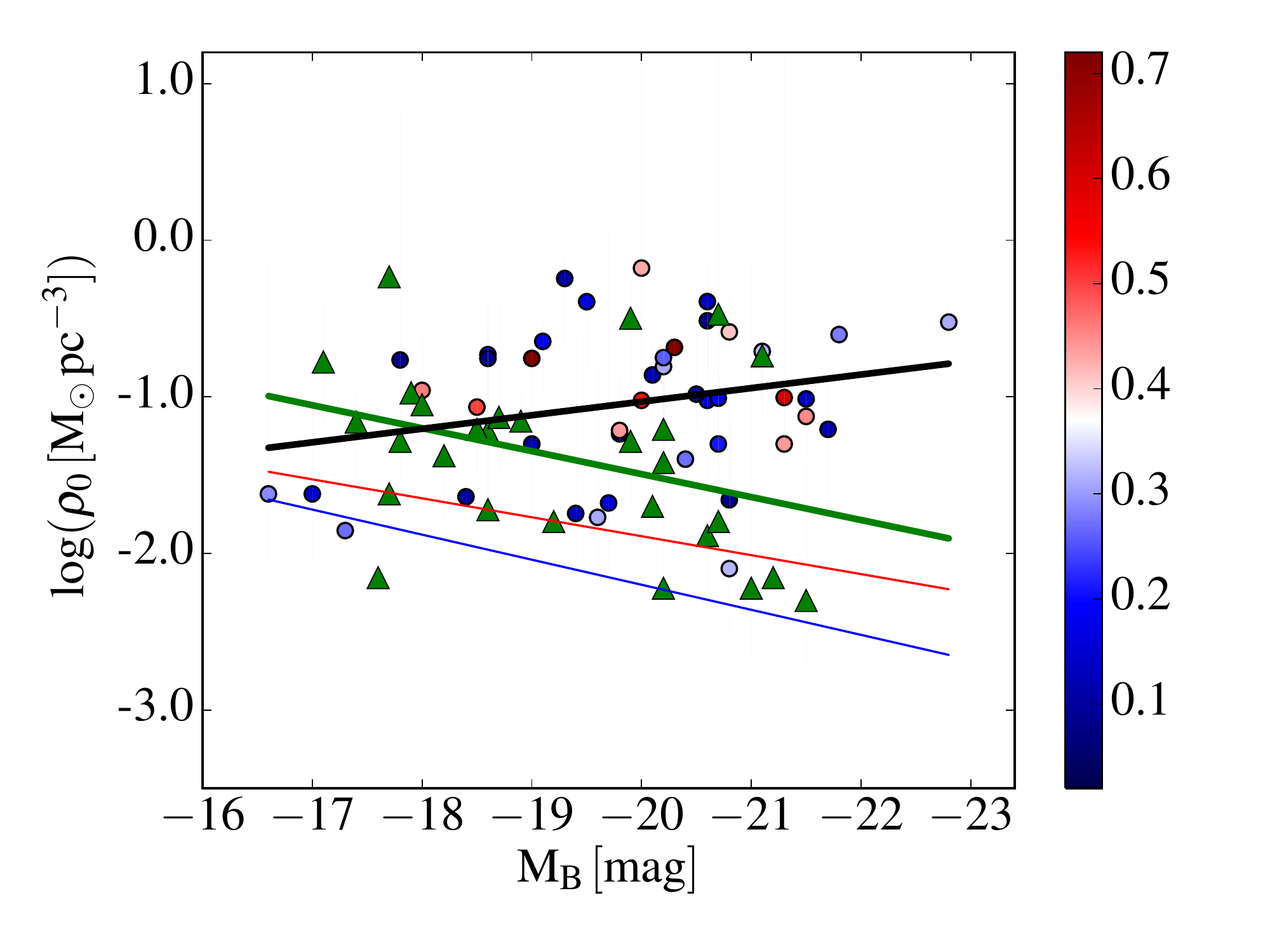}
\caption{Halo parameters (scaling radius on the left and central density on the right) versus absolute magnitude for the bulgeless (triangles) and bulge (dots) sub-samples. The colors of the dots represent the importance of the bulge (see color bar on the right). The thin blue  and red lines represent respectively the fit found by \citet{Kormendy+2004} and \citet{Toky+2014}, mainly for late-type galaxies.The thick green line represents the fit for the bulgeless and the thick black line for the bulge galaxies.}  
\label{rc_rho_BFM_mag_nobulge}
\end{center}
\end{figure*}

A comparison of the DM halo parameters with the absolute magnitudes is presented in Fig. \ref{rc_rho_BFM_mag_nobulge} for the BFM. 
Differences occur between bulgeless and bulge galaxies  when we compare the 2 parameters characterising the dark halo (scaling radius $r_0$ and central density $\rho_0$) to the absolute magnitudes $\rm M_B$. The bulgeless galaxies show a clearer correlation between r$_0$ and M$_B$ (left panel, green line) than bulge early-type spirals (left panel, black line). The main difference is found for $\rho_0$ that is anticorrelated to M$_B$ for bulgeless galaxies (right, green line), contrary to the correlation observed for the bulge galaxies (right, black line). Both trends for r$_0$ and $\rho_0$ for bulgeless galaxies were already found by \citet{Kormendy+2004} and \citet{Toky+2014}.
This shows that we should not generalize the scaling relations found previously to all morphological types. Clearly, DM is not distributed similarly in bulgeless and bulge systems. 
This either indicates intrinsic differences in the way DM is distributed in early-types and late-types or a problem in taking into account the luminous matter distribution of the bulge and the disc or a combination of both. Another possibility is that, in this work we are mainly probing the inner parts while many of the other studies were based on extended \Hi\, rotation curves.

In order to better understand the impact of using only optical rotation curves to constrain mass models rather than both optical and \Hi\, rotation curves, which generally extend well beyond the optical radius $D_ {25} / 2 $, we have used the \Hi\, rotation curves,  published in \citet{Toky+2014} and have artificially limited these rotation curves to their optical radius.
As shown in Appendix \ref{appendixC}, the optical rotation curves provide a good estimate of the mass model parameters when the plateau is reached, which is the case for 9 out of 15 galaxies: IC 2574, NGC 2403 (see Fig. \ref{ngc2403}), NGC 2841, NGC 3031, NGC 3621, NGC 55, NGC 7331, NGC 7793, NGC 925. The mass models offer a less satisfactory agreement when the optical curves are limited to their solid body rising part, which is the case for 6 out of 15 galaxies: DDO 154, NGC 2366 (see Fig. \ref{ngc2366}), NGC 247, NGC 300, NGC 3109 , NGC 3198. The tuning is particularly good when the constraints of the rotation curve come from internal regions (eg NGC 2403, NGC 7331, NGC 7793, NGC 925) and it is even the case to a lesser level when the plateau of the rotation curve is barely achieved (eg NGC 55, NGC 3621) or when it is not fully achieved (e.g. NGC 247). The case of NGC 3621 is interesting because the plateau is reached although the rotation curve is still slightly increasing beyond the optical radius. In addition, galaxies with decreasing rotation curves already observed within the optical radius show very good agreement (eg NGC 2841, NGC 3031, NGC 7331, NGC 7793). Indeed, decreasing rotation curves are generally observed for early-type galaxies, whose central regions are dominated by the presence of a bulge that optical data can constrain well.

\subsection{Bar influence}
\label{subsect:bar}
 
\begin{figure}
\begin{center}
	\includegraphics[width=8.5cm]{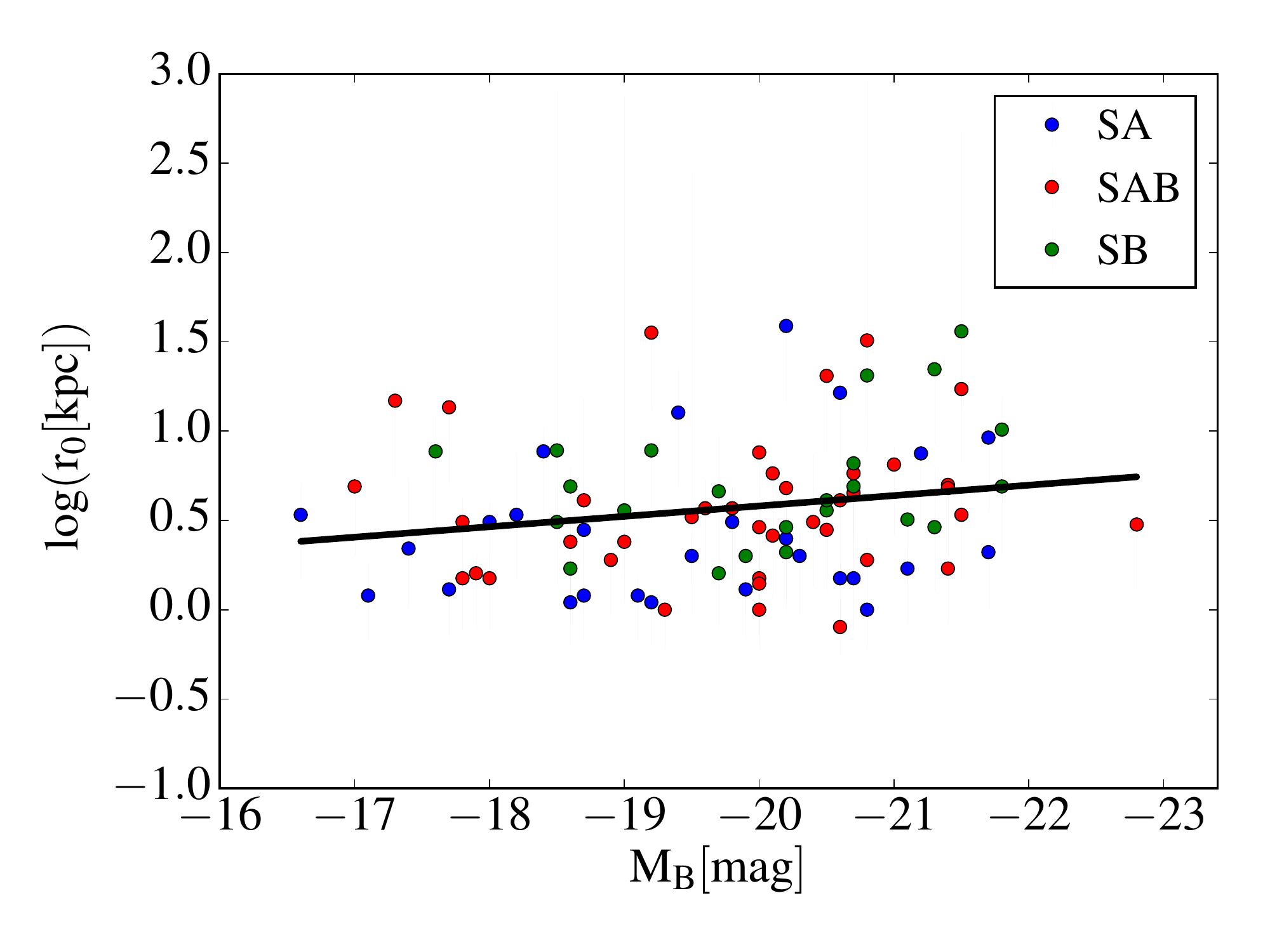}
	\includegraphics[width=8.5cm]{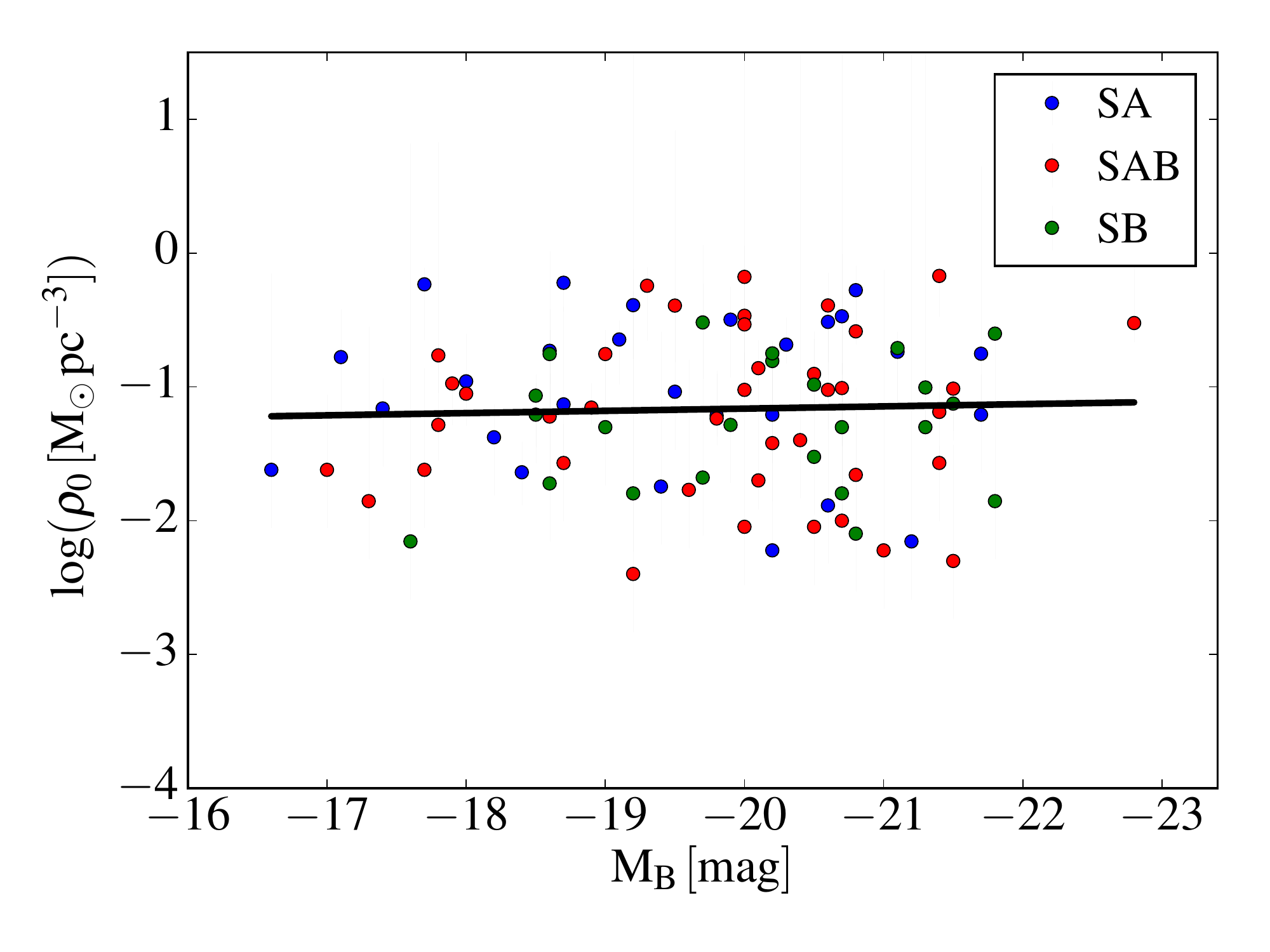}
\caption{Halo core radius (top) and central halo density (bottom) versus the absolute magnitude for the whole sample. The blue dots indicate non barred galaxies (SA); the red dots represent moderately barred galaxies (SAB) and the green dots display barred galaxies (SB).}
\label{fig:bar}
\end{center}
\end{figure}

To study the effect of the bar, we split the sample in non-barred (SA), moderately barred (SAB) and barred (SB) galaxies: the first sample of 31 is composed of galaxies with no bar, the second set of 46 galaxies is composed of systems with moderate bars and the third sample of 28 barred galaxies. 
In Fig. \ref{fig:bar}, we show the correlations between the halo core radius r$_0$ (top) or the central halo density $\rho_0$ (bottom) and the M$_B$ using different colors for non-barred galaxies (SA, blue), moderately barred galaxies (SAB, red) and barred galaxies (SB, black). These plots show that these correlations do not depend on the fact that a galaxy has developed a bar or not; the scatter around the 
fits is roughly identical for the three categories of galaxies.

\section{Summary and Conclusion}
\label{sect:conclusion}

In this work, we have studied mass models using H$\alpha$ rotation curves and mid-infrared photometry on a sample of 121 galaxies which cover all morphological types. Luminosity profiles have been decomposed using a flat (disc, bar, etc.) and a spherical (bulge) component. We used a pseudo-isothermal sphere (ISO) and a Navarro-Frenk-White model (NFW) to describe the dark matter halos and used various fitting strategies: a best fit model (BFM), a maximum disc model (MDM) and a M/L fixed using colors.

The two BFM models (ISO and NFW) used to describe the distribution of the DM halo within the visible discs are both acceptable, with ISO giving somewhat better results. For ISO with maximum disc fit, the M/L values are similar to the M/L obtained from the (W$_1$-W$_2$) color. This suggests that discs tend to be ``maximum''.
The ISO (MDM), ISO and NFW with M/L fixed from colors give also reasonable fits for almost all galaxies.

Previous works considered in this paper for comparison were carried on samples of galaxies mainly composed of late morphological types. This difference in galaxy populations with our sample, which covers all morphological types, may explain the disagreement between our results and those published in the literature.
For the ISO model, the trend between the dark halo parameters and the absolute magnitude found for no bulge galaxies in this paper (less luminous galaxies tend to have smaller core radius and higher central density) is in agreement with previous works \citep{Kormendy+2004,Toky+2014} but not the relations found for bulge galaxies. For bulge galaxies, less luminous galaxies tend to have smaller core radius and smaller central density.
However, we find that the trend between $\rho_0$ and $r_0$ does not depend on morphological types.
For the NFW profile, we can see that the halo is more concentrated for early type galaxies than for late type galaxies whatever technique is used (BFM or fixed M/L), again showing a difference in DM properties between early and late morphological types.
Therefore, the relation between the dark halo parameters and the luminosity of the galaxies seems dependent on the morphological types. 
We also checked whether the presence of a bar could impact the observed DM correlations and found not convincing trend.

In order to ensure that our results are not affected by the fact that we only consider optical rotation curves, we have analyzed the sample of \cite{Toky+2014} by truncating their \Hi\, rotation curves to the optical radius. We found that mass models are in good agreement when the plateau is (barely) reached within this radius, which is usually the case for the relatively early types of galaxies.

While the effect of adiabatic contraction of the halo by the disc is to steepen cores into cusps \citep{Dutton+2005}, observations continue to argue that dark matter profiles are more core-like. Stellar feedback at high redshift may flatten the cusps in the less massive galaxies by removing low angular momentum material leading to potential fluctuations decreasing the central density \citep{NavarroEke+1996} but those processes are not effective enough for massive galaxies with deeper potential well \citep{MacLow+1999}.

\section{Acknowledgments}

We thank Carlos Barbosa for the methods of the decomposition of surface brightness profiles.
Most of the research of MK was done while she was having a PhD Scholarship from the Science faculty of the University of Cape Town. 
CC's work is based upon research supported by the South African Research Chairs Initiative (SARChI) of the Department of Science 
and Technology (DST),  the Square Kilometer Array South Africa (SKA SA) and the National Research Foundation (NRF).
We acknowledge financial support from ``Programme National de Cosmologie et Galaxies'' (PNCG) funded by CNRS/INSU-IN2P3-INP, CEA and CNES, France.





\bibliographystyle{mnras}
\bibliography{exemple_biblio}

\onecolumn
\appendix
\section{Tables} 
\label{appendixA}
We list here the global properties and mass models parameters of 8 of the sample galaxies. The remaining galaxies are available in the online version. 
\input{latex_photometry_paper}



\section{Surface brightness profile and mass models} 
\label{appendixB}
We present the photometry, the luminosity profiles and the mass models of 2 of the 121 galaxies. The remaining galaxies are available in the online version.

\begin{figure*}
\includegraphics[width=10.0cm]{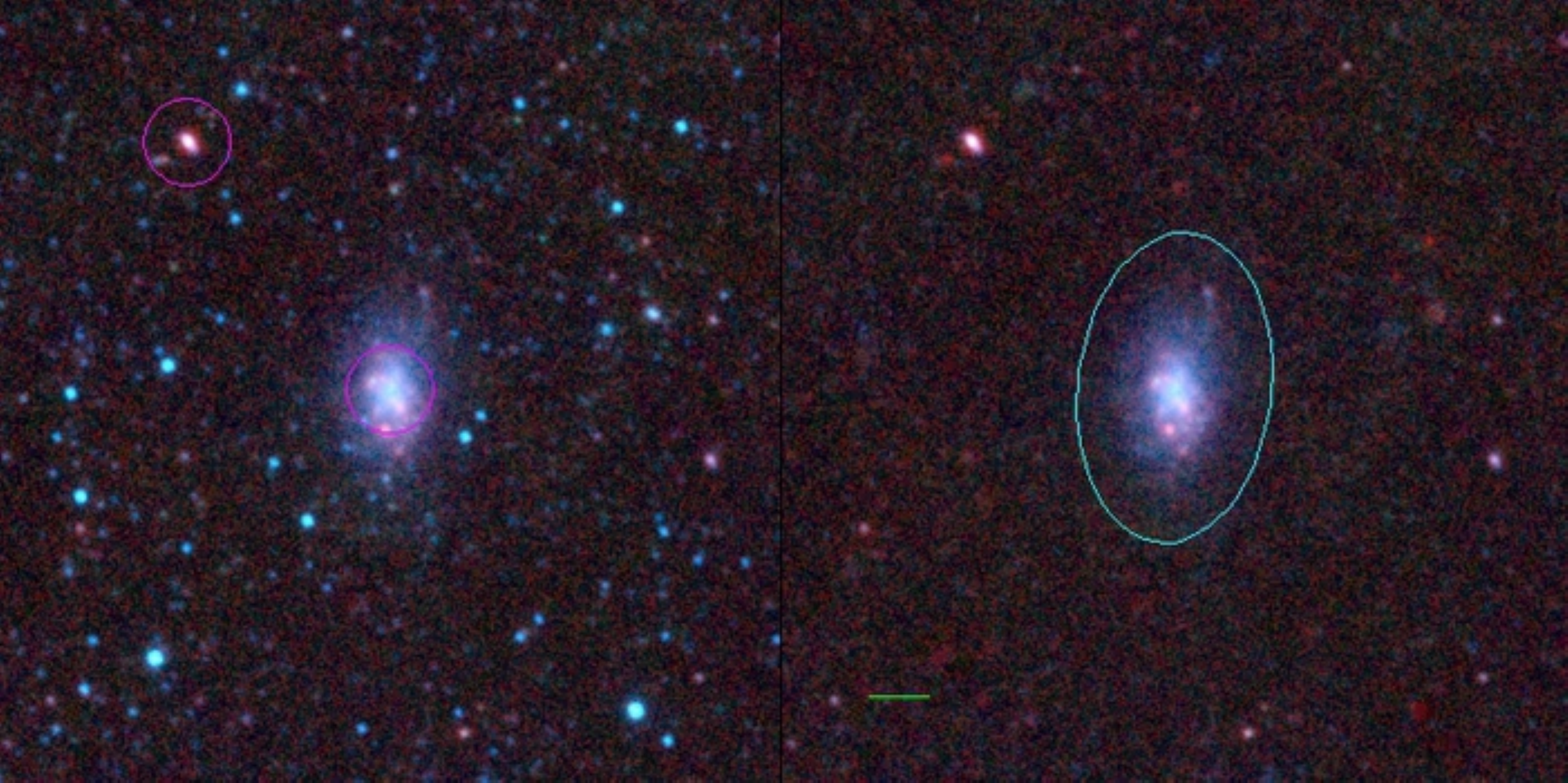}
\includegraphics[width=6.5cm]{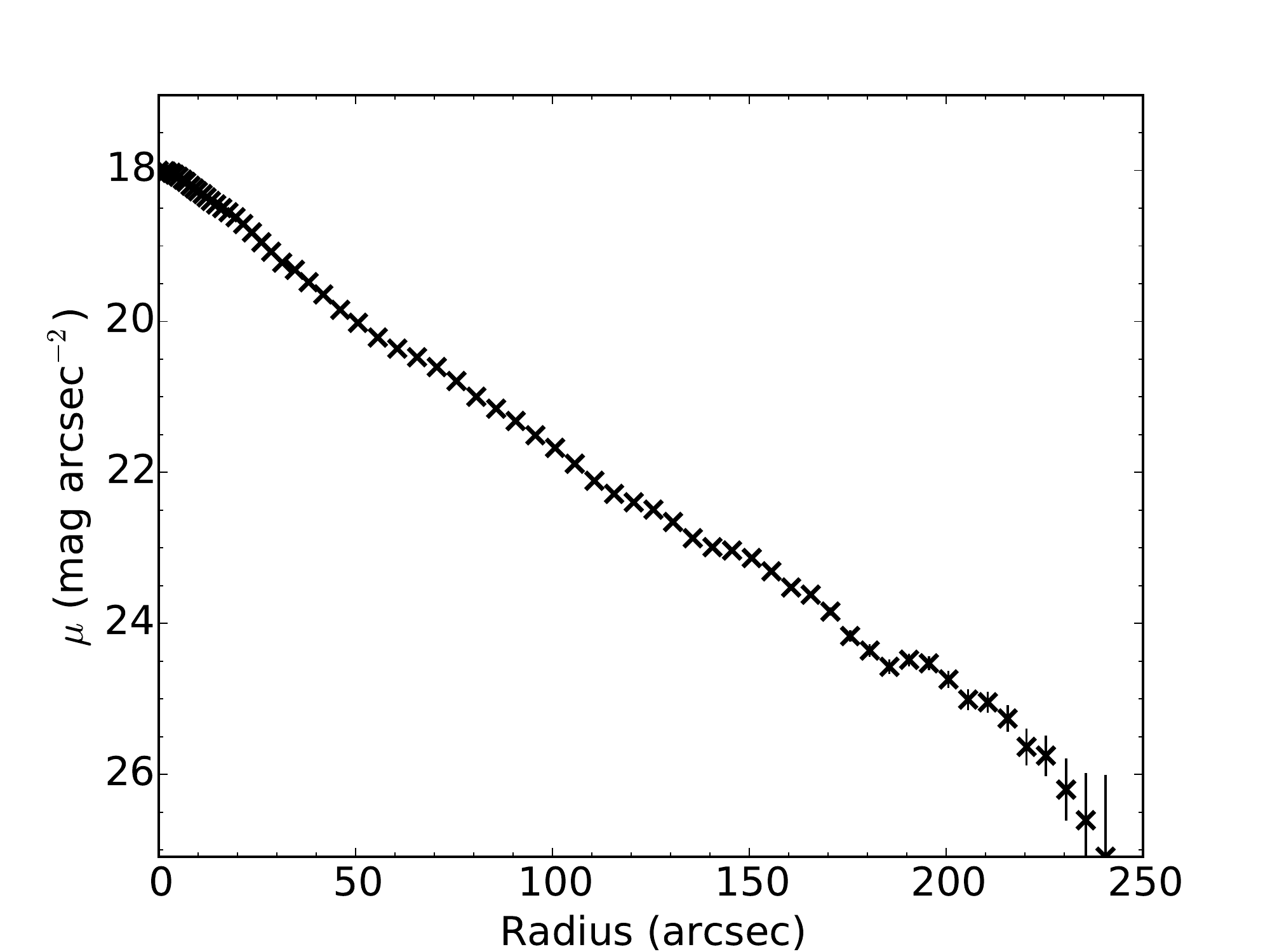}\\
\hspace*{-0.00cm} \includegraphics[width=0.35\textwidth]{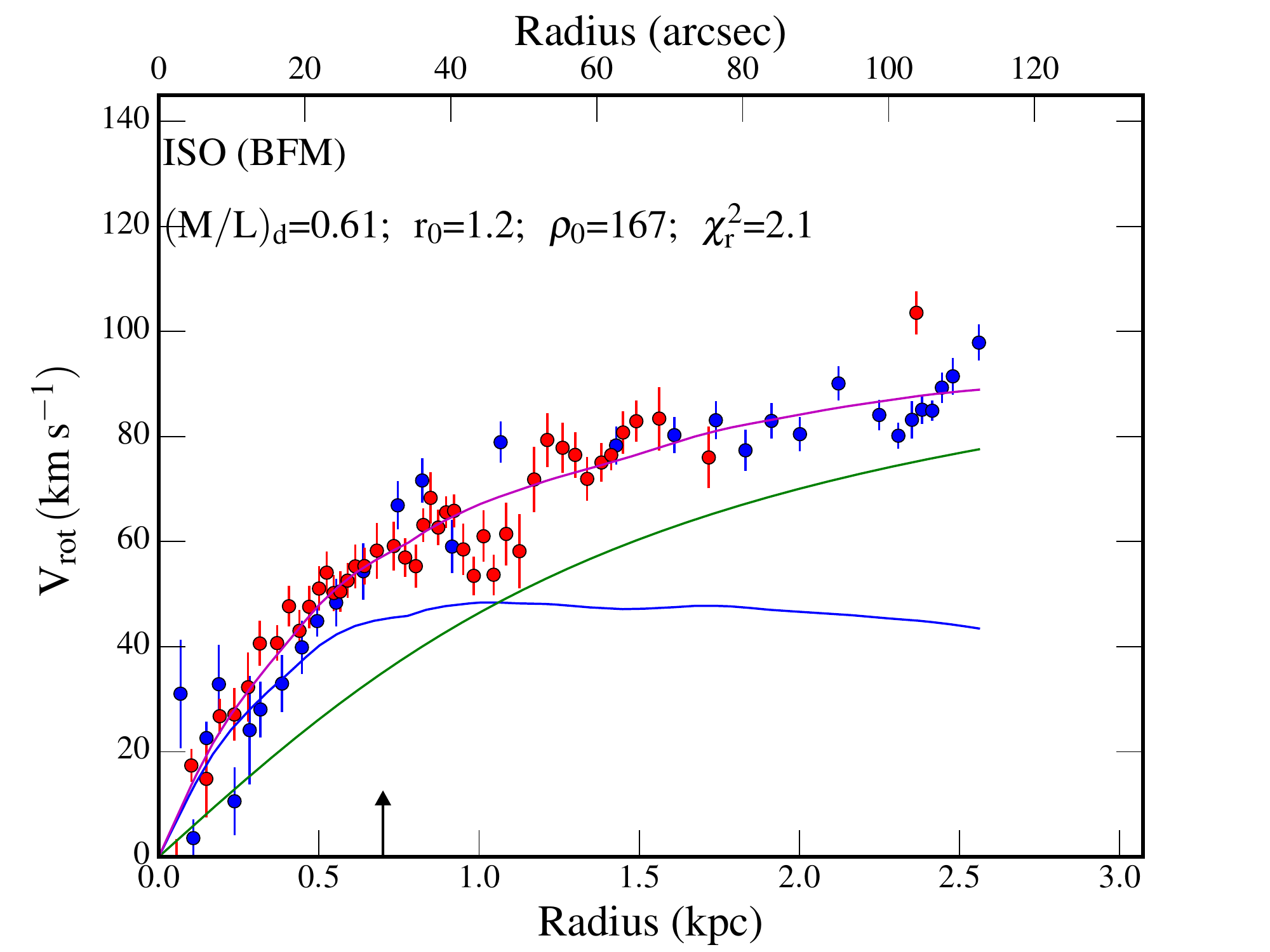}
\hspace*{-0.75cm} \includegraphics[width=0.35\textwidth]{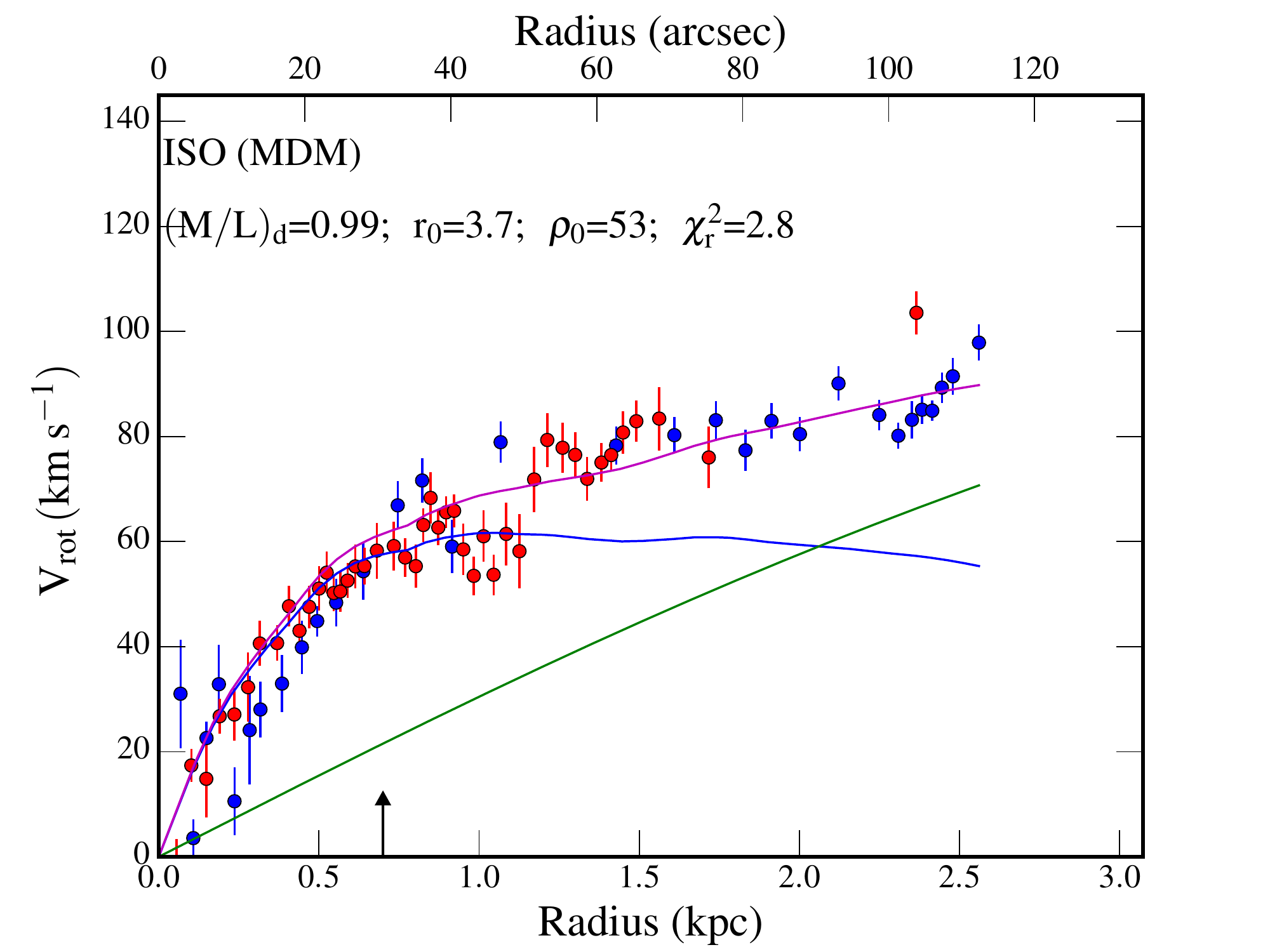}
\hspace*{-0.75cm} \includegraphics[width=0.35\textwidth]{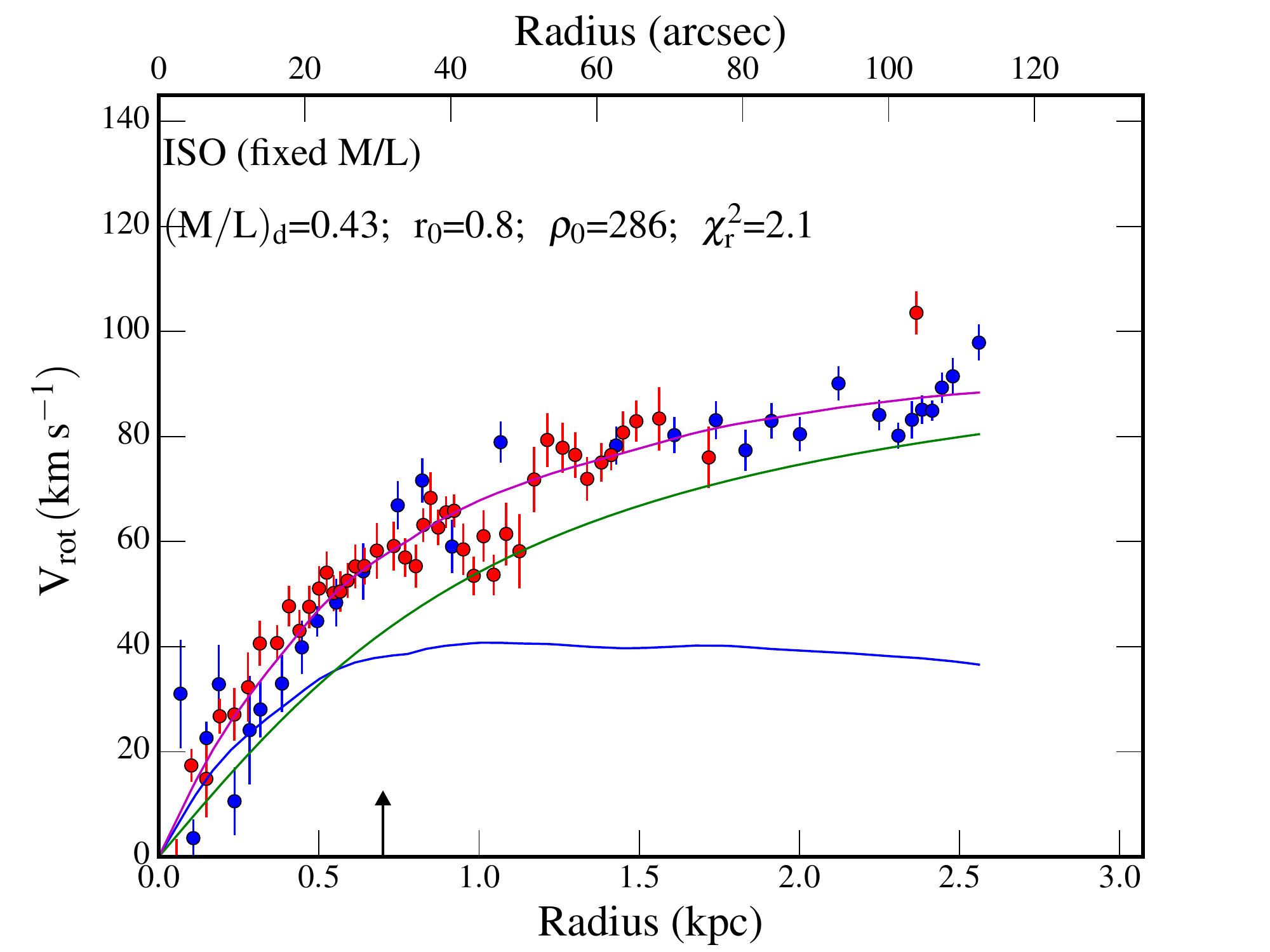}\\
\hspace*{-0.00cm} \includegraphics[width=0.35\textwidth]{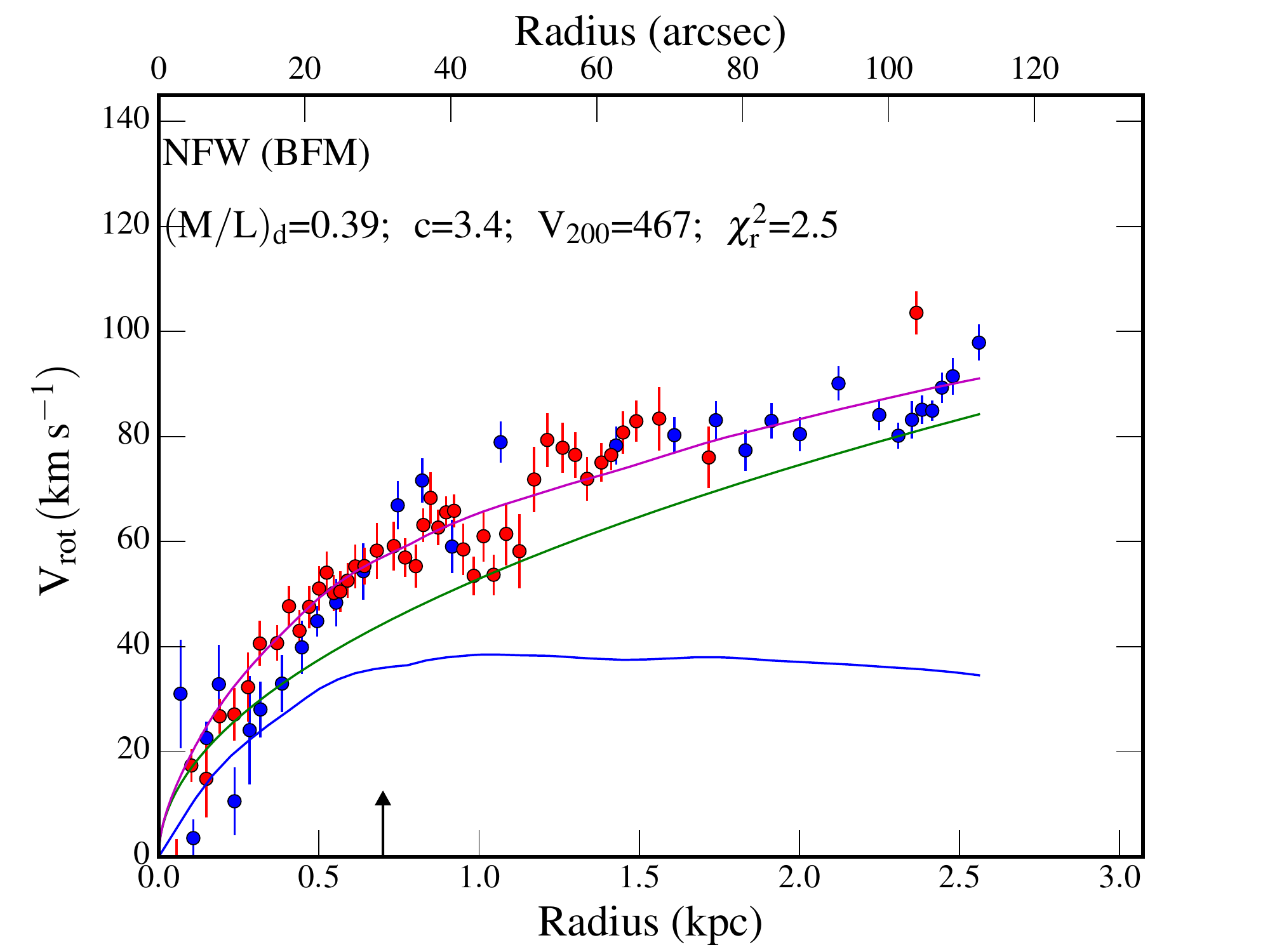}
\hspace*{-0.25cm} \vspace{-1.25cm} \includegraphics[width=0.31\textwidth]{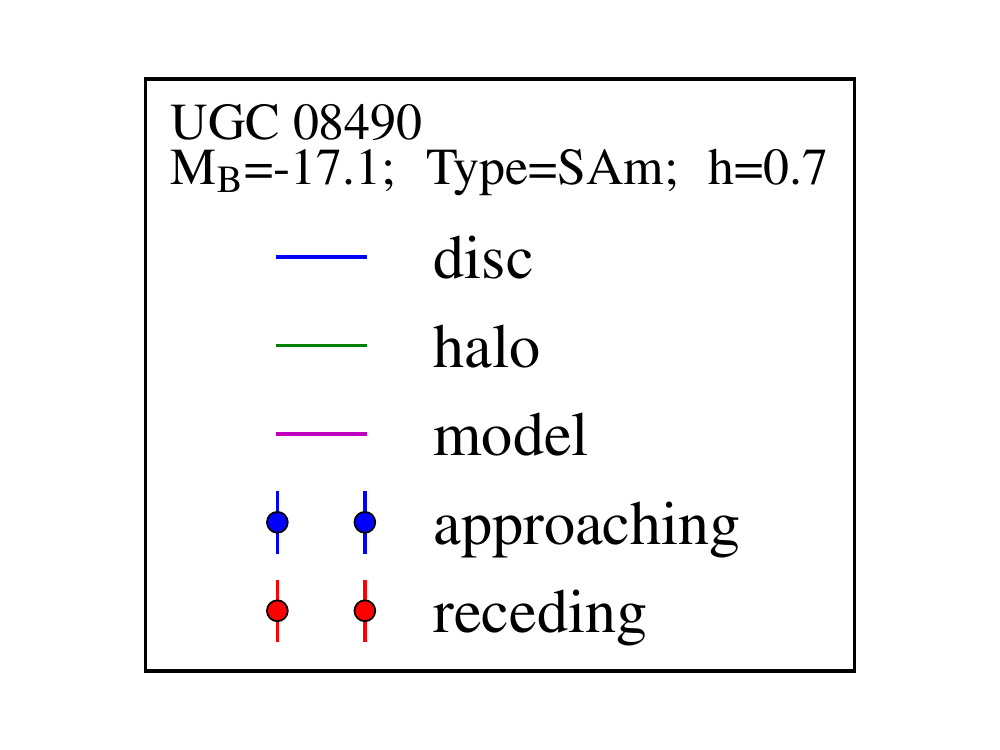} \vspace{1.25cm} \hspace*{-0.5cm}
\hspace*{-0.00cm} \includegraphics[width=0.35\textwidth]{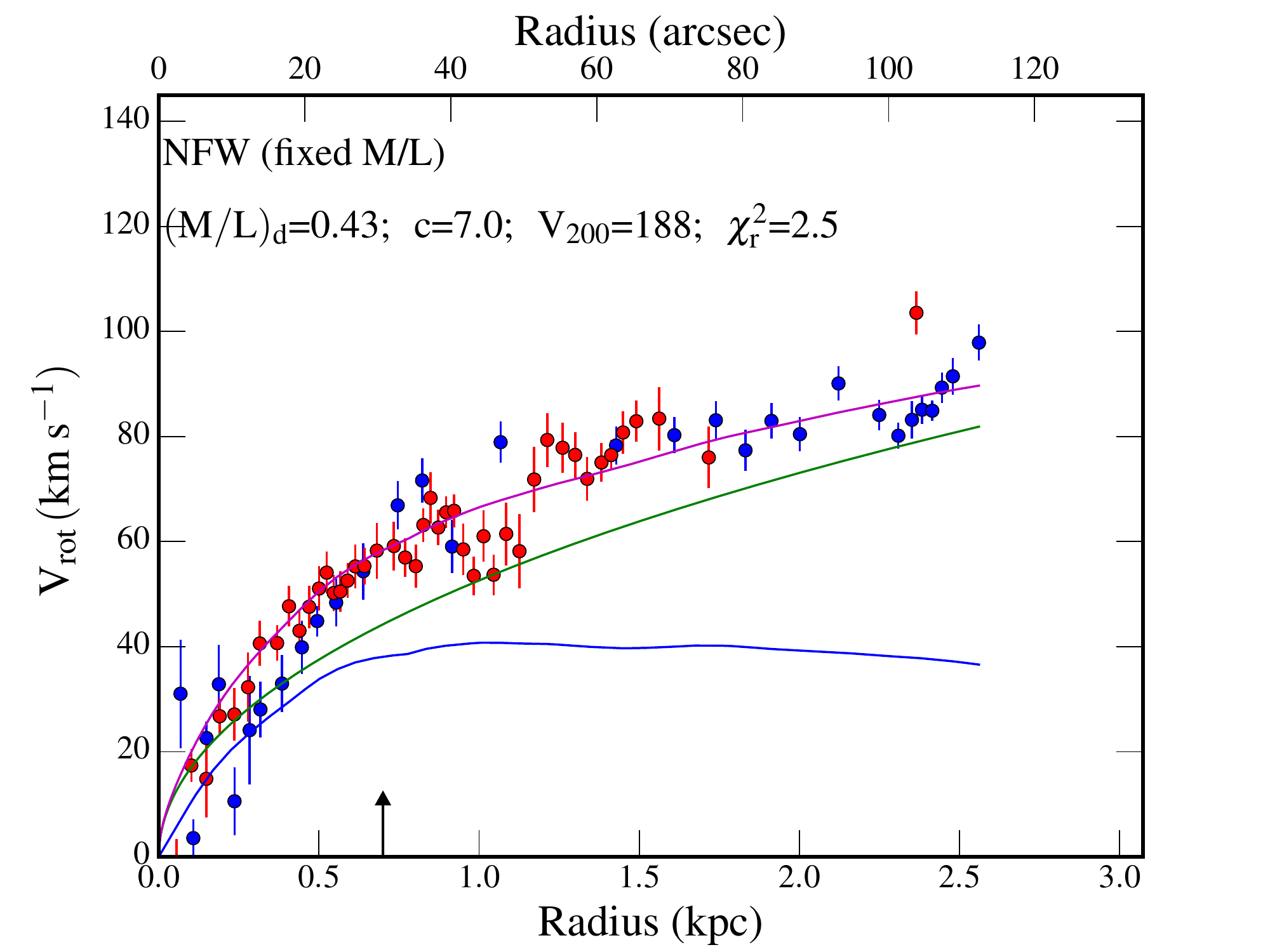}\\
\caption{Example of surface brightness profile decomposition and mass models for the galaxy UGC 8490. First line - (Left panel) WISE surface brightness image at 3.4 $\mu$m. Left side: image showing the field and galaxy. Right side: image after stars removed; the circle represents the 1 $\sigma$ isophotal ellipse, used for integrated photometry and the green line shows 1 arcmin in length. (Right panel) Luminosity profile decomposition corresponding to the left image. Lines 2 and 3 - Mass models.  Second line: pseudo-isothermal sphere density profiles (ISO). Third line: Navarro, Frenk \& White density profiles (NFW). First column: Best Fit Model (BFM). Second column: Maximum disc Model (MDM) for line 2 (ISO model). Third column: Mass-to-Light ratio M/L fixed using WISE W$_1$-W$_2$ color.  The name of the galaxy, its B-band absolute magnitude, morphological type and disc scale length have been indicated in the insert located line 3-column 2. For each model, the fitted parameters and the reduced $\chi^2$ have been indicated in each sub-panel.}
\label{massmodel1}
\end{figure*}

\begin{figure*}
\includegraphics[width=10.0cm]{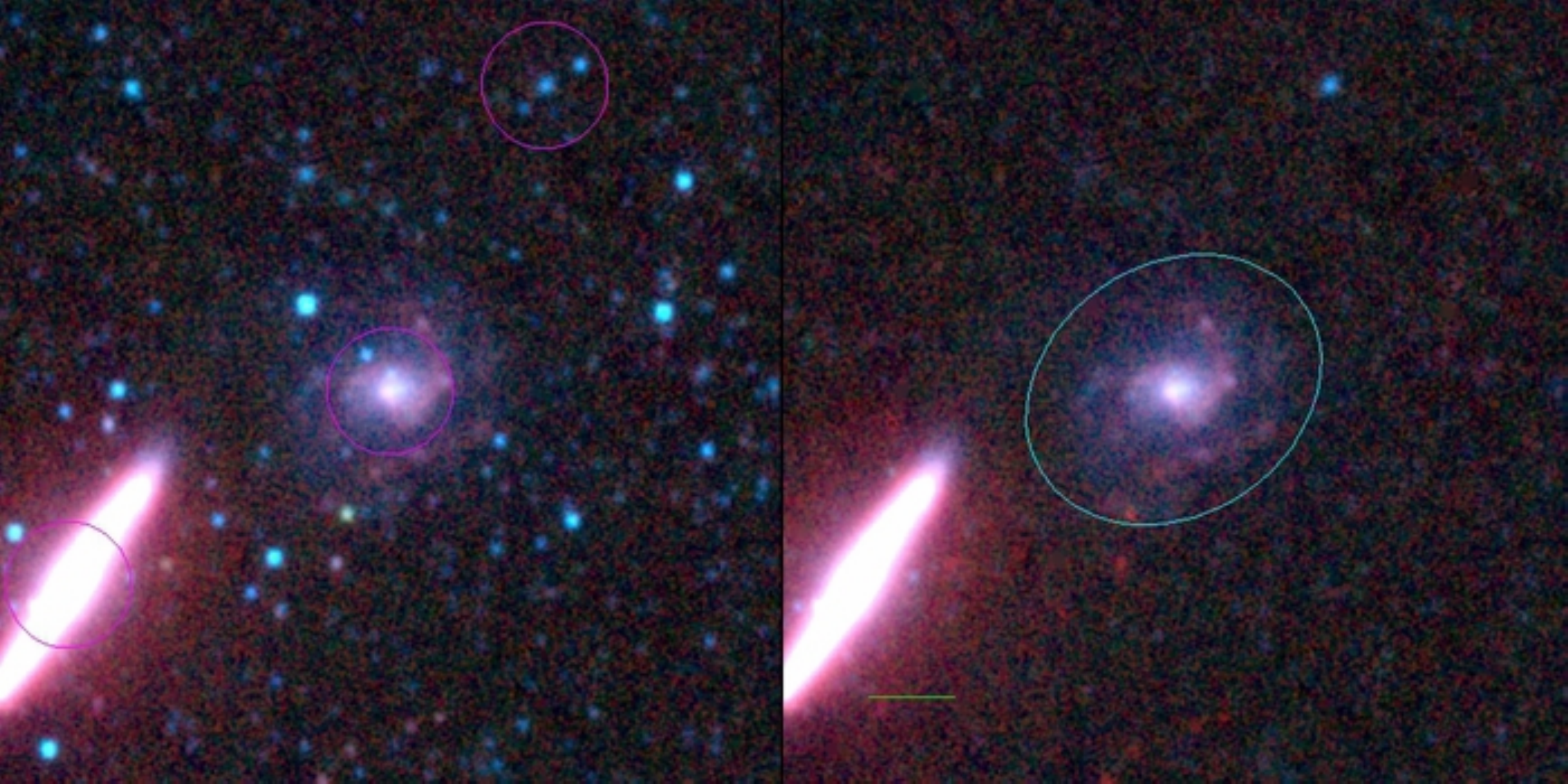}
\includegraphics[width=6.5cm]{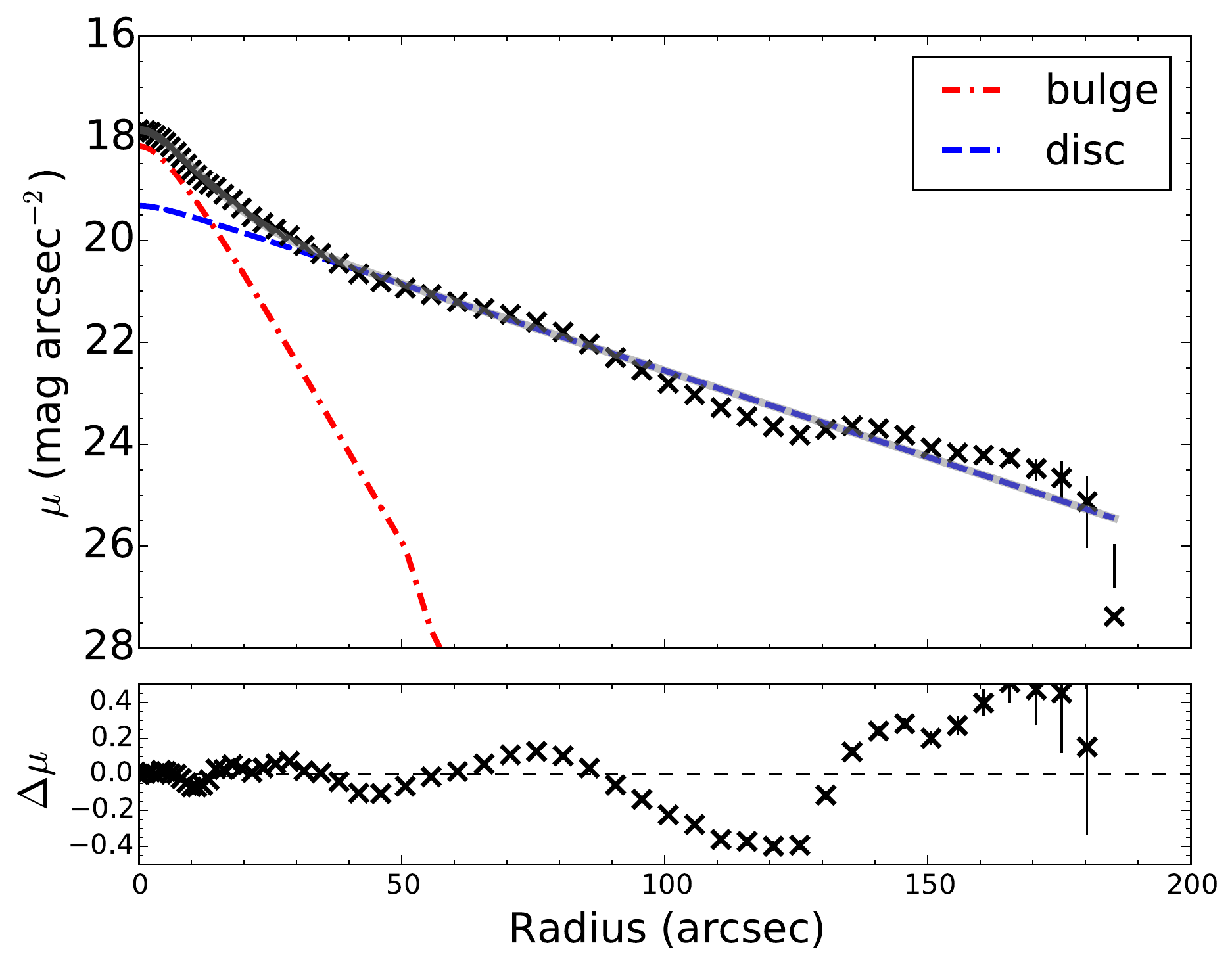}\\
\hspace*{-0.00cm} \includegraphics[width=0.35\textwidth]{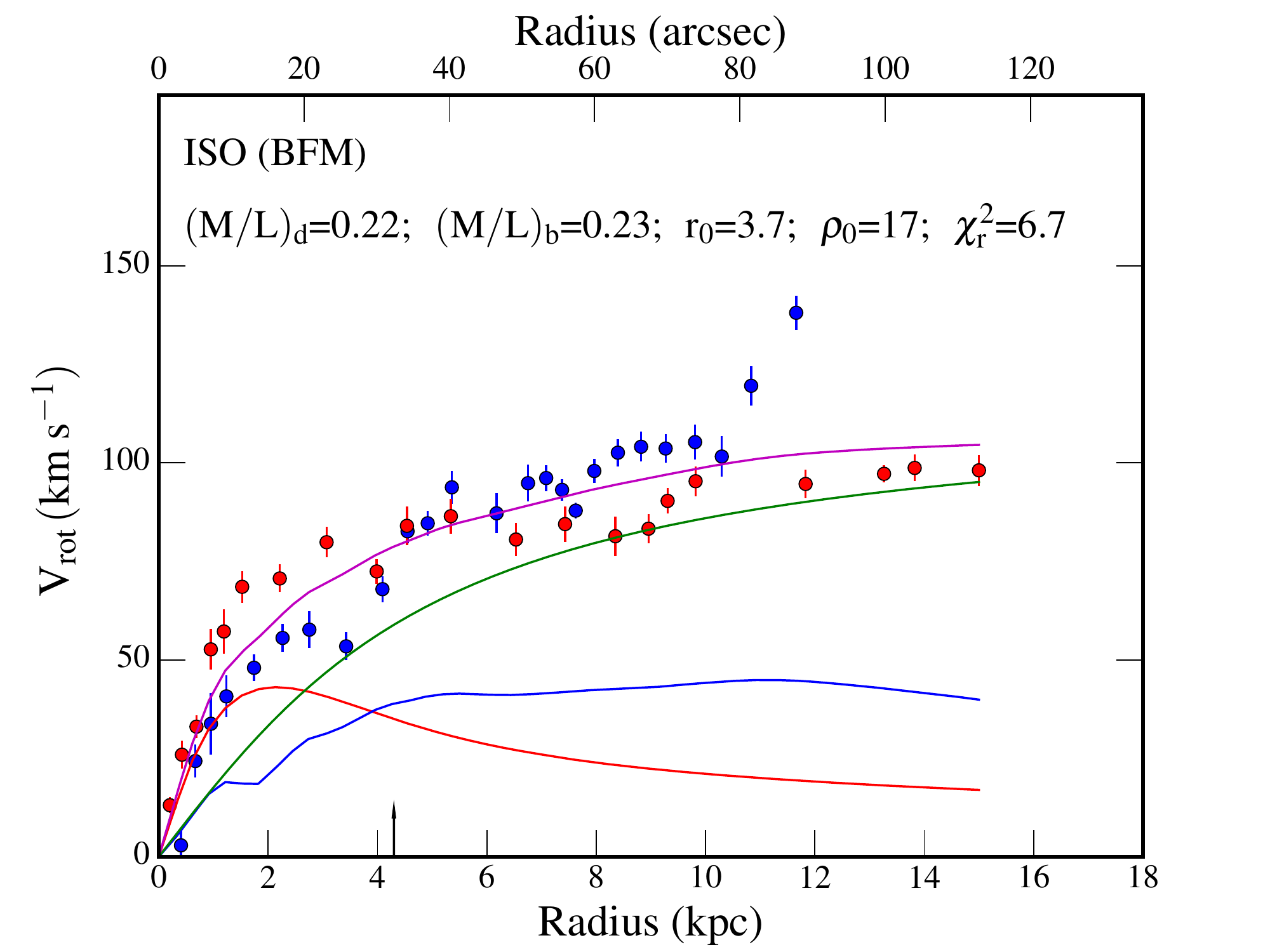}
\hspace*{-0.75cm} \includegraphics[width=0.35\textwidth]{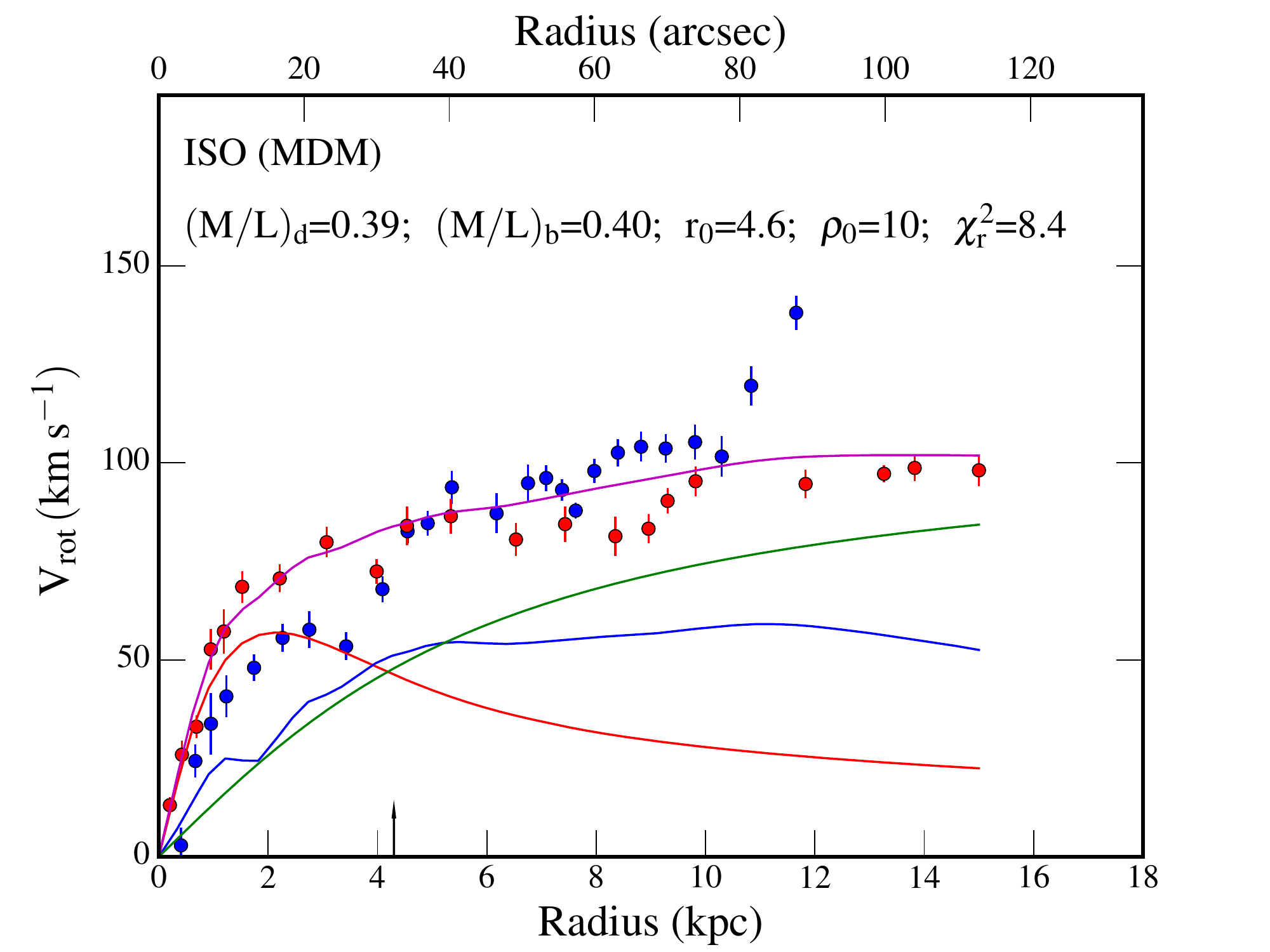}
\hspace*{-0.75cm} \includegraphics[width=0.35\textwidth]{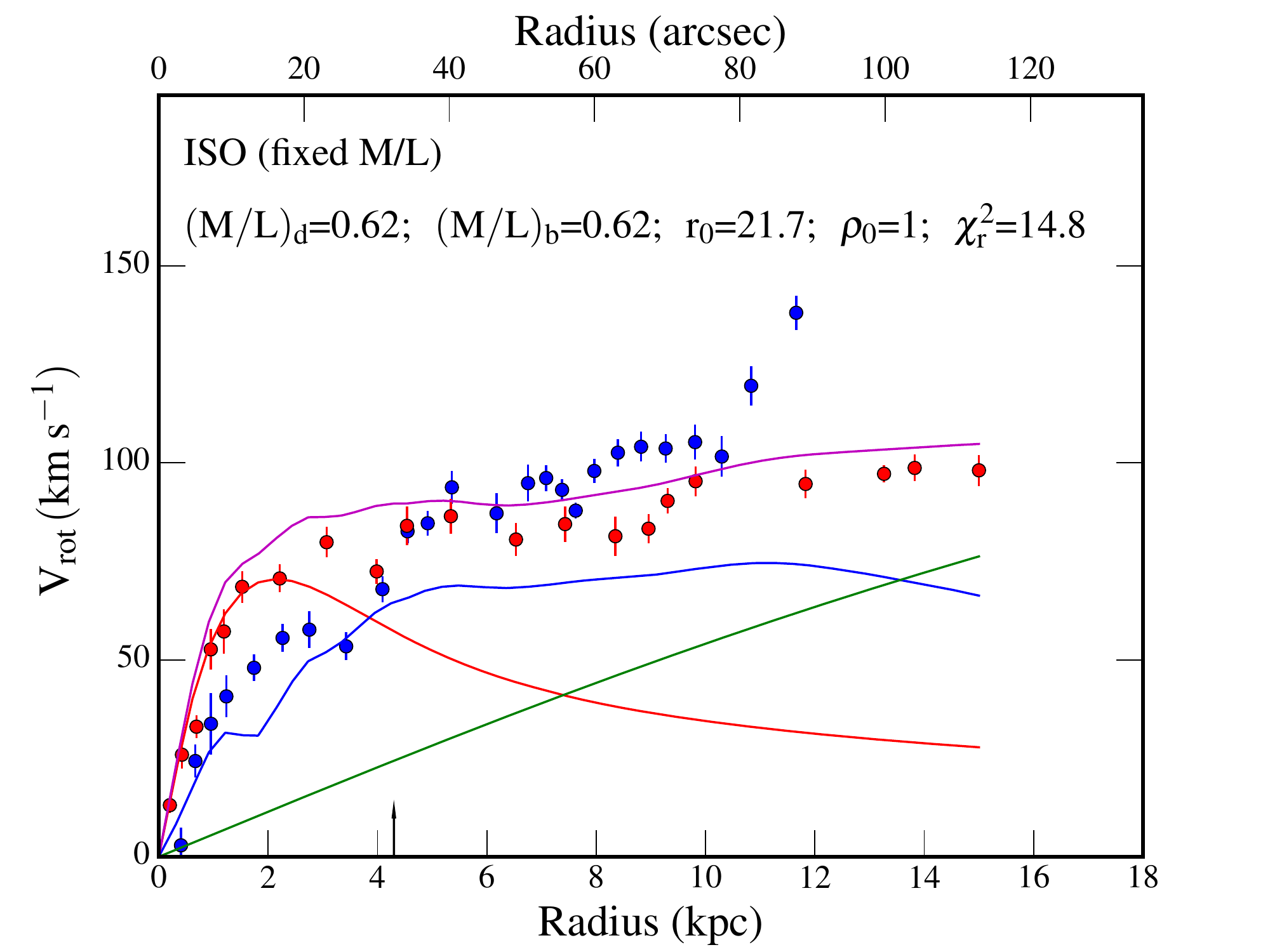}\\
\hspace*{-0.00cm} \includegraphics[width=0.35\textwidth]{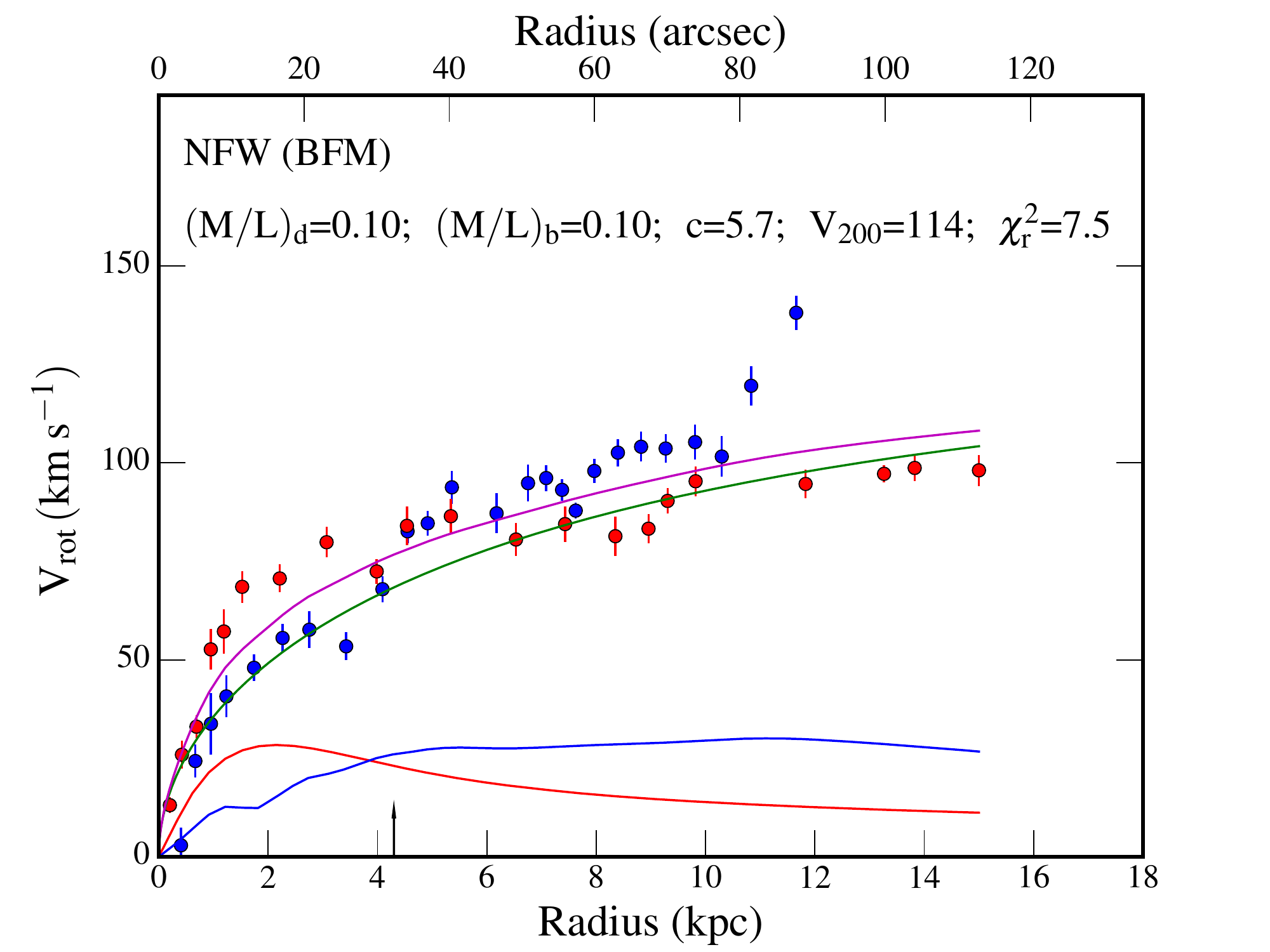}
\hspace*{-0.25cm} \vspace{-1.25cm} \includegraphics[width=0.31\textwidth]{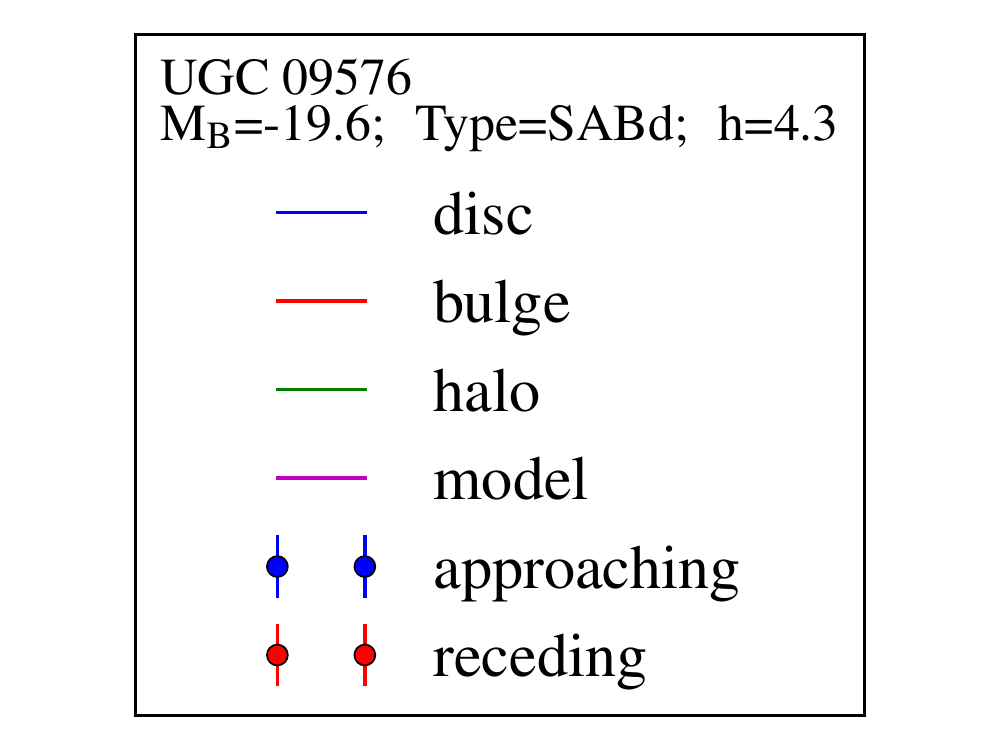} \vspace{1.25cm} \hspace*{-0.5cm}
\hspace*{-0.00cm} \includegraphics[width=0.35\textwidth]{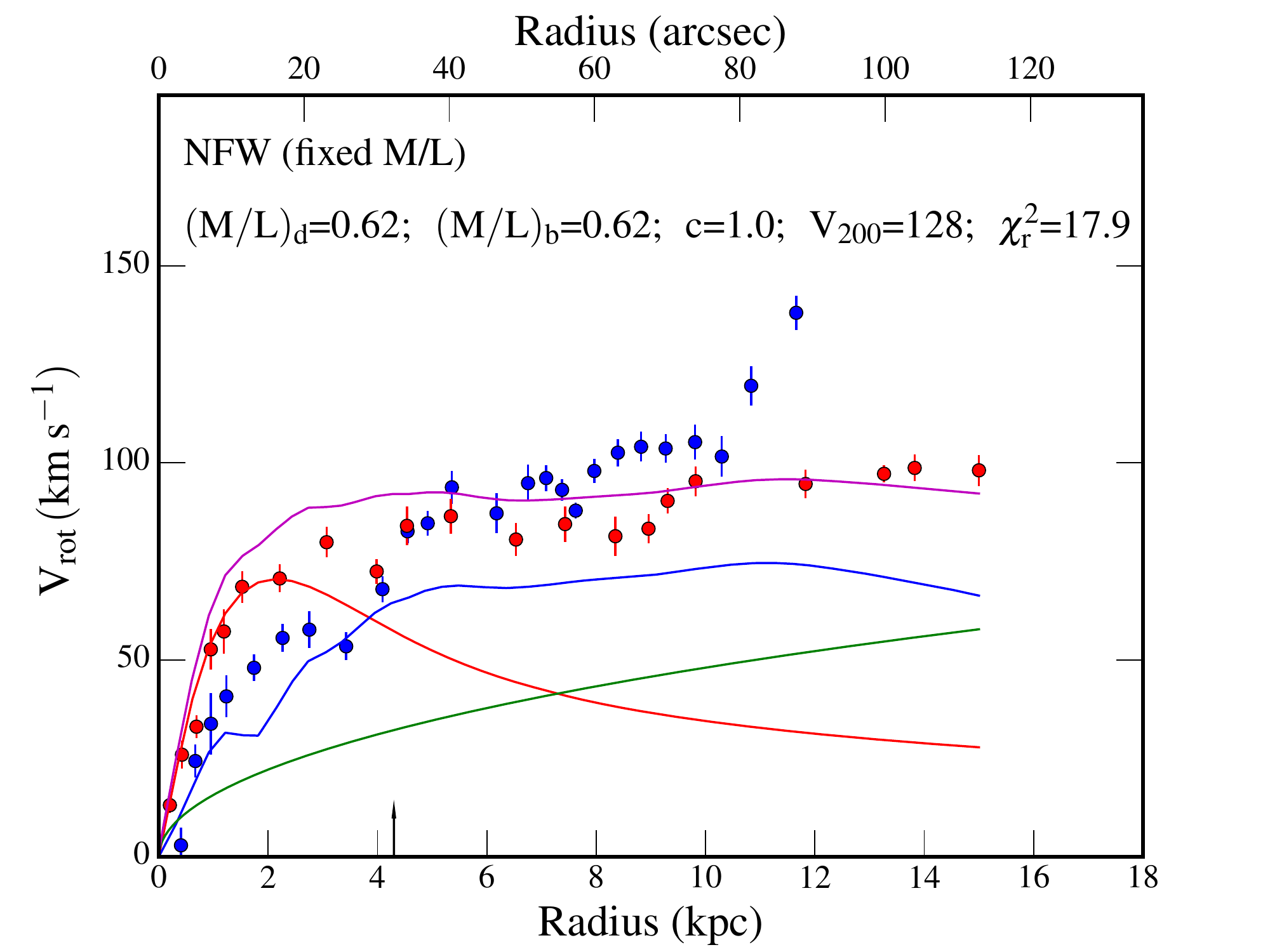}\\
\caption{Example of surface brightness profile decomposition and mass models for the galaxy UGC 9576. Lines and colors are same as in Fig. \ref{massmodel1}.}
\label{massmodel2}
\end{figure*}

\onecolumn

\section{Mass models derived from \Hi\, and optical radii-limited rotation curves taken from Randriamampandry et al. 2014} 
\label{appendixC}
This appendix contains the comparison between the mass models derived from \Hi\, and optical rotation curves limited to the solid body rising parts for 2 out of the \citet{Toky+2014}'s 15 galaxies. The remaining galaxies are available in the online version. 
\newpage
\begin{figure}
\includegraphics[width=0.47\textwidth]{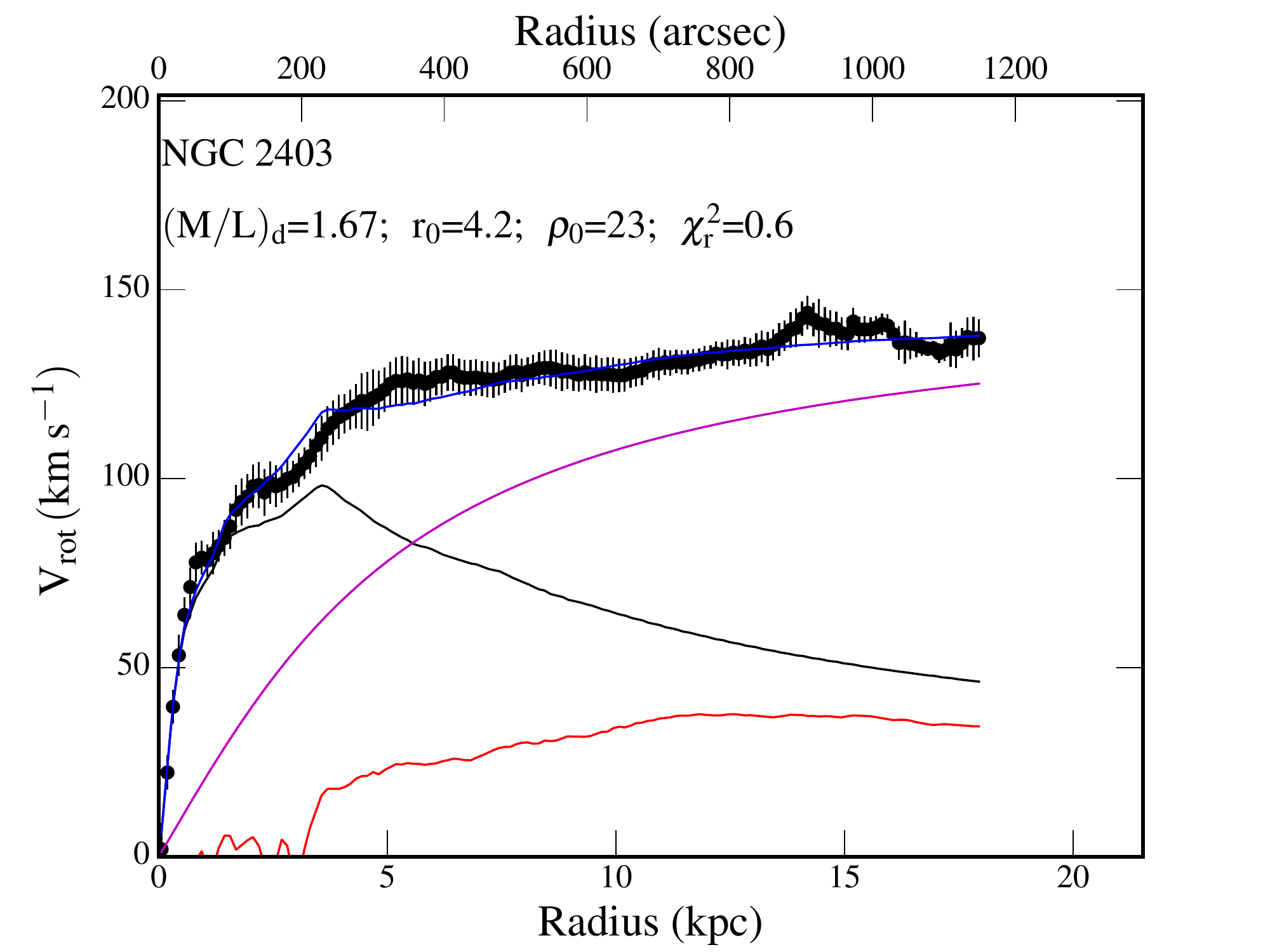}\quad\includegraphics[width=0.47\textwidth]{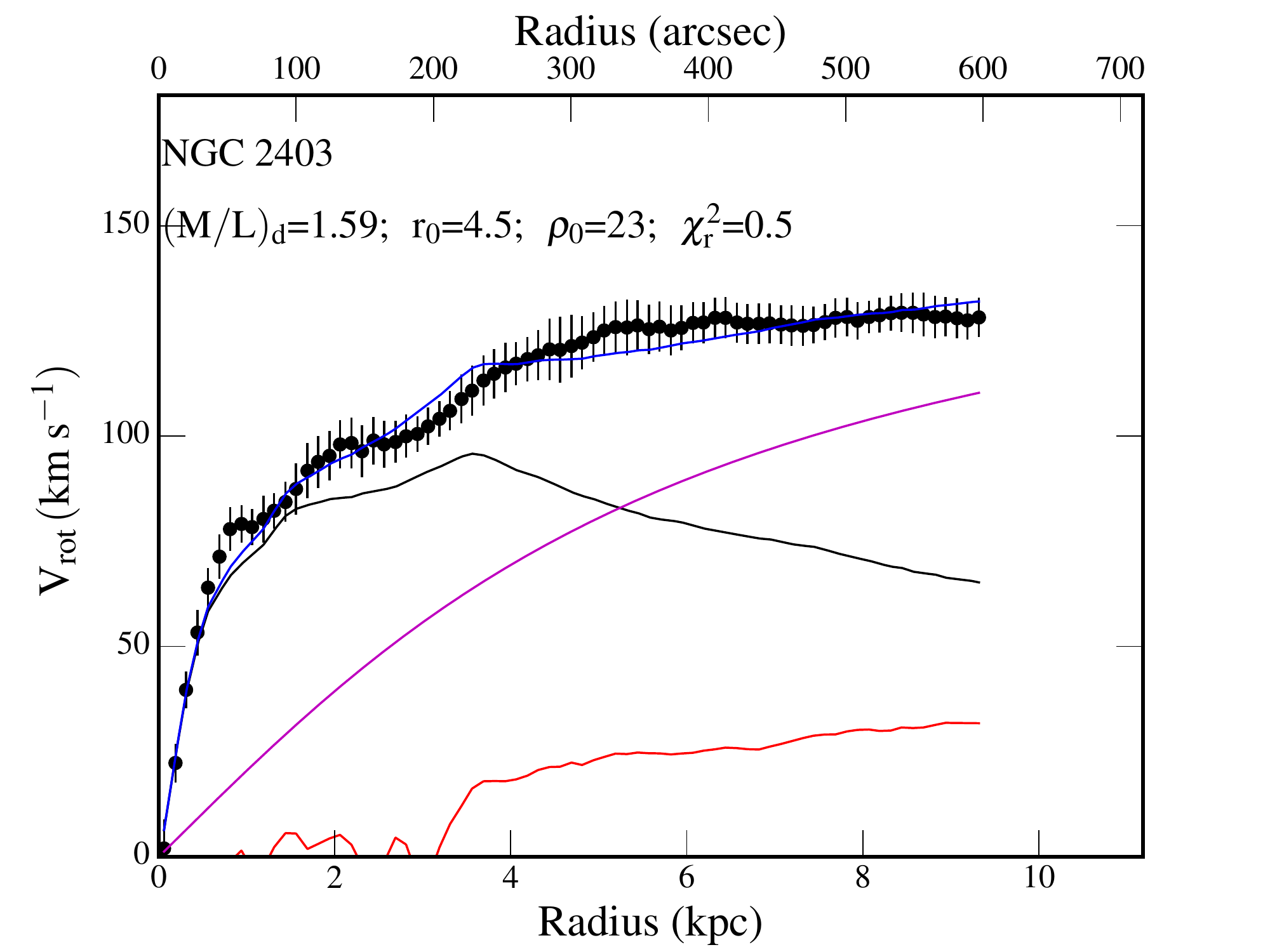}
\caption{Example of mass model for the galaxy NGC 2403. The left panel is obtained using the \Hi\, rotation curve and right panel using the rotation within the optical radius D$_{25} / 2$ kpc. The blue, magenta, black, green, and red lines represent the model, the halo, the disc, the bulge and the gas component respectively. The remaining mass models are available online.}
\label{ngc2403}
\end{figure}
\begin{figure}
\includegraphics[width=0.47\textwidth]{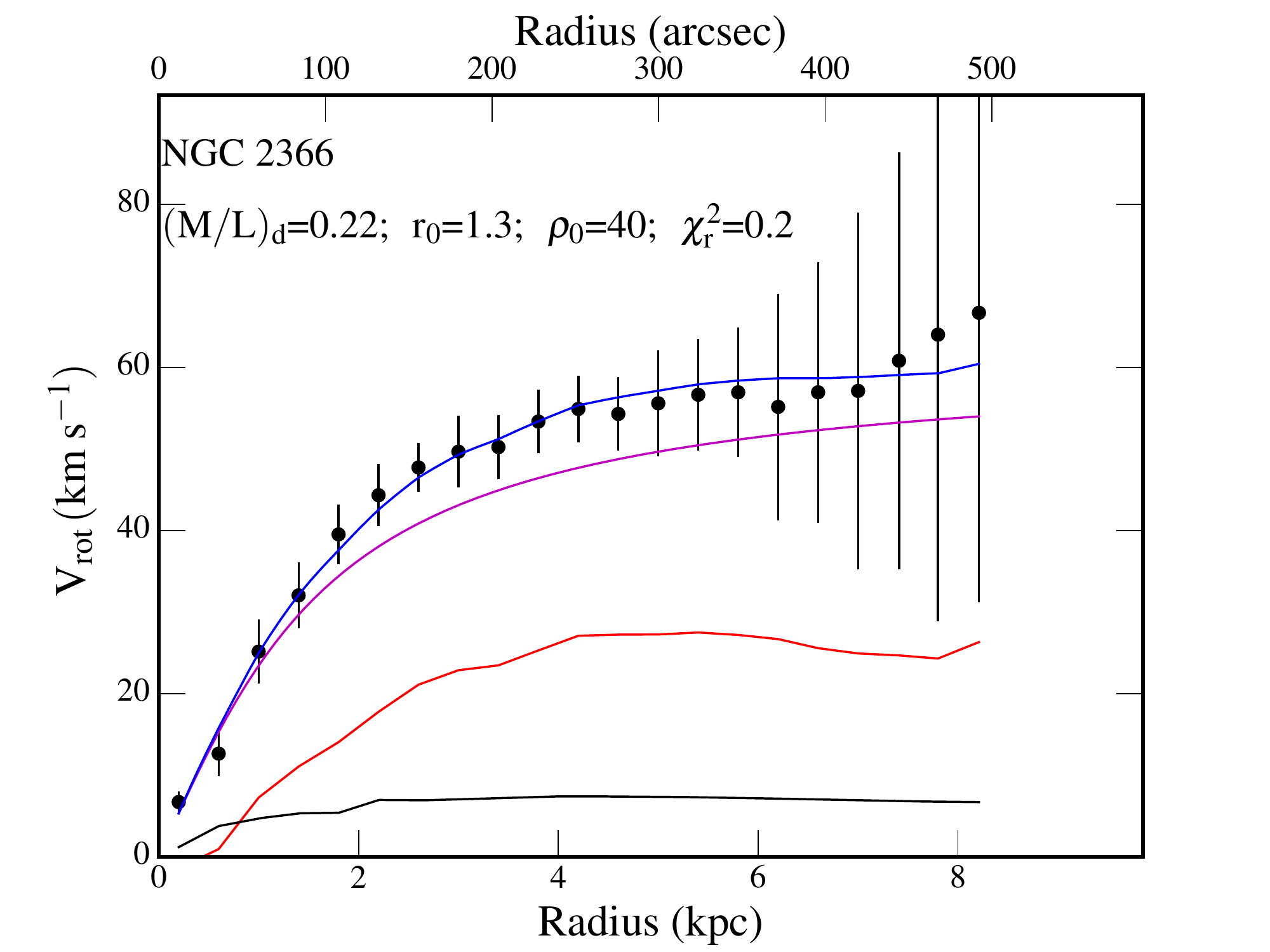}\quad\includegraphics[width=0.47\textwidth]{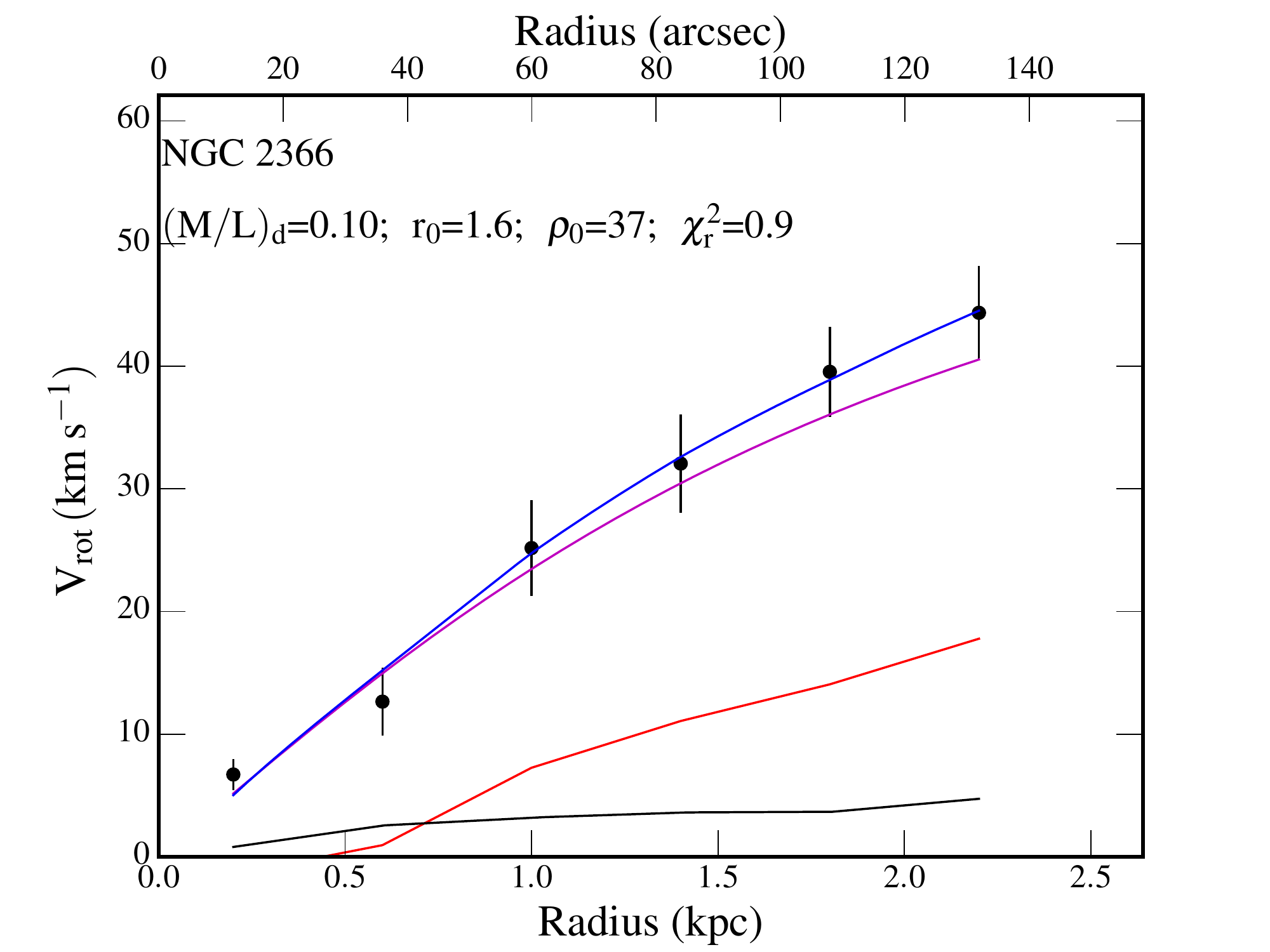}
\caption{Example of mass model for the galaxy NGC 2366. Lines and colors are same as in Fig. \ref{ngc2403}.}
\label{ngc2366}
\end{figure}


\bsp	
\label{lastpage}
\end{document}

%% file: latex_photometry_paper.tex
\newgeometry{left=3cm,right=3cm,top=3cm,bottom=3cm}{\small}

\captionsetup{width=1.3\textwidth}				
\begin{table*}
\caption[Global properties]{Global properties: (1) Name of the galaxy in the UGC catalogue. (2): Morphological type taken from the RC3 catalogue (except for galaxies UGC 3521, 3708, 3709, 3915, 4393, 10652, 11466 for which the morphological types are taken from Epinat et al. (2008)); (3): Absolute B-magnitude from Epinat et al. (2008); (4): Central surface brightness from the observed data in mag arcsec$^{-2}$; (5): Isophotal radius at the limiting surface brightness of 25 mag arcsec$^{-2}$ normalised by the disc scale length; (6): The last radius of the rotation curve normalised by the disc scale length; (7): Central surface brightness of the disc in mag arcsec$^{-2}$; (8): Disc scale length of the disc component in kpc; (9): Luminosity of the disc  in unit of 10$^8$\ L$\odot$ calculated at the isophotal radius; (10): Surface brightness of the bulge at the effective radius in mag arcsec$^{-2}$; (11): Effective radius of the bulge in kpc; (12): S\'{e}rsic index of the bulge; (13): Luminosity of the bulge in units of 10$^8$\ L$\odot$, derived at the isophotal radius; (14): Classification flag of the rotation curves: 1 and 2 correspond respectively to very high and high quality rotation curves, while 3 represents poor quality rotation curves.}
\label{tab:photometry}
\begin{tabular}{c c c c c c c c c c c c c c}

\hline
\multicolumn{1}{c }{\textbf{Galaxy}} & \multicolumn{5}{c}{\textbf{}} & \multicolumn{3}{c}{\textbf{Disc}} & \multicolumn{4}{c}{\textbf{Bulge}} & \multicolumn{1}{c}{\textbf{RC}} \\
\cmidrule(lr){1-1}
\cmidrule(lr){2-6}
\cmidrule(ll){7-9}
\cmidrule(ll){10-13}
\cmidrule(ll){14-14}
\multicolumn{1}{c}{\textbf{$\rm UGC$}} & \textbf{$\rm type$} & \textbf{$\rm Mag$} & \textbf{$\mu_0$$_{obs}$} & \textbf{$\rm R_{25}/h$} & \textbf{$\rm R_{last}/h$} & \textbf{$\mu_0$}  & \textbf{$\rm h$} & \textbf{$L_D$} & \textbf{$\mu_e$} & \textbf{$r_e$ } & \textbf{$\rm n$} &\textbf{$L_B$} & \textbf{flag} \\ 
\multicolumn{1}{c}{(1)} & {(2)} & {(3)} & {(4)} & {(5)} & {(6)} & {(7)} & {(8)} & {(9)} & {(10)} & {(11)} & {(12)}& {(13)} &  {(14)} \\ 
\cmidrule(lr){1-1}
\cmidrule(lr){2-6}
\cmidrule(ll){7-9}
\cmidrule(ll){10-13}
\cmidrule(ll){14-14}

00089 & $\rm SBa$ & $-21.5$ & $14.3$ & $2.8$ & $1.9$ & $16.8$ & $5.1$ & $245.4$ & $14.29$ & $1.1$ & $0.46$ & $203.2$ & $1$ \\ 
00094 & $\rm SAab$ & $-20.4$ & $16.9$ & $1.5$ & $1.7$ & $20.6$ & $6.8$ & $12.1$ & $18.26$ & $3.6$ & $1.08$ & $68.9$ & $1$ \\ 
00508 & $\rm SB(r)ab$ & $-21.8$ & $14.9$ & $2.8$ & $2.4$ & $17.3$ & $9.2$ & $392.6$ & $15.50$ & $1.4$ & $1.09$ & $151.6$ & $1$ \\ 
00528 & $\rm SABb$ & $-19.6$ & $15.0$ & $3.9$ & $1.8$ & $15.9$ & $1.1$ & $29.9$ & $15.40$ & $0.2$ & $0.45$ & $2.8$ & $1$ \\ 
00763 & $\rm SABm$ & $-18.9$ & $17.7$ & $3.5$ & $6.0$ & $17.6$ & $1.5$ & $8.5$ & $-$ & $-$ & $-$ & $-$ & $1$ \\ 
01256 & $\rm SBcd$ & $-18.9$ & $17.6$ & $4.6$ & $5.3$ & $17.5$ & $1.6$ & $10.1$ & $-$ & $-$ & $-$ & $-$ & $3$ \\ 
01317 & $\rm SAB(r)c$ & $-21.5$ & $15.5$ & $5.4$ & $5.6$ & $15.3$ & $4.3$ & $447.4$ & $-$ & $-$ & $-$ & $-$ & $1$ \\ 
01437 & $\rm SABbc$ & $-21.8$ & $15.6$ & $2.6$ & $4.9$ & $16.6$ & $5.3$ & $255.6$ & $16.05$ & $1.1$ & $0.55$ & $41.0$ & $1$ \\ 
\hline
\end{tabular}
\end{table*}

\setcounter{table}{1}

\begin{table*}

\caption[Parameters of mass models using the Best Fit Model (BFM) and fixed M/L techniques with the pseudo-isothermal (ISO) model]{Parameters of mass models using the Best Fit Model (BFM) and fixed M/L techniques with the pseudo-isothermal (ISO) model: (1) Name of the galaxy in the UGC catalogue, those marked with an asterisk (*) correspond to galaxies for which the presence of dark matter is not necessary. The columns (2) to (6) and (7) to (10) show respectively the BFM parameters, and the fixed M/L parameters for the ISO model. (2): M/L of the disc in  M$_{\odot}$/L$_{\odot}$; (3): M/L of the bulge in M$_{\odot}$/L$_{\odot}$. (4) \& (8): the core radius of the DM halo in kpc; (5) \& (9): the Central density of the DM halo in $10^{-3}$ M$_{\odot}$/pc$^3$; (6) \& (10): the reduced $\chi^2$; (7): M/L derived using the W1-W2 colour in units of M$_{\odot}$/L$_{\odot}$.}

\label{tab:iso}
\begin{tabular}{c c c c l c c c c l}
\hline
\multicolumn{1}{c }{\textbf{Galaxy}} & \multicolumn{5}{c}{\textbf{ISO (BFM)}} & \multicolumn{4}{c}{\textbf{ISO with fixed M/L}} \\  
\cmidrule(lr){1-1}
\cmidrule(lr){2-6}
\cmidrule(ll){7-10}
\multicolumn{1}{c}{\textbf{$\rm UGC$}} & \textbf{$\rm M/L$ $\rm Disc$} & \textbf{$\rm M/L$ $\rm Bulge$} & \textbf{$r_0$} & \textbf{$\rho_0$} & \textbf{$\chi^2$} & \textbf{$\rm M/L$ } & \textbf{$r_0$} & \textbf{$\rho_0$} & \textbf{$\chi^2$} \\ 
\multicolumn{1}{c}{(1)} & {(2)} & {(3)} &{(4)} & {(5)} & {(6)} & {(7)} & {(8)} & {(9)} & {(10)} \\ 
\cmidrule(lr){1-1}
\cmidrule(lr){2-6}
\cmidrule(ll){7-10}

00089 & $0.10_{0.01}^{0.01}$ & $0.30_{0.20}^{0.13}$ & $36.2_{35.7}^{93.8}$ & $75_{35}^{24}$ & $21.7$ & $0.30$ & $>39.2$ & $47_{47}^{252}$ & $23.1$ \\ 
00094 & $0.10_{0.01}^{0.08}$ & $0.62_{0.52}^{0.26}$ & $0.6_{0.1}^{0.2}$ & $750_{650}^{1}$ & $7.1$ & $0.45$ & $0.8_{0.3}^{0.1}$ & $750_{450}^{1}$ & $7.6$ \\ 
00508 & $0.64_{0.54}^{0.13}$ & $0.66_{0.56}^{1.34}$ & $4.9_{4.4}^{1.0}$ & $250_{20}^{49}$ & $1.4$ & $0.69$ & $5.1_{3.5}^{0.6}$ & $231_{31}^{168}$ & $1.3$ \\ 
00528 & $0.10_{0.01}^{0.01}$ & $0.10_{0.01}^{0.01}$ & $0.5_{0.1}^{9.5}$ & $0_{1}^{1}$ & $4.2$ & $0.24$ & $0.5_{0.1}^{9.5}$ & $0_{1}^{1}$ & $53.9$ \\ 
00763 & $0.16_{0.06}^{0.06}$ & $-$ & $1.9_{1.4}^{0.3}$ & $70_{20}^{29}$ & $2.4$ & $0.60$ & $17.5_{17.5}^{7.5}$ & $5_{5}^{44}$ & $7.5$ \\ 
01317 & $0.18_{0.08}^{0.01}$ & $-$ & $17.2_{16.7}^{0.8}$ & $5_{5}^{10}$ & $6.5$ & $0.33$ & $0.5_{0.1}^{17.5}$ & $0_{1}^{1}$ & $117.2$ \\ 
01437 & $0.14_{0.04}^{0.02}$ & $0.36_{0.23}^{0.55}$ & $0.8_{0.3}^{0.1}$ & $750_{550}^{1}$ & $8.3$ & $0.23$ & $0.7_{0.2}^{0.1}$ & $750_{750}^{1}$ & $11.9$ \\ 
01736 & $0.10_{0.01}^{0.01}$ & $0.10_{0.01}^{1.90}$ & $2.6_{2.1}^{0.6}$ & $138_{38}^{67}$ & $5.4$ & $0.51$ & $3.9_{3.4}^{0.1}$ & $48_{38}^{101}$ & $9.6$ \\ 

\hline
\end{tabular}
\end{table*}

\setcounter{table}{2}

\begin{table*}
\caption[Parameters of mass models using the maximum disc model technique with the pseudo-isothermal (ISO)]{Parameters of mass models using the maximum disc model (MDM) technique with the pseudo-isothermal (ISO) model: (1) Name of the galaxy in the UGC catalogue; (2) M/L of the disc in  M$_{\odot}$/L$_{\odot}$; (3) M/L of the bulge in  M$_{\odot}$/L$_{\odot}$; (4) Core radius of the DM halo in kpc; (5) Central density of the DM halo in $10^{-3}$ M$_{\odot}$/pc$^3$; (6) The reduced $\chi^2$.}
\label{tab:maximumdisc }
\begin{tabular}{c c c c l c }
\hline
\multicolumn{1}{c }{\textbf{Galaxy}} & \multicolumn{5}{c}{\textbf{Maximum  Disc   Model}} \\  
\cmidrule(lr){1-1}
\cmidrule(lr){2-6}
\multicolumn{1}{c}{\textbf{$\rm UGC$}} & \textbf{$\rm M/L$ $\rm  Disc $} & \textbf{$\rm M/L$ $\rm Bulge$} & \textbf{$r_0$} & \textbf{$\rho_0$} & \textbf{$\chi^2$} \\ 
\multicolumn{1}{c}{(1)} & {(2)} & {(3)} & {(4)} & {(5)} & {(6)} \\ 
\cmidrule(lr){1-1}
\cmidrule(lr){2-6}

00089 & $0.29_{0.19}^{0.01}$ & $0.32_{0.22}^{0.13}$ & $8.4_{7.9}^{41.6}$ & $50_{50}^{50}$ & $27.9$ \\ 
00094 & $0.61_{0.51}^{0.01}$ & $0.65_{0.45}^{0.25}$ & $0.5_{0.1}^{0.3}$ & $733_{713}^{16}$ & $9.0$ \\ 
00508 & $0.93_{0.83}^{0.01}$ & $0.94_{0.74}^{1.06}$ & $5.8_{4.8}^{0.8}$ & $166_{146}^{233}$ & $1.8$ \\ 
00528 & $0.10_{0.01}^{0.01}$ & $0.10_{0.01}^{0.01}$ & $0.5_{0.1}^{9.5}$ & $0_{1}^{13}$ & $5.1$ \\ 
00763 & $0.37_{0.27}^{0.01}$ & $-$ & $2.8_{2.3}^{0.1}$ & $30_{30}^{39}$ & $3.1$ \\ 
01317 & $0.18_{0.08}^{0.01}$ & $-$ & $7.9_{6.3}^{10.1}$ & $9_{9}^{6}$ & $7.8$ \\ 
01437 & $0.27_{0.17}^{0.01}$ & $0.46_{0.26}^{0.51}$ & $0.5_{0.1}^{4.7}$ & $321_{321}^{428}$ & $12.0$ \\ 
01736 & $0.61_{0.21}^{0.01}$ & $0.62_{0.22}^{0.99}$ & $10.0_{9.5}^{0.1}$ & $19_{18}^{81}$ & $11.3$ \\ 

\hline
\end{tabular}
\end{table*}

\restoregeometry

\setcounter{table}{3}
\begin{table*}
\caption[Parameters of mass models using the Best Fit Model (BFM) and fixed M/L techniques with the Navarro-Frenk-White model (NFW)]{Parameters of mass models using the Best Fit Model (BFM) and fixed M/L techniques with the Navarro-Frenk-White model (NFW): (1) Name of the galaxy in the UGC catalogue; the columns (2) to (6) and (7) to (10) show respectively the BFM parameters, and the fixed M/L parameters for the ISO model. (2): M/L of the disc  in  M$_{\odot}$/L$_{\odot}$; (3): M/L of the bulge in M$_{\odot}$/L$_{\odot}$. (4) \& (8): the central halo concentration index; (5) \& (9): the halo velocity in $\rm km\ s^{-1}$; (6) \& (10): the reduced $\chi^2$; (7): M/L derived using the W1-W2 colour in units of M$_{\odot}$/L$_{\odot}$.}
\label{tab:nfw}
\begin{tabular}{c c c c c c c c c c}

\hline
\multicolumn{1}{c }{\textbf{Galaxy}} & \multicolumn{5}{c}{\textbf{NFW (BFM)}} & \multicolumn{4}{c}{\textbf{NFW with fixed M/L}} \\  
\cmidrule(lr){1-1}
\cmidrule(lr){2-6}
\cmidrule(ll){7-10}
\multicolumn{1}{c}{\textbf{$\rm UGC$}} & \textbf{$\rm M/L$  $\rm Disc $} & \textbf{$\rm M/L$ $\rm Bulge$} & \textbf{$\rm c$} & \textbf{$\rm V_{200}$} & \textbf{$\chi^2$} & \textbf{$\rm M/L$ } & \textbf{$\rm c$} & \textbf{$\rm V_{200}$} & \textbf{$\chi^2$} \\ 
\multicolumn{1}{c}{(1)} & {(2)} & {(3)} & {(4)} & {(5)} & {(6)} & {(7)} & {(8)} & {(9)} & {(10)} \\ 
\cmidrule(lr){1-1}
\cmidrule(lr){2-6}
\cmidrule(ll){7-10}

00089 & $0.10_{0.01}^{0.01}$ & $0.10_{0.01}^{0.34}$ & $79.5_{78.5}^{0.1}$ & $128.3_{118.3}^{371.7}$ & $22.3$ & $0.30$ & $6.4_{5.4}^{0.1}$ & $409.3_{408.3}^{90.7}$ & $27.6$ \\ 
00094 & $0.10_{0.01}^{0.10}$ & $0.13_{0.03}^{0.71}$ & $48.4_{44.4}^{1.9}$ & $99.3_{19.3}^{0.7}$ & $4.2$ & $0.45$ & $54.5_{19.5}^{2.5}$ & $70.5_{20.5}^{10.2}$ & $5.3$ \\ 
00508 & $0.41_{0.31}^{0.25}$ & $0.45_{0.34}^{1.55}$ & $20.6_{19.6}^{4.4}$ & $377.3_{27.3}^{122.7}$ & $1.3$ & $0.69$ & $15.0_{10.0}^{2.5}$ & $427.3_{127.3}^{72.7}$ & $1.4$ \\ 
00528 & $0.10_{0.01}^{0.01}$ & $0.10_{0.01}^{0.01}$ & $100.0_{99.9}^{0.1}$ & $5.5_{4.5}^{44.5}$ & $4.1$ & $0.24$ & $1.0_{0.1}^{6.0}$ & $1.0_{0.1}^{1.1}$ & $53.9$ \\ 
00763 & $0.10_{0.01}^{0.01}$ & $-$ & $10.5_{9.5}^{0.6}$ & $93.6_{73.6}^{106.4}$ & $3.0$ & $0.60$ & $1.0_{0.1}^{0.1}$ & $248.6_{218.6}^{8.4}$ & $11.8$ \\ 
01317 & $0.10_{0.01}^{0.01}$ & $-$ & $15.2_{14.2}^{0.6}$ & $126.4_{106.4}^{73.6}$ & $5.5$ & $0.33$ & $100.0_{99.0}^{0.1}$ & $1.0_{0.1}^{0.1}$ & $117.2$ \\ 
01437 & $0.10_{0.01}^{0.01}$ & $0.10_{0.01}^{0.82}$ & $53.2_{51.1}^{0.1}$ & $95.0_{35.0}^{105.0}$ & $3.4$ & $0.23$ & $70.8_{69.8}^{4.5}$ & $63.5_{54.5}^{86.5}$ & $5.6$ \\ 
01736 & $0.10_{0.01}^{0.01}$ & $0.10_{0.01}^{0.90}$ & $4.5_{3.5}^{3.0}$ & $500.0_{300.0}^{0.1}$ & $6.5$ & $0.51$ & $2.6_{1.6}^{0.8}$ & $500.0_{400.0}^{0.1}$ & $10.4$ \\ 

\hline
\end{tabular}
\end{table*}

\restoregeometry

%% file: paper.bbl
\begin{thebibliography}{}
\makeatletter
\relax
\def\mn@urlcharsother{\let\do\@makeother \do\$\do\&\do\#\do\^\do\_\do\%\do\~}
\def\mn@doi{\begingroup\mn@urlcharsother \@ifnextchar [ {\mn@doi@}
  {\mn@doi@[]}}
\def\mn@doi@[#1]#2{\def\@tempa{#1}\ifx\@tempa\@empty \href
  {http://dx.doi.org/#2} {doi:#2}\else \href {http://dx.doi.org/#2} {#1}\fi
  \endgroup}
\def\mn@eprint#1#2{\mn@eprint@#1:#2::\@nil}
\def\mn@eprint@arXiv#1{\href {http://arxiv.org/abs/#1} {{\tt arXiv:#1}}}
\def\mn@eprint@dblp#1{\href {http://dblp.uni-trier.de/rec/bibtex/#1.xml}
  {dblp:#1}}
\def\mn@eprint@#1:#2:#3:#4\@nil{\def\@tempa {#1}\def\@tempb {#2}\def\@tempc
  {#3}\ifx \@tempc \@empty \let \@tempc \@tempb \let \@tempb \@tempa \fi \ifx
  \@tempb \@empty \def\@tempb {arXiv}\fi \@ifundefined
  {mn@eprint@\@tempb}{\@tempb:\@tempc}{\expandafter \expandafter \csname
  mn@eprint@\@tempb\endcsname \expandafter{\@tempc}}}

\bibitem[\protect\citeauthoryear{{Amram}, {Balkowski}, {Boulesteix}, {Cayatte},
  {Marcelin}  \& {Sullivan}}{{Amram} et~al.}{1996}]{Amram+1996}
{Amram} P.,  {Balkowski} C.,  {Boulesteix} J.,  {Cayatte} V.,  {Marcelin} M.,
  {Sullivan} III W.~T.,  1996, \aap, \href
  {http://adsabs.harvard.edu/abs/1996A%26A...310..737A} {310, 737}

\bibitem[\protect\citeauthoryear{{Barbosa} et~al.,}{{Barbosa}
  et~al.}{2015}]{Barbosa+2015}
{Barbosa} C.~E.,  et~al., 2015, \mn@doi [\mnras] {10.1093/mnras/stv1685}, \href
  {http://adsabs.harvard.edu/abs/2015MNRAS.453.2965B} {453, 2965}

\bibitem[\protect\citeauthoryear{{Begeman}}{{Begeman}}{1987}]{Begeman+1987}
{Begeman} K.~G.,  1987, PhD thesis, , Kapteyn Institute, (1987)

\bibitem[\protect\citeauthoryear{{Bell} \& {de Jong}}{{Bell} \& {de
  Jong}}{2001}]{Bell+2001}
{Bell} E.~F.,  {de Jong} R.~S.,  2001, \mn@doi [\apj] {10.1086/319728}, \href
  {http://adsabs.harvard.edu/abs/2001ApJ...550..212B} {550, 212}

\bibitem[\protect\citeauthoryear{{Binney} \& {Tremaine}}{{Binney} \&
  {Tremaine}}{2008}]{Binney+2008}
{Binney} J.,  {Tremaine} S.,  2008, {Galactic Dynamics: Second Edition}.
Princeton University Press

\bibitem[\protect\citeauthoryear{{Blais-Ouellette}, {Carignan}, {Amram}  \&
  {C{\^o}t{\'e}}}{{Blais-Ouellette} et~al.}{1999}]{Ouellette+1999}
{Blais-Ouellette} S.,  {Carignan} C.,  {Amram} P.,   {C{\^o}t{\'e}} S.,  1999,
  \mn@doi [\aj] {10.1086/301066}, \href
  {http://adsabs.harvard.edu/abs/1999AJ....118.2123B} {118, 2123}

\bibitem[\protect\citeauthoryear{{Blais-Ouellette}, {Amram}  \&
  {Carignan}}{{Blais-Ouellette} et~al.}{2001}]{Ouellette+2001}
{Blais-Ouellette} S.,  {Amram} P.,   {Carignan} C.,  2001, \mn@doi [\aj]
  {10.1086/319944}, \href {http://adsabs.harvard.edu/abs/2001AJ....121.1952B}
  {121, 1952}

\bibitem[\protect\citeauthoryear{{Bosma}}{{Bosma}}{1981}]{Bosma+1981}
{Bosma} A.,  1981, \mn@doi [\aj] {10.1086/113063}, \href
  {http://adsabs.harvard.edu/abs/1981AJ.....86.1825B} {86, 1825}

\bibitem[\protect\citeauthoryear{{Bottema} \& {Pesta{\~n}a}}{{Bottema} \&
  {Pesta{\~n}a}}{2015}]{Bottema+2015}
{Bottema} R.,  {Pesta{\~n}a} J.~L.~G.,  2015, \mn@doi [\mnras]
  {10.1093/mnras/stv182}, \href
  {http://adsabs.harvard.edu/abs/2015MNRAS.448.2566B} {448, 2566}

\bibitem[\protect\citeauthoryear{{Buchhorn}}{{Buchhorn}}{1992}]{Buchhorn1992}
{Buchhorn} M.,  1992, PhD thesis, , Australian National Univ., (1992)

\bibitem[\protect\citeauthoryear{{Bullock}, {Kolatt}, {Sigad}, {Somerville},
  {Kravtsov}, {Klypin}, {Primack}  \& {Dekel}}{{Bullock}
  et~al.}{2001}]{Bullock+2001}
{Bullock} J.~S.,  {Kolatt} T.~S.,  {Sigad} Y.,  {Somerville} R.~S.,  {Kravtsov}
  A.~V.,  {Klypin} A.~A.,  {Primack} J.~R.,   {Dekel} A.,  2001, \mn@doi
  [\mnras] {10.1046/j.1365-8711.2001.04068.x}, \href
  {http://adsabs.harvard.edu/abs/2001MNRAS.321..559B} {321, 559}

\bibitem[\protect\citeauthoryear{{Carignan} \& {Freeman}}{{Carignan} \&
  {Freeman}}{1985}]{Carignan+1985}
{Carignan} C.,  {Freeman} K.~C.,  1985, \mn@doi [\apj] {10.1086/163316}, \href
  {http://adsabs.harvard.edu/abs/1985ApJ...294..494C} {294, 494}

\bibitem[\protect\citeauthoryear{{Carignan} \& {Freeman}}{{Carignan} \&
  {Freeman}}{1988}]{Carignan+1988}
{Carignan} C.,  {Freeman} K.~C.,  1988, \mn@doi [\apjl] {10.1086/185260}, \href
  {http://adsabs.harvard.edu/abs/1988ApJ...332L..33C} {332, L33}

\bibitem[\protect\citeauthoryear{{Carignan}, {Charbonneau}, {Boulanger}  \&
  {Viallefond}}{{Carignan} et~al.}{1990}]{Carignan+1990}
{Carignan} C.,  {Charbonneau} P.,  {Boulanger} F.,   {Viallefond} F.,  1990,
  \aap, \href {http://adsabs.harvard.edu/abs/1990A%26A...234...43C} {234, 43}

\bibitem[\protect\citeauthoryear{{Cluver} et~al.,}{{Cluver}
  et~al.}{2014}]{Cluver+2014}
{Cluver} M.~E.,  et~al., 2014, \mn@doi [\apj] {10.1088/0004-637X/782/2/90},
  \href {http://adsabs.harvard.edu/abs/2014ApJ...782...90C} {782, 90}

\bibitem[\protect\citeauthoryear{{Cole} \& {Lacey}}{{Cole} \&
  {Lacey}}{1996}]{Cole+1996}
{Cole} S.,  {Lacey} C.,  1996, \mn@doi [\mnras] {10.1093/mnras/281.2.716},
  \href {http://adsabs.harvard.edu/abs/1996MNRAS.281..716C} {281, 716}

\bibitem[\protect\citeauthoryear{{Courteau} \& {Rix}}{{Courteau} \&
  {Rix}}{1999}]{Courteau+1999}
{Courteau} S.,  {Rix} H.-W.,  1999, \mn@doi [\apj] {10.1086/306872}, \href
  {http://adsabs.harvard.edu/abs/1999ApJ...513..561C} {513, 561}

\bibitem[\protect\citeauthoryear{{Di Cintio}, {Brook}, {Macci{\`o}}, {Stinson},
  {Knebe}, {Dutton}  \& {Wadsley}}{{Di Cintio} et~al.}{2014a}]{DiCintio+2014a}
{Di Cintio} A.,  {Brook} C.~B.,  {Macci{\`o}} A.~V.,  {Stinson} G.~S.,  {Knebe}
  A.,  {Dutton} A.~A.,   {Wadsley} J.,  2014a, \mn@doi [\mnras]
  {10.1093/mnras/stt1891}, \href
  {http://adsabs.harvard.edu/abs/2014MNRAS.437..415D} {437, 415}

\bibitem[\protect\citeauthoryear{{Di Cintio}, {Brook}, {Dutton}, {Macci{\`o}},
  {Stinson}  \& {Knebe}}{{Di Cintio} et~al.}{2014b}]{DiCintio+2014b}
{Di Cintio} A.,  {Brook} C.~B.,  {Dutton} A.~A.,  {Macci{\`o}} A.~V.,
  {Stinson} G.~S.,   {Knebe} A.,  2014b, \mn@doi [\mnras]
  {10.1093/mnras/stu729}, \href
  {http://adsabs.harvard.edu/abs/2014MNRAS.441.2986D} {441, 2986}

\bibitem[\protect\citeauthoryear{{Dicaire} et~al.,}{{Dicaire}
  et~al.}{2008}]{Dicaire+2008}
{Dicaire} I.,  et~al., 2008, \mn@doi [\mnras]
  {10.1111/j.1365-2966.2008.12868.x}, \href
  {http://adsabs.harvard.edu/abs/2008MNRAS.385..553D} {385, 553}

\bibitem[\protect\citeauthoryear{{Dutton}, {Courteau}, {de Jong}  \&
  {Carignan}}{{Dutton} et~al.}{2005}]{Dutton+2005}
{Dutton} A.~A.,  {Courteau} S.,  {de Jong} R.,   {Carignan} C.,  2005, \mn@doi
  [\apj] {10.1086/426375}, \href
  {http://adsabs.harvard.edu/abs/2005ApJ...619..218D} {619, 218}

\bibitem[\protect\citeauthoryear{{Epinat} et~al.,}{{Epinat}
  et~al.}{2008a}]{Epinat+2008b}
{Epinat} B.,  et~al., 2008a, \mn@doi [\mnras]
  {10.1111/j.1365-2966.2008.13422.x}, \href
  {http://adsabs.harvard.edu/abs/2008MNRAS.388..500E} {388, 500}

\bibitem[\protect\citeauthoryear{{Epinat}, {Amram}  \& {Marcelin}}{{Epinat}
  et~al.}{2008b}]{Epinat+2008a}
{Epinat} B.,  {Amram} P.,   {Marcelin} M.,  2008b, \mn@doi [\mnras]
  {10.1111/j.1365-2966.2008.13796.x}, \href
  {http://adsabs.harvard.edu/abs/2008MNRAS.390..466E} {390, 466}

\bibitem[\protect\citeauthoryear{{Freeman}}{{Freeman}}{1970}]{Freeman+1970}
{Freeman} K.~C.,  1970, \mn@doi [\apj] {10.1086/150474}, \href
  {http://adsabs.harvard.edu/abs/1970ApJ...160..811F} {160, 811}

\bibitem[\protect\citeauthoryear{{Garrido}, {Marcelin}, {Amram}  \&
  {Boulesteix}}{{Garrido} et~al.}{2002}]{Garrido+2002}
{Garrido} O.,  {Marcelin} M.,  {Amram} P.,   {Boulesteix} J.,  2002, \mn@doi
  [\aap] {10.1051/0004-6361:20020479}, \href
  {http://adsabs.harvard.edu/abs/2002A%26A...387..821G} {387, 821}

\bibitem[\protect\citeauthoryear{{Garrido}, {Marcelin}, {Amram}  \&
  {Boissin}}{{Garrido} et~al.}{2003}]{Garrido+2003}
{Garrido} O.,  {Marcelin} M.,  {Amram} P.,   {Boissin} O.,  2003, \mn@doi
  [\aap] {10.1051/0004-6361:20021784}, \href
  {http://adsabs.harvard.edu/abs/2003A%26A...399...51G} {399, 51}

\bibitem[\protect\citeauthoryear{{Garrido}, {Marcelin}  \& {Amram}}{{Garrido}
  et~al.}{2004}]{Garrido+2004}
{Garrido} O.,  {Marcelin} M.,   {Amram} P.,  2004, \mn@doi [\mnras]
  {10.1111/j.1365-2966.2004.07483.x}, \href
  {http://adsabs.harvard.edu/abs/2004MNRAS.349..225G} {349, 225}

\bibitem[\protect\citeauthoryear{{Garrido}, {Marcelin}, {Amram}, {Balkowski},
  {Gach}  \& {Boulesteix}}{{Garrido} et~al.}{2005}]{Garrido+2005}
{Garrido} O.,  {Marcelin} M.,  {Amram} P.,  {Balkowski} C.,  {Gach} J.~L.,
  {Boulesteix} J.,  2005, \mn@doi [\mnras] {10.1111/j.1365-2966.2005.09274.x},
  \href {http://adsabs.harvard.edu/abs/2005MNRAS.362..127G} {362, 127}

\bibitem[\protect\citeauthoryear{{Gentile}, {Salucci}, {Klein}, {Vergani}  \&
  {Kalberla}}{{Gentile} et~al.}{2004}]{Gentile+2004}
{Gentile} G.,  {Salucci} P.,  {Klein} U.,  {Vergani} D.,   {Kalberla} P.,
  2004, \mn@doi [\mnras] {10.1111/j.1365-2966.2004.07836.x}, \href
  {http://adsabs.harvard.edu/abs/2004MNRAS.351..903G} {351, 903}

\bibitem[\protect\citeauthoryear{{Hohl}}{{Hohl}}{1971}]{Hohl+1971}
{Hohl} F.,  1971, \mn@doi [\apj] {10.1086/151091}, \href
  {http://adsabs.harvard.edu/abs/1971ApJ...168..343H} {168, 343}

\bibitem[\protect\citeauthoryear{{Jarrett} et~al.,}{{Jarrett}
  et~al.}{2012}]{Jarrett+2012}
{Jarrett} T.~H.,  et~al., 2012, \mn@doi [\aj] {10.1088/0004-6256/144/2/68},
  \href {http://adsabs.harvard.edu/abs/2012AJ....144...68J} {144, 68}

\bibitem[\protect\citeauthoryear{{Jarrett} et~al.,}{{Jarrett}
  et~al.}{2013}]{Jarrett+2013}
{Jarrett} T.~H.,  et~al., 2013, \mn@doi [\aj] {10.1088/0004-6256/145/1/6},
  \href {http://adsabs.harvard.edu/abs/2013AJ....145....6J} {145, 6}

\bibitem[\protect\citeauthoryear{{Jobin} \& {Carignan}}{{Jobin} \&
  {Carignan}}{1990}]{Jobin+1990}
{Jobin} M.,  {Carignan} C.,  1990, \mn@doi [\aj] {10.1086/115548}, \href
  {http://adsabs.harvard.edu/abs/1990AJ....100..648J} {100, 648}

\bibitem[\protect\citeauthoryear{{Kalnajs}}{{Kalnajs}}{1983a}]{Kalnajs1983}
{Kalnajs} A.,  1983a, in {Athanassoula} E.,  ed.,  IAU Symposium Vol. 100,
  Internal Kinematics and Dynamics of Galaxies. pp 87--88

\bibitem[\protect\citeauthoryear{{Kalnajs}}{{Kalnajs}}{1983b}]{Kalnajs+1983}
{Kalnajs} A.,  1983b, in {Athanassoula} E.,  ed.,  IAU Symposium Vol. 100,
  Internal Kinematics and Dynamics of Galaxies. pp 87--88

\bibitem[\protect\citeauthoryear{{Kassin}, {de Jong}  \& {Pogge}}{{Kassin}
  et~al.}{2006a}]{Kassin+2006a}
{Kassin} S.~A.,  {de Jong} R.~S.,   {Pogge} R.~W.,  2006a, \mn@doi [\apjs]
  {10.1086/498394}, \href {http://adsabs.harvard.edu/abs/2006ApJS..162...80K}
  {162, 80}

\bibitem[\protect\citeauthoryear{{Kassin}, {de Jong}  \& {Weiner}}{{Kassin}
  et~al.}{2006b}]{Kassin+2006b}
{Kassin} S.~A.,  {de Jong} R.~S.,   {Weiner} B.~J.,  2006b, \mn@doi [\apj]
  {10.1086/502959}, \href {http://adsabs.harvard.edu/abs/2006ApJ...643..804K}
  {643, 804}

\bibitem[\protect\citeauthoryear{{Katz}, {Lelli}, {McGaugh}, {Di Cintio},
  {Brook}  \& {Schombert}}{{Katz} et~al.}{2017}]{Katz+2017}
{Katz} H.,  {Lelli} F.,  {McGaugh} S.~S.,  {Di Cintio} A.,  {Brook} C.~B.,
  {Schombert} J.~M.,  2017, \mn@doi [\mnras] {10.1093/mnras/stw3101}, \href
  {http://adsabs.harvard.edu/abs/2017MNRAS.466.1648K} {466, 1648}

\bibitem[\protect\citeauthoryear{{Kent}}{{Kent}}{1986}]{Kent1986}
{Kent} S.~M.,  1986, \mn@doi [\aj] {10.1086/114106}, \href
  {http://adsabs.harvard.edu/abs/1986AJ.....91.1301K} {91, 1301}

\bibitem[\protect\citeauthoryear{{Kettlety} et~al.,}{{Kettlety}
  et~al.}{2018}]{Kettlety+2018}
{Kettlety} T.,  et~al., 2018, \mn@doi [\mnras] {10.1093/mnras/stx2379}, \href
  {http://adsabs.harvard.edu/abs/2018MNRAS.473..776K} {473, 776}

\bibitem[\protect\citeauthoryear{{Kormendy} \& {Freeman}}{{Kormendy} \&
  {Freeman}}{2004}]{Kormendy+2004}
{Kormendy} J.,  {Freeman} K.~C.,  2004, in {Ryder} S.,  {Pisano} D.,  {Walker}
  M.,   {Freeman} K.,  eds,  IAU Symposium Vol. 220, Dark Matter in Galaxies.
  p.~377 (\mn@eprint {} {astro-ph/0407321})

\bibitem[\protect\citeauthoryear{{Kravtsov}, {Klypin}, {Bullock}  \&
  {Primack}}{{Kravtsov} et~al.}{1998}]{Kravtsov+1998}
{Kravtsov} A.~V.,  {Klypin} A.~A.,  {Bullock} J.~S.,   {Primack} J.~R.,  1998,
  \mn@doi [\apj] {10.1086/305884}, \href
  {http://adsabs.harvard.edu/abs/1998ApJ...502...48K} {502, 48}

\bibitem[\protect\citeauthoryear{{Lelli}, {McGaugh}  \& {Schombert}}{{Lelli}
  et~al.}{2016}]{Lelli+2016}
{Lelli} F.,  {McGaugh} S.~S.,   {Schombert} J.~M.,  2016, preprint, \href
  {http://adsabs.harvard.edu/abs/2016arXiv160609251L} {} (\mn@eprint {arXiv}
  {1606.09251})

\bibitem[\protect\citeauthoryear{{Mac Low} \& {Ferrara}}{{Mac Low} \&
  {Ferrara}}{1999}]{MacLow+1999}
{Mac Low} M.-M.,  {Ferrara} A.,  1999, \mn@doi [\apj] {10.1086/306832}, \href
  {http://adsabs.harvard.edu/abs/1999ApJ...513..142M} {513, 142}

\bibitem[\protect\citeauthoryear{{Moore}, {Governato}, {Quinn}, {Stadel}  \&
  {Lake}}{{Moore} et~al.}{1998}]{Moore+1998}
{Moore} B.,  {Governato} F.,  {Quinn} T.,  {Stadel} J.,   {Lake} G.,  1998,
  \mn@doi [\apjl] {10.1086/311333}, \href
  {http://adsabs.harvard.edu/abs/1998ApJ...499L...5M} {499, L5}

\bibitem[\protect\citeauthoryear{{Navarro}, {Eke}  \& {Frenk}}{{Navarro}
  et~al.}{1996a}]{NavarroEke+1996}
{Navarro} J.~F.,  {Eke} V.~R.,   {Frenk} C.~S.,  1996a, \mn@doi [\mnras]
  {10.1093/mnras/283.3.L72}, \href
  {http://adsabs.harvard.edu/abs/1996MNRAS.283L..72N} {283, L72}

\bibitem[\protect\citeauthoryear{{Navarro}, {Frenk}  \& {White}}{{Navarro}
  et~al.}{1996b}]{Navarro+1996}
{Navarro} J.~F.,  {Frenk} C.~S.,   {White} S.~D.~M.,  1996b, \mn@doi [\apj]
  {10.1086/177173}, \href {http://adsabs.harvard.edu/abs/1996ApJ...462..563N}
  {462, 563}

\bibitem[\protect\citeauthoryear{{Navarro}, {Frenk}  \& {White}}{{Navarro}
  et~al.}{1996c}]{Navarro+1996A}
{Navarro} J.~F.,  {Frenk} C.~S.,   {White} S.~D.~M.,  1996c, \mn@doi [\apj]
  {10.1086/177173}, \href {http://adsabs.harvard.edu/abs/1996ApJ...462..563N}
  {462, 563}

\bibitem[\protect\citeauthoryear{{Noordermeer}}{{Noordermeer}}{2006}]{Noordermeer+2006}
{Noordermeer} E.,  2006, PhD thesis, Groningen: Rijksuniversiteit

\bibitem[\protect\citeauthoryear{{Oh}, {de Blok}, {Brinks}, {Walter}  \&
  {Kennicutt}}{{Oh} et~al.}{2011}]{OH+2011}
{Oh} S.-H.,  {de Blok} W.~J.~G.,  {Brinks} E.,  {Walter} F.,   {Kennicutt} Jr.
  R.~C.,  2011, \mn@doi [\aj] {10.1088/0004-6256/141/6/193}, \href
  {http://adsabs.harvard.edu/abs/2011AJ....141..193O} {141, 193}

\bibitem[\protect\citeauthoryear{{Ostriker} \& {Peebles}}{{Ostriker} \&
  {Peebles}}{1973}]{Ostriker+1973}
{Ostriker} J.~P.,  {Peebles} P.~J.~E.,  1973, \mn@doi [\apj] {10.1086/152513},
  \href {http://adsabs.harvard.edu/abs/1973ApJ...186..467O} {186, 467}

\bibitem[\protect\citeauthoryear{{Randriamampandry} \&
  {Carignan}}{{Randriamampandry} \& {Carignan}}{2014}]{Toky+2014}
{Randriamampandry} T.~H.,  {Carignan} C.,  2014, \mn@doi [\mnras]
  {10.1093/mnras/stu100}, \href
  {http://adsabs.harvard.edu/abs/2014MNRAS.439.2132R} {439, 2132}

\bibitem[\protect\citeauthoryear{{Randriamampandry}, {Combes}, {Carignan}  \&
  {Deg}}{{Randriamampandry} et~al.}{2015}]{Randriamampandry+2015}
{Randriamampandry} T.~H.,  {Combes} F.,  {Carignan} C.,   {Deg} N.,  2015,
  \mn@doi [\mnras] {10.1093/mnras/stv2147}, \href
  {http://adsabs.harvard.edu/abs/2015MNRAS.454.3743R} {454, 3743}

\bibitem[\protect\citeauthoryear{{Sackett}}{{Sackett}}{1997}]{Sackett+1997}
{Sackett} P.~D.,  1997, \mn@doi [\apj] {10.1086/304223}, \href
  {http://adsabs.harvard.edu/abs/1997ApJ...483..103S} {483, 103}

\bibitem[\protect\citeauthoryear{{Salucci}}{{Salucci}}{2001}]{Salucci+2001}
{Salucci} P.,  2001, \mn@doi [\mnras] {10.1046/j.1365-8711.2001.04076.x}, \href
  {http://adsabs.harvard.edu/abs/2001MNRAS.320L...1S} {320, L1}

\bibitem[\protect\citeauthoryear{{Spano}, {Marcelin}, {Amram}, {Carignan},
  {Epinat}  \& {Hernandez}}{{Spano} et~al.}{2008}]{Spano+2008}
{Spano} M.,  {Marcelin} M.,  {Amram} P.,  {Carignan} C.,  {Epinat} B.,
  {Hernandez} O.,  2008, \mn@doi [\mnras] {10.1111/j.1365-2966.2007.12545.x},
  \href {http://adsabs.harvard.edu/abs/2008MNRAS.383..297S} {383, 297}

\bibitem[\protect\citeauthoryear{{Verheijen}}{{Verheijen}}{2001}]{Verheijen+2001}
{Verheijen} M.~A.~W.,  2001, \mn@doi [\apj] {10.1086/323887}, \href
  {http://adsabs.harvard.edu/abs/2001ApJ...563..694V} {563, 694}

\bibitem[\protect\citeauthoryear{{Weinberg} \& {Katz}}{{Weinberg} \&
  {Katz}}{2002}]{Weinberg+2002}
{Weinberg} M.~D.,  {Katz} N.,  2002, \mn@doi [\apj] {10.1086/343847}, \href
  {http://adsabs.harvard.edu/abs/2002ApJ...580..627W} {580, 627}

\bibitem[\protect\citeauthoryear{{de Blok} \& {Bosma}}{{de Blok} \&
  {Bosma}}{2002}]{Block+2002}
{de Blok} W.~J.~G.,  {Bosma} A.,  2002, \mn@doi [\aap]
  {10.1051/0004-6361:20020080}, \href
  {http://adsabs.harvard.edu/abs/2002A%26A...385..816D} {385, 816}

\bibitem[\protect\citeauthoryear{{de Blok}, {McGaugh}, {Bosma}  \& {Rubin}}{{de
  Blok} et~al.}{2001}]{Blok+2001}
{de Blok} W.~J.~G.,  {McGaugh} S.~S.,  {Bosma} A.,   {Rubin} V.~C.,  2001,
  \mn@doi [\apjl] {10.1086/320262}, \href
  {http://adsabs.harvard.edu/abs/2001ApJ...552L..23D} {552, L23}

\bibitem[\protect\citeauthoryear{{de Blok}, {Walter}, {Brinks}, {Trachternach},
  {Oh}  \& {Kennicutt}}{{de Blok} et~al.}{2008}]{Blok+2008}
{de Blok} W.~J.~G.,  {Walter} F.,  {Brinks} E.,  {Trachternach} C.,  {Oh}
  S.-H.,   {Kennicutt} Jr. R.~C.,  2008, \mn@doi [\aj]
  {10.1088/0004-6256/136/6/2648}, \href
  {http://adsabs.harvard.edu/abs/2008AJ....136.2648D} {136, 2648}

\bibitem[\protect\citeauthoryear{{de Denus-Baillargeon}, {Hernandez},
  {Boissier}, {Amram}  \& {Carignan}}{{de Denus-Baillargeon}
  et~al.}{2013}]{deDenus-Baillargeon+2013}
{de Denus-Baillargeon} M.-M.,  {Hernandez} O.,  {Boissier} S.,  {Amram} P.,
  {Carignan} C.,  2013, \mn@doi [\apj] {10.1088/0004-637X/773/2/173}, \href
  {http://cdsads.u-strasbg.fr/abs/2013ApJ...773..173D} {773, 173}

\bibitem[\protect\citeauthoryear{{de Jong} \& {van der Kruit}}{{de Jong} \&
  {van der Kruit}}{1995}]{Jong+1995}
{de Jong} R.~S.,  {van der Kruit} P.~C.,  1995, VizieR Online Data Catalog,
  \href {http://adsabs.harvard.edu/abs/1995yCat..41060451D} {410}

\bibitem[\protect\citeauthoryear{{van Albada}, {Bahcall}, {Begeman}  \&
  {Sancisi}}{{van Albada} et~al.}{1985}]{vanAlbada+1985}
{van Albada} T.~S.,  {Bahcall} J.~N.,  {Begeman} K.,   {Sancisi} R.,  1985,
  \mn@doi [\apj] {10.1086/163375}, \href
  {http://adsabs.harvard.edu/abs/1985ApJ...295..305V} {295, 305}

\makeatother
\end{thebibliography}
